 \documentclass{elsart}

\usepackage{graphicx}

\begin{document}


  
\begin{frontmatter}
   \title{Effect of Autonomous Driving on Traffic Breakdown in Mixed Traffic Flow: A Critical Mini-Review }
 
\author{Boris S. Kerner $^1$}

 \address{$^1$
Physics of Transport and Traffic, University Duisburg-Essen,
47048 Duisburg, Germany}


\maketitle

\begin{abstract}
In this mini-review, a critical analysis of the   effect of autonomous driving vehicles
on traffic breakdown in mixed traffic flow consisting of    randomly distributed human driving  
and autonomous driving vehicles is made.
Autonomous   vehicles based on classical (standard) adaptive cruise control 
 (ACC) in a vehicle
and on  an ACC in the framework of three-phase traffic theory
 (TPACC -- {\bf T}hree-traffic-{\bf P}hase ACC) introduced recently [Phys. Rev. E   97  (2018) 042303]
  are considered.    Due to the particular
 importance of characteristics of traffic breakdown (transition from free traffic flow to congested traffic)
for almost all approaches to traffic control and management in traffic networks, the basis of this critical review is a study of the effect of autonomous   vehicles on the probability of traffic breakdown and on stochastic highway capacity
in mixed traffic flow. 
We show that within a wide range of dynamic parameters  of classical ACC, the ACC-vehicles   
can deteriorate the traffic system considerably while initiating
	traffic breakdown and reducing highway capacity at a   bottleneck. Contrarily, in the same range of parameters of dynamic rules
of TPACC, the	TPACC-vehicles  either do not effect on traffic characteristics or sometimes 
	can even improve them. 
 To understand    physical reasons for the effect
	of classical ACC- and TPACC-vehicles
	on traffic breakdown, we introduce
	 a model of ACC that  can be considered a combination of   dynamic features of classical ACC and TPACC.
	With the use of this model,   we find how the amplitude of a local speed disturbance caused by the ACC
	in a vicinity of a bottleneck and the probability of traffic breakdown
	depend on the dynamic parameters of the ACC.
	To emphasize that the deterioration of the characteristics of mixed traffic flow through
classical ACC-vehicles is not associated with a well-known effect of   string instability
of platoons of autonomous vehicles,   we limit by a consideration of only such classical ACC-vehicles
whose platoon satisfies condition for string stability. 
\end{abstract}
\end{frontmatter}
\tableofcontents
\newpage

\section{Introduction
\label{Introduction}}

It is commonly assumed that future  vehicular traffic  
 is a mixed
  traffic flow consisting of random distributed human  driving and autonomous  (automated) driving vehicles
	(see, e.g.,~\cite{Ioannou,Ioannou1996,Ioannou2006,IoannouChien2002A,Levine1966A_Aut,Liang1999A_Aut,Liang2000A_Aut,Meyer2014A,Bengler2014A,Maurer2015A,Swaroop1996A_Aut,Swaroop2001A_Aut}, \cite{Varaiya1993A,Lin2009A,Martinez2007A,Brummelen2018A,AHS,AHS2,EuroAHS,GermanAHS,fail_Shladover1995A,Rajamani2012A_Aut,Davis2004B9,Davis2014C,Davis2015A_Int1}).
		There exist a large series of papers by the well-known and massive $\lq\lq$Automated Highway System'' 
	project involving the US government and a large number of 
	transportation researchers~\cite{AHS,AHS2}, EU projects~\cite{EuroAHS},
	and projects made in Germany~\cite{GermanAHS}.
 A consortium of researches   
	  all over the world performed extensive and pioneering research into autonomous and automated driving vehicle systems  (see   references to these extensive  research, for example,
	in reviews and books by Ioannou~\cite{Ioannou},
Ioannou and Sun~\cite{Ioannou1996},
Ioannou and Kosmatopoulos~\cite{Ioannou2006}, Shladover~\cite{fail_Shladover1995A},  Rajamani~\cite{Rajamani2012A_Aut},
Meyer and   Beiker~\cite{Meyer2014A},
 Bengler et al.~\cite{Bengler2014A}, and Van Brummelen et al.~\cite{Brummelen2018A}).
	
	An autonomous   driving vehicle is a self-driving vehicle
	that can move without a driver. Autonomous driving is realized through the use
 an automated  system in a vehicle: The automated system has control over the vehicle in traffic flow.
	For this reason, autonomous   driving vehicle
	is also called automated driving (or automatic driving) vehicle.
	It should be noted that in the engineering science the terms   
	{\it autonomous   driving}  
	and {\it automated   driving} are not synonyms.
	There are two reason for this. While an autonomous   driving vehicle should be able to
	move without a driver in the vehicle, there are several different levels of automation associated with
	automated driving. The levels  include, for example, a level of    $\lq\lq$conditional automation'' in which the driver must be present to provide any corrections when needed and a level of $\lq\lq$full automation''
	in which the automated vehicle system is in complete control of the vehicle and human presence is no longer needed. Additionally, in contrast with autonomous driving vehicle that moves fully autonomous from other vehicles, it is often assumed that  automated   driving
	can be supported by so-called cooperative driving 
	that can be realized through a diverse variety of cooperative automated systems  like
	vehicle-to-vehicle (V2V) communication (ad-hog vehicle networks) and 
	vehicle-to-infrastructure (V2X) communication.
	However,  to study  
	  dynamic strategies for future reliable autonomous   driving that should increase
		network capacity and traffic safety,
	in this article we limit a consideration of an automated vehicle system that
	is in complete control of the vehicle  as well as
	we assume that there are  no cooperative vehicle systems that can support automated driving.
	In other words, for the subject discussed in this article
	there is no difference  between the terms {\it autonomous driving} and
	 {\it automated driving}.

  As mentioned, autonomous  driving vehicles  
	should considerably
enhance highway  capacity. Highway capacity
 is limited by   traffic breakdown at road bottlenecks. Traffic breakdown is a transition from free flow at a bottleneck to congested traffic at the bottleneck (see, e.g., reviews and books~\cite{May,Manual2000,Manual2010,Helbing2001,Haight1963A,Gazis2002,Gartner,Gartner2,ElefteriadouBook2014_Int1},
\cite{Da,Sch,Brockfeld2003,Bellomo,Ferrara2018A,Leu,Mahnke,MahnkeKLub2009A,Wid,Wh2,Treiber_Int1,Schadschneider2011},
\cite{Saifuzzaman2015A,Pa1983,New,Nagel2003A,Nagatani_R,KernerBook,Kerner2009,Kerner2017A}).
It has been found that  highway capacity exhibits a stochastic nature~\cite{Elefteriadou1995A,Persaud1998B}:
At the same flow rate in free flow at a   bottleneck traffic breakdown can occur but it 
should not necessarily occur. In further empirical studies of this
probabilistic traffic breakdown~\cite{Elefteriadou1995A,Persaud1998B}, it has been found that empirical traffic breakdown exhibits the nucleation nature~\cite{Kerner1998C,Kerner1998B,Kerner1999B,Kerner1999A,Kerner1999C,Kerner2000,Kerner2001,Kerner2002A,Kerner2002B,Waves}:
Traffic breakdown can be induced by a time-limited localized congested pattern reaching a
highway bottleneck.  Because empirical traffic breakdown
in free flow  at a bottleneck is the probabilistic
phenomenon that exhibits the empirical nucleation nature,   
 the probability of traffic breakdown in free flow at the bottleneck is
 one of the main characteristics of the traffic stream.

  As known   
(see, e.g., reviews and 
books~\cite{May,Manual2000,Manual2010,ElefteriadouBook2014_Int1,KernerBook,Kerner2009,Kerner2017A}), most important features of traffic breakdown in free flow at an on-ramp bottleneck on a single-lane road  are qualitatively the same as those in highly heterogeneous traffic flow consisting of very different types of vehicles on multi-lane road with different types of road bottlenecks.   In particular, this conclusion
 is related to the empirical flow-rate dependence of the breakdown probability~\cite{ElefteriadouBook2014_Int1,Kerner2017A}. Therefore, to find the effect of different features of the dynamics of autonomous driving vehicles in  mixed traffic flow on the probability of traffic breakdown at a road bottleneck, it is sufficient to study  a simple case of  mixed vehicular traffic where traffic  consists only  of two types of vehicles
(human driving and autonomous driving vehicles) moving on a single-lane road with an on-ramp bottleneck.

On the single-lane road, no  vehicles can pass.
  For this reason, the effect of autonomous driving on traffic breakdown and highway capacity  
	can be understood through
	an analysis of  
	an adaptive cruise control (ACC) in a vehicle:
   An ACC-vehicle follows the preceding vehicle
	(that can be either
	a human driving vehicle or an  ACC-vehicle) automatically based on 
	some ACC dynamics rules of motion (see, 
	e.g.,~\cite{Ioannou,Ioannou1996,Ioannou2006,IoannouChien2002A,Levine1966A_Aut,Liang1999A_Aut,Liang2000A_Aut,Swaroop1996A_Aut,Swaroop2001A_Aut,Rajamani2012A_Aut,Davis2004B9,Davis2014C}).
	
	In~\cite{KernerPat1,KernerPat2,KernerPat3,KernerPat4} 
	the author introduced a strategy of ACC in the framework of the three-phase 
	theory
	called TPACC -- Three-traffic-Phase ACC   (for a review, see~\cite{Kerner2017E,Kerner2018C,Kerner2019C}). One of the most important features of
	TPACC is the existence of the indifference zone in car-following of the three-phase theory.
In this review article, we the use of 
a simple TPACC model~\cite{Kerner2018C,Kerner2019C} we will show that the TPACC strategy can exhibit the following important
advantages in comparison with the classical (standard) ACC strategy: 
\begin{description}
\item (i)
 The mean amplitude of speed disturbances
at a road bottleneck occurring through TPACC-vehicle can be considerably smaller than that 
introduced by classical  ACC-vehicles at the same model parameters.
\item  (ii) In mixed traffic flow with TPACC-vehicles   the probability of traffic breakdown at a road bottleneck
can be considerably smaller than 
in mixed traffic flow with classical ACC-vehicles. 
\end{description}
	We explain the physics of the improving of the traffic stream  through TPACC-vehicles.

The main objective of this review paper is a critical comparison between   autonomous   vehicles based on classical ACC
and on   TPACC.
Characteristics of traffic breakdown and stochastic highway capacity are particular  important   
for almost all approaches to traffic control and management in traffic networks. For this reason,
  the critical comparison between different dynamical characteristics of autonomous vehicles
is based on a study of the effect of autonomous   vehicles on the probability of traffic breakdown and on stochastic highway capacity
in mixed traffic flow  consisting of randomly distributed human driving  
and autonomous driving vehicles. 
 
   To make this critical analysis clear, in this review paper we introduce 
	 a model of ACC that  can be considered a combination of   dynamic features of classical ACC and TPACC.
	With the use of this model,   we find how the amplitude of a local speed disturbance caused by the ACC
	in a vicinity of a bottleneck and the probability of traffic breakdown
	depend on the dynamic parameters of the ACC.
	To emphasize that the deterioration of the characteristics of mixed traffic flow through
classical ACC-vehicles is not associated with a well-known effect of   string instability
of platoons of autonomous vehicles,   we limit by a consideration of only such classical ACC-vehicles
whose platoon satisfies condition for string stability.

		The article is organized as follows: First, we discuss the empirical nucleation nature
		  of traffic breakdown and associated stochastic highway capacity as well as
			the   consequences the empirical nucleation nature of traffic breakdown for the evaluation of performance of autonomous driving
in mixed traffic flow  through the use of traffic simulations 
		(Sec.~\ref{Capacity_S}). In this section we will also
		consider the reason for the failure of standard approaches for simulations of mixed traffic flow.
		Classical (standard) ACC strategy is the subject of Sec.~\ref{Classical_ACC_S}.
The strategy to autonomous driving
 in the framework of the three-phase theory called TPACC is discussed in Sec.~\ref{TPACC_St_S}.
The effect of   classical ACC and TPACC on traffic breakdown   at a bottleneck is studied in Sec.~\ref{Prob_S}.
The dependence of characteristics of traffic breakdown on   time headway of classical ACC and TPACC  
 is the subject of Sec.~\ref{ACC_Cl_Param_Prob_S}.
The influence  of dynamic rules of autonomous driving on speed disturbances   at the bottleneck
is considered in Sec.\ref{Dyn_Rules_S}.
The effect of platoons of autonomous driving vehicles
 on the probability of traffic breakdown in mixed traffic flow is discussed in Sec.~\ref{Prob_Platoon_S}.
 Traffic stream flow characteristics of mixed traffic flow are discussed in Sec.~\ref{Stream_S}.
	In discussion  (Sec.~\ref{Dis_S}), we formulate paper conclusions  (Sec.~\ref{Conl_S}), consider
	the applicability of the TPACC model for a reliable analysis of  
	some features of future autonomous driving in mixed traffic flow
	    (Sec.~\ref{Value}) and	discuss a question
			whether vehicular traffic in networks consisting of  only autonomous vehicles is real option in the future
			(Sec.~\ref{Mixed_S}). 
	In Appendix~\ref{KKl_Model_Ap}, we present  
	the Kerner-Klenov stochastic microscopic three-phase model   for  
  human driving vehicles~\cite{KKl,KKl2003A,KKl2009A}
  used for all simulations of   mixed traffic flow made in the paper, in Appendix~\ref{Cla_ACC_S}
	we explain simulations of the classical ACC model,
	in Appendix~\ref{App_TPACC_Model} we consider a discrete TPACC model used for simulations of 
	TPACC model, in Appendix~\ref{Models_ACC_TP_Sec} we present a discrete version of 
	a ACC-model that can be considered a combination of the classical ACC- and TPACC-models, in  Appendix~\ref{Models_Bott_Sec}
	we consider a model of vehicle merging at an on-ramp bottleneck,
	and in Appendix~\ref{Boundary_Ini_Con_S} we consider
	boundary   conditions used for simulations of mixed traffic flow.
	
 \section{Nucleation Nature of Traffic Breakdown and Its Consequences for Evaluation of Performance of Autonomous Driving
in Mixed Traffic Flow  
\label{Capacity_S}}

\subsection{Three-Phase Traffic Theory}

Three-phase traffic theory    
 is a  framework for understanding of states of {\it empirical} traffic flow
 in three traffic phases~\cite{KernerBook,Kerner2017A,Kerner1998C,Kerner1998B,Kerner1999B,Kerner1999A,Kerner1999C,Kerner2000,Kerner2001,Kerner2002A,Kerner2002B}:
\begin{description}
\item 1. Free flow (F).
\item 2. Synchronized flow (S).
\item 3. Wide-moving jam (J).
\end{description}
The 
synchronized flow  and   wide-moving jam traffic phases belong to congested traffic.

The synchronized flow      
   and wide moving jam  phases in congested traffic are defined through the use of the following {\it macroscopic  empirical   criteria} [J] and [S]  for, respectively, the wide-moving jam 
	and the synchronized flow   traffic phases in congested traffic~\cite{KernerBook,Kerner2009,Kerner2017A,Kerner2018D,Kerner2018A,Kerner2017B,Kerner_Review_Int1,Kerner_Review2_Int1,Kerner_Review3_Int1}:
  \begin{itemize}
  \item {[J]} 
  The   wide moving jam  phase 
  is defined as follows. A wide moving jam is a moving jam   that maintains the mean velocity of the downstream  front of the jam  as the jam propagates. 
 Vehicles accelerate
  within the downstream jam front   from low speed states
  (sometimes as low as zero) inside
  the jam to higher speeds downstream of the jam.
  A wide moving jam maintains the mean velocity of the downstream jam front, even as it propagates through other different (possibly very complex) traffic states of free flow and synchronized flow or highway bottlenecks.  
  This is a characteristic feature of the  wide moving jam phase. 
  \item {[S]}
  The synchronized flow
   phase
  is defined as follows. In contrast to the
   wide moving jam
  traffic phase, the  downstream front of the
   synchronized flow  phase
  does {\it not} maintain the mean velocity  of the downstream front. In particular,
  the downstream front of synchronized flow   is often {\it fixed} at a   
   bottleneck.
  In other words, the
    synchronized flow traffic phase does not show the characteristic feature [J] of the wide moving jam  phase. 
		\item	The phase definitions [S] and [J] mean that if in a set of empirical traffic data
   congested traffic states associated with  the wide moving jam traffic 
	phase have been identified through the use of the criterion [J]\footnote{If the set of empirical data is related to microscopic traffic data in which single vehicle speeds and time headway between vehicles have been measured 
	(the data is often called  as single vehicle data), then the wide moving jam traffic 
	phase in the microscopic data can be identified through a microscopic criterion for the   wide moving jam traffic 
	phase that can be found in Sec.~2.6 of the book~\cite{Kerner2009}.  
After    the wide moving jam traffic 
	phase has been identified   in the microscopic empirical traffic data of congested traffic,
	then with certainty all  remaining congested states in the empirical microscopic  data
	of congested traffic   are related to the synchronized flow phase~\cite{Kerner2009,KKH,KKH1,KKHR}.
	}, then with certainty all  remaining congested states in the empirical data set are related to the synchronized flow phase.
  \end{itemize}
The downstream front of synchronized flow separates synchronized flow upstream from free flow downstream.
  Within the downstream front of  synchronized flow vehicles accelerate from lower speeds in synchronized flow upstream of the front to higher speeds in free flow downstream of the front.

\subsection{Empirical Nucleation Nature of Traffic Breakdown  
\label{Capacity_Sub}}

In all empirical data measured
over years in different countries, traffic breakdown
at a highway bottleneck is a phase transition from free flow (F) to the
synchronized flow
phase (S) of congested traffic (F$\rightarrow$S transition).
 Therefore, traffic breakdown at a highway bottleneck and 
an F$\rightarrow$S transition at the bottleneck are synonyms.

\begin{figure}
\begin{center}
\includegraphics*[scale=.6]{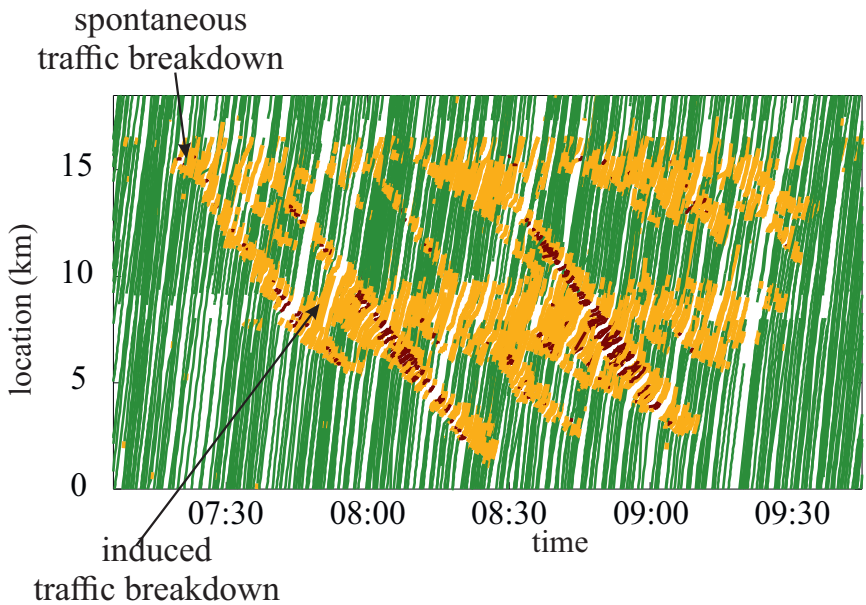}
\caption{(Color  online) Empirical spontaneous and empirical induced
traffic breakdown (F$\rightarrow$S transition) with complex pattern formations in probe vehicle data:
(a) Schema of highway section;
there are two bottlenecks  
 at which
	traffic breakdown is   observed
	(explanations of these bottlenecks can be found in Secs.~2.3.2 and~9.2.1 of~\cite{KernerBook}): (i) A downstream bottleneck
	caused by an off-ramp within an intersection $\lq\lq$Nordwestkreuz Frankfurt" 
	(effective bottleneck location at about 15 km); (ii) an upstream bottleneck caused by 
	on-ramps within at an intersection $\lq\lq$Bad Homburger Kreuz"
	(effective bottleneck location at about 9 km). (b) Overview of phase transitions with pattern formation;
  green, yellow, and red parts on trajectories in (b) are related to   the
free flow, synchronized flow, and wide moving jam traffic phases.  To distinguish between the three traffic phases,
a model for the detection of phase transition points along  a vehicle trajectory of Ref.~\cite{KernerRSch201310_nu}
has been applied;  parameters of the model, in which some speed and time thresholds for the different traffic phases
have been used,
  are related to Table 2 of Ref.~\cite{KernerRSch201310_nu}. 
Adapted from~\cite{Kerner2019AA}.
\label{Empirical_PRE_10} } 
\end{center}
\end{figure}
 
Probably the most important empirical feature of the synchronized flow is that
the F$\rightarrow$S transition (traffic breakdown)
 exhibits the nucleation nature~\cite{Kerner1998C,Kerner1998B,Kerner1999B,Kerner1999A}\cite{Kerner1999C,Kerner2000,Kerner2001,Kerner2002A,Kerner2002B}. This means that  traffic breakdown occurs    in a metastable   free flow
with respect to an  F$\rightarrow$S transition at the bottleneck.  Indeed, empirical data shows that
 empirical traffic breakdown (F$\rightarrow$S transition) at a highway bottleneck can be either {\it spontaneous}
 or {\it induced} traffic breakdown.
\begin{itemize}
\item  As shown and proven in~\cite{KernerBook,Kerner2009,Kerner2017A,Waves},
the occurrence of empirical induced traffic breakdown
at the bottleneck is the empirical proof of the empirical nucleation nature of traffic breakdown
(F$\rightarrow$S transition).
\end{itemize}
An empirical example of both spontaneous
and induced traffic breakdowns occurring at different bottlenecks is shown in Fig.~\ref{Empirical_PRE_10}.

\subsection{Explanation of Traffic Breakdown Nucleation in Three-Phase Traffic Theory \label{Simp_Ex_Sec}}

 The main reason of the three-phase theory is the explanation of the empirical nucleation nature of traffic breakdown (F$\rightarrow$S transition) at the bottleneck. To reach this goal, in congested traffic a new traffic phase called synchronized flow has been introduced~\cite{KernerBook,Kerner2009,Kerner2017A,Kerner2018D,Kerner2018A,Kerner2017B,Kerner_Review_Int1,Kerner_Review2_Int1,Kerner_Review3_Int1}. The basic feature of the synchronized flow traffic phase formulated in the three-phase theory leads to the nucleation nature of the F$\rightarrow$S transition. 
In this sense, the synchronized flow traffic phase, which ensures the nucleation nature of the F$\rightarrow$S transition at a highway bottleneck, and the three-phase traffic theory can be considered synonymous.

In the three-phase traffic theory, the empirical nucleation nature of traffic breakdown
is associated with the metastability of free flow with respect to traffic breakdown
 (F$\rightarrow$S transition) at a highway bottleneck.  
The metastability of free flow with respect to traffic breakdown
  (F$\rightarrow$S transition) is as follows~\cite{KernerBook,Kerner2017A,Kerner1998C,Kerner1998B,Kerner1999B,Kerner1999A,Kerner1999C,Kerner2000,Kerner2001,Kerner2002A,Kerner2002B,Waves}: 
There can be many speed (density, flow rate)
disturbances in free flow at the bottleneck.
Amplitudes of the disturbances
 can be very different. When a disturbance occurs randomly whose
 amplitude is larger than a critical one,
then traffic breakdown occurs. Such a disturbance resulting in traffic breakdown is
called {\it nucleus} for the breakdown. Otherwise, if the disturbance amplitude is smaller than the critical
one, the disturbance decays. As a result, no traffic breakdown occurs.

As emphasized
 in~\cite{Kerner2017A,Kerner2018D,Kerner2018A,Kerner2017B,Kerner_Review_Int1,Kerner_Review2_Int1,Kerner_Review3_Int1}, the   metastability of free flow 
  with respect to traffic breakdown
	(F$\rightarrow$S transition)  at a highway bottleneck   
	has   no   sense for    standard  traffic and transportation theories.
	Reviews of the  standard  traffic and transportation theories
	can be found in~\cite{May,Manual2000,Manual2010,Helbing2001,Haight1963A,Gazis2002,Gartner,Gartner2}, \cite{ElefteriadouBook2014_Int1,Da,Sch,Brockfeld2003,Bellomo,Ferrara2018A,Leu,Mahnke,MahnkeKLub2009A,Wid,Wh2}, \cite{Treiber_Int1,Schadschneider2011,Saifuzzaman2015A,Pa1983,New,Nagel2003A,Nagatani_R,Ashton1966,Drew1968}. The criticism of the  standard  traffic and transportation theories
	has been made in reviews and
	books~\cite{KernerBook,Kerner2009,Kerner2017A,Kerner2018D,Kerner2018A,Kerner2017B,Kerner_Review_Int1,Kerner_Review2_Int1,Kerner_Review3_Int1}. 
	In particular, in~\cite{Kerner2017A,Kerner2018D,Kerner2018A,Kerner2017B},
	\cite{Kerner_Review_Int1,Kerner_Review2_Int1,Kerner_Review3_Int1}
			it has been shown that the three-phase theory is incommensurable with 
all earlier traffic flow theories and models.
	The term $\lq\lq$incommensurable"  has been introduced by Kuhn~\cite{Kuhn2012}    to explain a paradigm shift in a scientific field.

\subsection{Understanding Stochastic Highway Capacity \label{Under_C_Sec}}

\begin{figure}
\begin{center}
\includegraphics*[scale=.7]{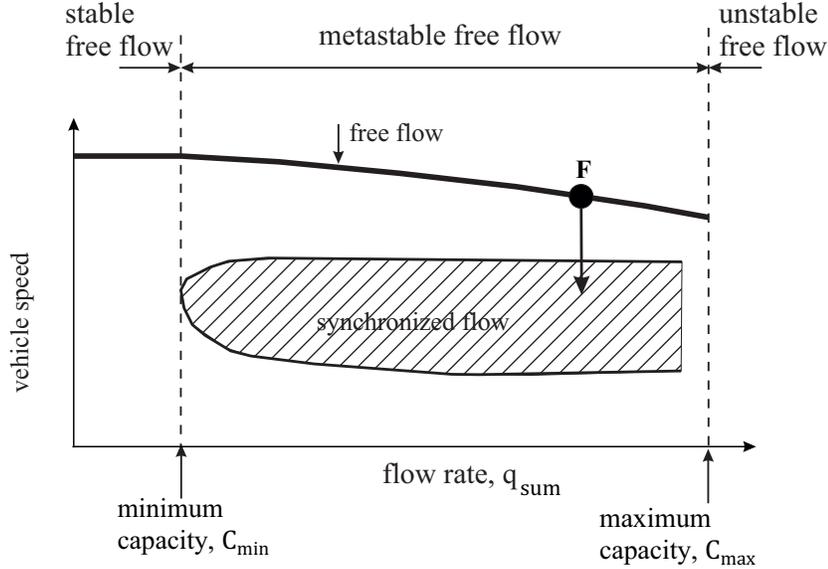}
\caption{Qualitative explanation of the empirical metastability of
free flow    with respect to 
traffic breakdown  (F$\rightarrow$S transition) at   bottleneck in
three-phase theory~\cite{KernerBook,Kerner2017A,Kerner1998C,Kerner1998B,Kerner1999B,Kerner1999A,Kerner1999C,Kerner2000,Kerner2001,Kerner2002A,Kerner2002B}.  Qualitative Z-speed--flow-rate characteristic for traffic breakdown;
 F -- free flow, S   -- synchronized flow (two-dimensional (2D) hatched region). Arrow F$\rightarrow$S illustrates
symbolically one of possible F$\rightarrow$S transitions 
occurring in a metastable state of free flow when a nucleus
appears in the free flow at the bottleneck.
Adapted from~\cite{Kerner2017A}.
\label{Breakdown_nuc_simple2} } 
\end{center}
\end{figure}

 One of the most important   
 consequences of the empirical   nucleation nature of   traffic breakdown at a highway bottleneck
is as follows~\cite{KernerBook,Kerner2017A}:  
 The metastability of free flow
with respect to traffic breakdown (F$\rightarrow$S transition) is realized within a flow rate range 
 (Fig.~\ref{Breakdown_nuc_simple2})
\begin{equation}
C_{\rm min} \leq q_{\rm sum} < C_{\rm max},
 \label{Cap_q_F_Int_Met}
 \end{equation}
where $q_{\rm sum}$ is the flow rate in free flow at a highway bottleneck,
$C_{\rm min}$ is a minimum highway capacity, $C_{\rm max}$
is a maximum highway capacity.

	This conclusion of empirical observations of
traffic breakdown
(F$\rightarrow$S transition)
 at the bottleneck
  leads  to the following definition of 
 stochastic highway capacity of free flow at a highway bottleneck made  in the three-phase 
theory~\cite{KernerBook}:
  \begin{itemize}
	\item {\it At any time instant}, there are the infinite
number of  highway capacities $C$ of free flow at the bottleneck. 
The range of the infinite number of  highway capacities 
 is limited by the minimum highway capacity $C_{\rm min}$
 and the maximum highway capacity $C_{\rm max}$  (Fig.~\ref{Breakdown_nuc_simple2}):
   \begin{equation}
 C_{\rm min}\leq C \leq C_{\rm max},
 \label{meta_F_QFS}
\end{equation}
where
 $C_{\rm min}< C_{\rm max}$.  
The existence of an infinite number of highway capacities   at any time instant  means that highway capacity is stochastic.  
\end{itemize}
 
 The physical sense of this capacity definition is as follows.
 Highway capacity is limited by traffic breakdown (F$\rightarrow$S transition)
 in an initial free flow at a highway bottleneck. In other words,
  any   flow rate $q_{\rm sum}$ in free flow at the bottleneck  at which traffic breakdown  can occur is highway capacity.  At any time instant, there are the infinite number of such highway capacities
 $C=q_{\rm sum}$
at which traffic breakdown {\it can} occur. These capacities satisfy 
conditions (\ref{meta_F_QFS}).   Thus, in the three-phase theory
 any flow rate $q_{\rm sum}$
 in free flow at a highway bottleneck that satisfies conditions
 \begin{equation}
 C_{\rm min}\leq q_{\rm sum}\leq C_{\rm max}
 \label{Cap_q_F_Int_in}
 \end{equation}
 is equal to one of the stochastic highway capacities of free flow at the bottleneck
          (Fig.~\ref{Breakdown_nuc_simple2}).
					A more detailed consideration of the nucleation nature of traffic breakdown and
	stochastic (probabilistic)
	features of the infinite number of the highway capacities
	can be found in~\cite{Kerner2017A,Kerner2018A,Kerner2017B}.
 
	\subsection{Paradigm Shift in Traffic and Transportation Science}

The fundamental change in the meaning of stochastic highway capacity made in the three-phase traffic theory can be considered
the paradigm shift in traffic and transportation science. This is because the meaning of highway capacity is the basis for the development of   methods for traffic control, management, and organization of a traffic network as well as ITS-applications.

The paradigm of standard traffic and transportation theories can be formulated as follows:  At a given time instant there is a value of highway capacity. This is also true when it is assumed that at any time instant there is a stochastic value of highway capacity. This means that when the flow rate at a bottleneck exceeds the capacity value at this time instant, traffic breakdown must occur at the bottleneck. This classical understanding of highway capacity is the basis for   standard methods for traffic control, management, and organization of a traffic network as well as ITS-applications (see  results of
standard traffic and transportation theories as well as  some results
 of their ITS-applications, for example, in
  reviews and books~\cite{May,Manual2000,Manual2010,Helbing2001,Haight1963A,Gazis2002,Gartner,Gartner2,ElefteriadouBook2014_Int1,Da,Sch,Brockfeld2003,Bellomo}, \cite{Ferrara2018A,Leu,Mahnke,MahnkeKLub2009A,Wid,Wh2,Treiber_Int1,Schadschneider2011}, \cite{Saifuzzaman2015A,Pa1983,Newell1963,New,Nagel2003A,Nagatani_R}, \cite{Ashton1966,Drew1968,Prigogine1971,Gerlough1975,Bell1997,Sheffi1984,Ran1996,Mannering1998,Brackstone1999,Mahmassani2001,Peeta2001}, \cite{Shvetsov2003,Maerivoet2005,Papageorgiou2008,Gartner2009,Rakha2009,Piccoli2009,DTA2011,Roess2014,Friesz2016,Hegyi2017}, \cite{Seo2017,Kachroo2018,Kessels2019,MATSim_Nagel2016,Barcelo2010,TRANSIMS,TRANSIMS_USA}).

In accordance with the understanding of stochastic highway capacity made in 
  the three-phase traffic theory~\cite{KernerBook,Kerner2017A,Kerner1998C,Kerner1998B,Kerner1999B,Kerner1999A,Kerner1999C,Kerner2000,Kerner2001,Kerner2002A,Kerner2002B} and briefly reviewed in Sec.~\ref{Under_C_Sec},
the new paradigm in traffic and transportation science can been formulated as follows: 
\begin{itemize}
\item {\it At any time instant} there is a range 
of the flow rate (\ref{Cap_q_F_Int_Met})
within which traffic breakdown  at the bottleneck can occur with some probability   but traffic breakdown does not
 necessarily occur (Fig.~\ref{Breakdown_nuc_simple2}). 
\end{itemize}

Thus, in contrast with the standard traffic theories, in the three-phase traffic theory
 there is a range of highway capacity values between a minimum and a maximum highway capacity 
at any time instant (Fig.~\ref{Breakdown_nuc_simple2}). When the flow rate at a bottleneck is inside this capacity range related to this time instant, traffic breakdown can occur at the bottleneck only with some probability, i.e., in some cases traffic breakdown occurs, in other cases traffic breakdown does not occur. The existence at any time instant of the infinite number of highway capacity values within the capacity range means that highway capacity is stochastic. Both the minimum and maximum highway capacities are also stochastic values. We can make the following conclusions:

\begin{itemize}
\item
This basic change in the meaning of stochastic
 highway capacity results from the empirical nucleation nature of traffic breakdown at a bottleneck.
\item 
Three-phase traffic theory is a theoretical framework that explains stochastic highway capacity by a capacity range
that exists at any time instant.  When at a time instant the flow rate at the bottleneck is inside the capacity range related to this time instant, traffic breakdown can occur with some probability    but traffic breakdown does not
 necessarily occur.
\end{itemize} 
To be consistent with the empirical nucleation nature of traffic breakdown, other methods (in comparison with the methods developed in accordance with the standard meaning of highway capacity) for traffic control and   management in traffic networks, other models for simulations of mixed traffic flow   as well as other models for the evaluation of ITS-applications should be developed.

	\subsection{Driver Behavioral Assumption of
	 Traffic Breakdown Nucleation in Three-Phase Traffic Theory \label{Simp_Dr_Sec}}
	
\begin{figure} 
\begin{center}
\includegraphics[scale=.6]{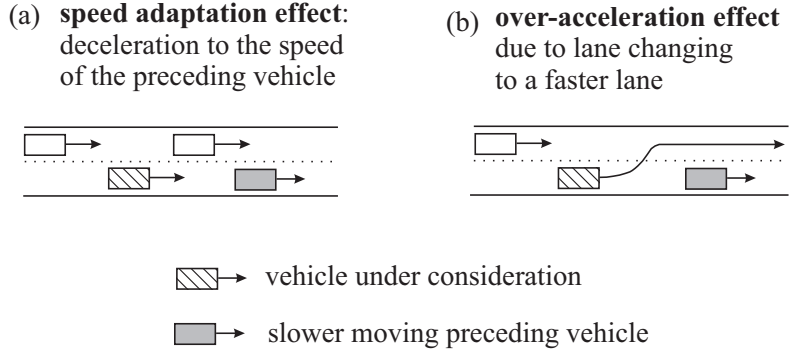}
\caption{Qualitative explanation of driver speed adaptation (a) and driver over-acceleration through the lane changing (b).
}
\label{Adap_OverAccel}
\end{center}
\end{figure}

 To answer a  question about a driver behavior that is responsible for
  the empirical nucleation nature of traffic breakdown (F$\rightarrow$S transition) at a highway bottleneck, we should mention that
when a driver approaches a slower moving preceding vehicle and the driver cannot immediately pass the slow vehicle,
the driver must decelerate to the speed of the slow moving preceding vehicle. This  driver deceleration can be called
{\it driver speed adaptation}  (Fig.~\ref{Adap_OverAccel} (a)). To    escape
  from this car-following  of the slow moving preceding vehicle, the driver searches for the opportunity to accelerate. We call    vehicle 
acceleration from the  car-following of the slow moving preceding vehicle as  {\it    
 over-acceleration}.

\begin{figure} 
\begin{center}
\includegraphics[scale=.6]{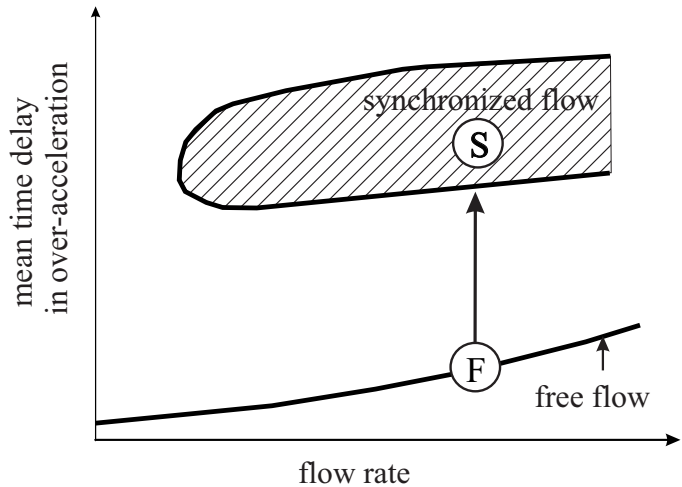}
\caption{Hypothesis of three-phase traffic theory about the discontinuous character of
the mean time delay in   over-acceleration. F is a state of free flow, S is a state of synchronized flow.
}
\label{Adap}
\end{center}
\end{figure}

To simplify a qualitative explanation of 
     over-acceleration, we  consider a vehicle approaching a slow preceding
	vehicle  moving in the right line of a multi-lane road\footnote{In general,     over-acceleration is a   vehicle  maneuver leading to a higher speed from initial car-following of  a slow moving preceding vehicle.
	The   over-acceleration is possible both on a single-lane and multi-lane roads.
	A mathematical model  of the   over-acceleration on a single-lane road 
	incorporated in the Kerner-Klenov model~\cite{KKl,KKl2003A,KKl2009A} used for all simulations in this paper 
	has been explained  in Sec.~5.10 of 
 the book~\cite{Kerner2017A}.}. If the vehicle under consideration cannot immediately pass the slow moving vehicle, the vehicle 
	must decelerate to a lower speed of the preceding vehicle
		($\lq\lq$speed adaptation" in Fig.~\ref{Adap_OverAccel} (a)).
 On the multi-lane road,    over-acceleration
leading to the vehicle escaping from the car-following  is often possible through lane changing to a faster lane
with the subsequent passing of the slow moving vehicle (Fig.~\ref{Adap_OverAccel} (b)). 
However,   after the driver has begun the speed adaptation to the speed of the slow moving preceding vehicle
(Fig.~\ref{Adap_OverAccel} (a))
there can be a waiting time    before the lane changing maneuver leading to vehicle acceleration with the subsequent
passing of the slow moving vehicle is successful. We call
this waiting time   as a {\it time delay in   over-acceleration}.

The vehicle density in   synchronized flow    is larger than
the density is in free flow at the same flow rate.
We can assume that the larger the vehicle density, the more difficult to 
 escape   from the car-following  of the slow moving preceding vehicle through   over-acceleration. Respectively,
  the larger the vehicle density, the longer should be the {\it mean} time delay in   over-acceleration.
Through a large synchronized flow density, vehicles prevent each other to accelerate to a higher speed.
Such a strong bunching of vehicle motion distinguishes synchronized flow from free flow: In free flow,   vehicle bunching is weak   and, therefore,
vehicles can much easier  
escape from the car-following of the slow moving preceding vehicle.
This qualitative consideration leads to the assumption
made in the three-phase traffic theory 
 about   the {\it discontinuous character} of the   mean  time delay in   over-acceleration (Fig.~\ref{Adap}): In synchronized flow, the   mean  time delay in  
 over-acceleration should be considerably longer than
it is in free flow.

Therefore, during traffic breakdown (F$\rightarrow$S transition) at a bottleneck
there should be a jump in the   mean  time delay in  over-acceleration from
a relatively short     mean  time delay in  over-acceleration in free flow to a considerably longer
 mean  time delay in  over-acceleration in synchronized flow
(up-arrow in the mean  time delay in   over-acceleration
 in Fig.~\ref{Adap}).
The assumption about the discontinuous character
of the the mean time delay in    over-acceleration (Fig.~\ref{Adap}) 
is consequent with the discontinuous character of the speed--flow-rate dependence
(Fig.~\ref{Breakdown_nuc_simple2}).

\begin{figure} 
\begin{center}
\includegraphics[scale=.8]{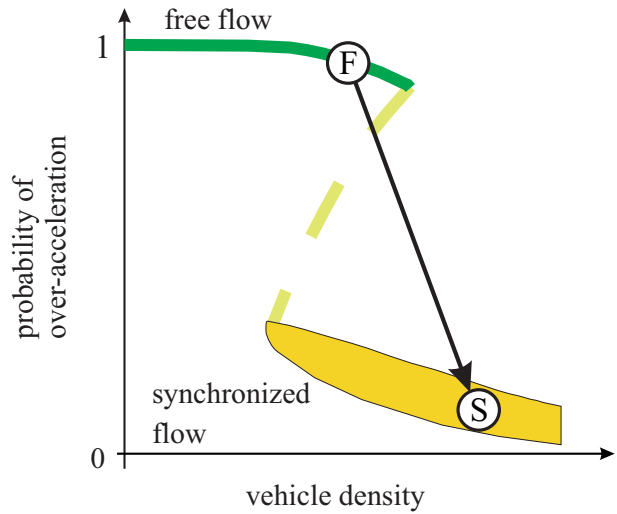}
\caption{Hypothesis of three-phase traffic theory about the discontinuous character of
the probability of   over-acceleration. F is a state of free flow, S is a state of synchronized flow.
Down-arrow shows qualitatively a drop in the probability  of   over-acceleration  
caused by traffic breakdown (F$\rightarrow$S transition) at a bottleneck.
}
\label{2Z}
\end{center}
\end{figure}

The hypothesis about the discontinuous character of the   mean  time delay in   over-acceleration
(Fig.~\ref{Adap}) is equivalent to  the hypothesis about  the discontinuous character of the probability that    
    over-acceleration is realized during a given time interval 	(probability  
 of   over-acceleration for short) (Fig.~\ref{2Z})~\cite{KernerBook,Kerner2009,Kerner2017A}.
Indeed,  the probability  
 of   over-acceleration 
 is the larger, the shorter the mean time delay in   over-acceleration (Fig.~\ref{Adap}). Therefore,
  traffic breakdown (F$\rightarrow$S transition) at the bottleneck that leads to a
  jump in the   mean  time delay in  over-acceleration (up-arrow  
 in Fig.~\ref{Adap}) is also accompanied by a drop in the probability  
 of   over-acceleration    (down-arrow in Fig.~\ref{2Z}).

It should be emphasized that the time delay in   over-acceleration should not be confounded with the well-known
driver reaction time.  
Indeed, for a safety lane changing a driver should
wait for the occurrence of a long enough (safety) time headway between two following each other vehicles in the neighborhood  
lane to which
the driver would like to change. Time headway between 
 vehicles in the neighborhood   lane do not depend on 
   the    reaction time of the driver that wants to change to this lane. However, the   waiting time for the safety lane changing
depends on time headway between 
 vehicles in the neighborhood faster  lane. In   turn, this   waiting time is  
the   time delay in   over-acceleration through lane changing to a faster lane (Fig.~\ref{Adap_OverAccel} (b)). 
\begin{itemize}
\item
Thus, we can conclude that even if the driver reaction time were negligible short,
the mean time delay in   over-acceleration  
is a finite value that exhibits the discontinuous character qualitatively   shown in Fig.~\ref{Adap}.
\end{itemize}

 We can also assume that the  discontinuous behavior of the mean
 time delay in   over-acceleration through lane changing to a faster lane shown in Fig.~\ref{Adap} can   remain for
autonomous driving vehicles, for which (in contrast to human driving vehicles)
the   reaction time on an unexpected deceleration of the preceding vehicle
	or on a sudden reduction in the spacing due to a merging vehicle can theoretically be made as short as zero.
Indeed, as for human driving vehicles rather than the  reaction time of an autonomous driving vehicle, 
values of
time headway between 
 vehicles in the neighborhood faster  lane determine the time delay in  over-acceleration
 for the safety lane changing
of  the autonomous driving vehicle ($\lq\lq$vehicle under consideration" in Fig.~\ref{Adap_OverAccel} (b)).

\subsection{Main Prediction of   Three-Phase Traffic Theory}
	
 The main prediction of the three-phase traffic theory is as follows.
\begin{itemize}
\item There is an S$\rightarrow$F instability in synchronized flow. 
The S$\rightarrow$F instability is a growing wave
of a local increase in the speed in synchronized flow. The uninterrupted growth of this wave
 leads to a transition from synchronized flow to free flow.
 The S$\rightarrow$F instability exhibits the nucleation nature: Only a large enough initial local 
increase in the speed in synchronized flow can lead to the
S$\rightarrow$F instability, whereas local disturbances of small enough local speed increase in synchronized flow decay. 
\item
The nucleation nature
	of the S$\rightarrow$F instability    governs
 the   metastability of free flow 
  with respect to 
 the   F$\rightarrow$S transition  at the bottleneck.
In its turn, the   metastability of free flow 
  with respect to 
 the   F$\rightarrow$S transition  at the bottleneck  explains the
empirical nucleation nature of traffic breakdown~\cite{KernerBook,Kerner2015C}.
 \end{itemize}

To explain the main prediction of the three-phase theory, we assume that traffic breakdown (F$\rightarrow$S transition) 
has occurred at a highway bottleneck. Due to this traffic breakdown, synchronized flow emerges at the bottleneck.
We assume that the downstream front of the synchronized flow is fixed at the bottleneck.
 In the synchronized flow,  many local speed disturbances with very different disturbance amplitudes can occur.

In the three-phase theory it has been proven that due to a finite time delay in    over-acceleration 
 discussed in Sec.~\ref{Simp_Dr_Sec}, a local speed increase within one of the
  local speed disturbances in synchronized flow     
can randomly begin to grow over time. As a result,  
 a growing wave of a local speed increase in synchronized flow appears. The development of such a traffic flow
  instability leads to the returning of free flow   at the bottleneck. This  instability
	is called as an S$\rightarrow$F instability~\cite{Kerner2015C}.

  Within a    speed wave of a local increase in  the  speed of synchronized flow,
there is 
a spatiotemporal competition between 
  over-acceleration and   speed adaptation. It has been found that
	a   growing  wave of a local speed increase in synchronized flow (S$\rightarrow$F instability) is realized
only  when within the   speed wave 
over-acceleration overcomes      speed adaptation. This can only be possible
 when the  amplitude  of a local speed increase in synchronized flow
is large enough. In contrast, when the amplitude of a  
local speed increase in synchronized flow is small enough,  a dissolving   wave of a local increase in the speed
in synchronized flow is realized. 
In other words, the S$\rightarrow$F instability exhibits the nucleation nature. 
  
 It must be emphasized that
	the S$\rightarrow$F instability~\cite{Kerner2017A,Kerner2015C} should not be confounded with the well-known 
  classical traffic flow instability introduced in 1950s--1960s~\cite{Chandler,Gazis1961A10,Herman1959,GH,KS1,KS2,KS3,Newell1961} and incorporated   in a huge number of traffic flow 
models (see, e.g., explanations of classical traffic flow instability and references in reviews and books~\cite{Helbing2001,Haight1963A,Gazis2002,Gartner,Gartner2,ElefteriadouBook2014_Int1,Sch,Brockfeld2003}, \cite{Leu,Wid,Treiber_Int1,Schadschneider2011,Saifuzzaman2015A}, \cite{Nagel2003A,Nagatani_R,Ashton1966,Drew1968}). Indeed,  the classical traffic flow instability
 leads to a growing wave of a local {\bf  decrease} in the speed in 
traffic flow 
(see, e.g.,~\cite{Helbing2001,Haight1963A,Gazis2002,Gartner,Gartner2,ElefteriadouBook2014_Int1,Sch,Brockfeld2003}, \cite{Leu,Wid,Treiber_Int1,Schadschneider2011,Saifuzzaman2015A}, \cite{Nagel2003A,Nagatani_R,Ashton1966,Drew1968}, \cite{Chandler,Gazis1961A10,Herman1959,GH,KS1,KS2,KS3,Newell1961,New2002}). In contrast with the classical traffic flow instability, the S$\rightarrow$F instability introduced in the three-phase traffic theory~\cite{Kerner2017A,Kerner2015C} leads to a growing wave of a local {\bf  increase} in the speed in synchronized flow.
	  
\subsection{Failure of Standard Approaches for Simulations of Mixed Traffic Flow \label{Failure_S}}

It should be mentioned that
the effect of  autonomous  (automated) vehicles  on mixed traffic flow   was intensively   considered already in 1990s-2000s 
in the works by Dharba and  Rajagopal~\cite{Dharba1999A},
Marsden {\it et al.}~\cite{Marsden2001A},
VanderWerf {\it et al.}~\cite{VanderWerf2001A,VanderWerf2002A},
Treiber and Helbing~\cite{TreiberH2001A},
Li {\it et al.}~\cite{Shrivastava2002A},
Kukuchi {\it et al.}~\cite{Kukuchi2003A},
Bose and   Ioannou~\cite{BoseIoannou2003A},
 Suzuki~\cite{Suzuki2003A}, 
Davis~\cite{Davis2004B9}, 
  Zhou and  Peng~\cite{Zhou2005A},
 van Arem {\it et al.}~\cite{vanArem2006},
Kesting {\it et al.}~\cite{Kesting2007A,Kesting2008A,Kesting2010A}; this is a subject of intensive
studies up to now (see, e.g.,~\cite{Davis2014C,Treiber_Int1,Shladover2012A,Ngoduy2012A,Ngoduy2013A,Papageorgiou2015B,Papageorgiou2015A,Papageorgiou2015C},
\cite{Talebpour2016A,Wang2017A,Mamouei2018A,Perraki2018A,Sharon2017A,HanAhn2018A}, \cite{ChenAhn2018A,Zhou2017B,Klawtanong2020A,Geng_Zhang2019A,H_B_Zhu2020A}, \cite{ZhouZhu2020A,Zhou_Ahn2020A,Danjue_Chen2019A,Zijia_Zhong2020A,Fangfang_Zheng2020A,Shuang_Jin2020A}, \cite{Zhou_Zhu2020A,ZhihongYao2019A,YeYamamoto2020A,ZhuZhang2018A,Wen-Xing2018A,YeYamamoto2018A,YeYamamoto2018B} and references there).

As proven in details  
in the books~\cite{KernerBook,Kerner2009,Kerner2017A}, classical (standard) traffic flow
theories and models of human  driving vehicles
cannot explain the empirical  nucleation nature of traffic breakdown 
(F$\rightarrow$S transition) at the bottleneck.
	\begin{itemize}
	\item Traffic and transportation theories, which are not consistent with the nucleation 
	nature of   traffic breakdown (F$\rightarrow$S transition)
	at highway bottlenecks, cannot be applied for  
	the development of reliable   traffic control, dynamic traffic assignment 
	as well as other  reliable ITS-applications in traffic and transportation networks. 
	 \end{itemize} 
This criticism on all standard traffic flow models has been explained in details in the book~\cite{Kerner2017A};
a brief explanation of this critical statement can also be found in~\cite{Kerner2018A,Kerner_Review_Int1,Kerner_Review2_Int1,Kerner_Review3_Int1}.

In almost all studies of mixed traffic flow that the author knows
models of {\it human driving} vehicles have been used that
cannot explain the nucleation nature of traffic breakdown 
(F$\rightarrow$S transition) at the bottleneck.
This criticism is also
related to traffic flow theories and models used for 
  studies of human  driving vehicles  in mixed traffic flow in 
	Refs.~\cite{Treiber_Int1,Dharba1999A,Marsden2001A,VanderWerf2001A,VanderWerf2002A,TreiberH2001A,Shrivastava2002A,Kukuchi2003A,BoseIoannou2003A}, \cite{Suzuki2003A,Zhou2005A,vanArem2006,Kesting2007A,Kesting2008A,Kesting2010A,Shladover2012A,Ngoduy2012A,Ngoduy2013A,Papageorgiou2015B}, \cite{Papageorgiou2015A,Papageorgiou2015C,Talebpour2016A,Wang2017A,Mamouei2018A,Perraki2018A,Sharon2017A,HanAhn2018A,ChenAhn2018A,Zhou2017B}, \cite{Klawtanong2020A,Geng_Zhang2019A,H_B_Zhu2020A,ZhouZhu2020A,Zhou_Ahn2020A,Danjue_Chen2019A,Zijia_Zhong2020A,Fangfang_Zheng2020A}, \cite{Shuang_Jin2020A,Zhou_Zhu2020A,ZhihongYao2019A,YeYamamoto2020A,ZhuZhang2018A,Wen-Xing2018A,YeYamamoto2018A,YeYamamoto2018B}. Because the standard traffic flow theories and
	models contradict the empirical  nucleation 
	nature of   traffic breakdown (F$\rightarrow$S transition), we can made the following conclusion.
		\begin{itemize}
	\item Applications of standard traffic flow theories and models of human  driving vehicles used, for example, in the 
  studies of mixed traffic flow in~\cite{Treiber_Int1,Dharba1999A,Marsden2001A,VanderWerf2001A,VanderWerf2002A,TreiberH2001A,Shrivastava2002A,Kukuchi2003A,BoseIoannou2003A}, \cite{Suzuki2003A,Zhou2005A,vanArem2006,Kesting2007A,Kesting2008A,Kesting2010A,Shladover2012A,Ngoduy2012A,Ngoduy2013A,Papageorgiou2015B}, \cite{Papageorgiou2015A,Papageorgiou2015C,Talebpour2016A,Wang2017A,Mamouei2018A,Perraki2018A,Sharon2017A,HanAhn2018A,ChenAhn2018A,Zhou2017B}, \cite{Klawtanong2020A,Geng_Zhang2019A,H_B_Zhu2020A,ZhouZhu2020A,Zhou_Ahn2020A,Danjue_Chen2019A,Zijia_Zhong2020A,Fangfang_Zheng2020A}, \cite{Shuang_Jin2020A,Zhou_Zhu2020A,ZhihongYao2019A,YeYamamoto2020A,ZhuZhang2018A,Wen-Xing2018A,YeYamamoto2018A,YeYamamoto2018B} 
	  lead to invalid   conclusions about the effect of
		autonomous driving vehicles on highway capacity and traffic breakdown in mixed traffic flow. 
	 \end{itemize}
For this reason,  in this review article we will not consider results of~\cite{Treiber_Int1,Dharba1999A,Marsden2001A,VanderWerf2001A,VanderWerf2002A,TreiberH2001A,Shrivastava2002A,Kukuchi2003A,BoseIoannou2003A}, \cite{Suzuki2003A,Zhou2005A,vanArem2006,Kesting2007A,Kesting2008A,Kesting2010A,Shladover2012A,Ngoduy2012A,Ngoduy2013A,Papageorgiou2015B}, \cite{Papageorgiou2015A,Papageorgiou2015C,Talebpour2016A,Wang2017A,Mamouei2018A,Perraki2018A,Sharon2017A,HanAhn2018A,ChenAhn2018A,Zhou2017B}, \cite{Klawtanong2020A,Geng_Zhang2019A,H_B_Zhu2020A,ZhouZhu2020A,Zhou_Ahn2020A,Danjue_Chen2019A,Zijia_Zhong2020A,Fangfang_Zheng2020A}, \cite{Shuang_Jin2020A,Zhou_Zhu2020A,ZhihongYao2019A,YeYamamoto2020A,ZhuZhang2018A,Wen-Xing2018A,YeYamamoto2018A,YeYamamoto2018B}
as well as   results of all other simulations in which
standard traffic flow models for the description of human driving vehicles in mixed traffic flow have been used, which are not consistent with the empirical nucleation nature of traffic breakdown (F$\rightarrow$S transition) at a highway bottleneck.

\section{Classical (Standard)   Strategy of Adaptive Cruise Control (ACC) \label{Classical_ACC_S}}

 In a classical (standard) ACC model, acceleration (deceleration) $a^{\rm (ACC)}$ of
the ACC vehicle is  determined  by   the space gap to the preceding vehicle $g$ 
and the relative speed $\Delta v=v_{\ell}-v$ measured by the ACC vehicle
 as well as by 
  a desired time headway
$\tau^{\rm (ACC)}_{\rm d}$ of the ACC-vehicle to the 
 preceding 
vehicle    (see, e.g.,~\cite{IoannouChien2002A,Levine1966A_Aut,Liang1999A_Aut,Liang2000A_Aut,Swaroop1996A_Aut,Swaroop2001A_Aut,Rajamani2012A_Aut,Davis2004B9,Davis2014C}): 
\begin{equation}
a^{\rm (ACC)} = K_{1}(g-v\tau^{\rm (ACC)}_{\rm d})+K_{2} (v_{\ell}-v),
 \label{ACC_General}
 \end{equation} 
where $v$ is the speed of the ACC-vehicle, $v_{\ell}$ is the speed of the preceding vehicle; here and below
  $v$, $v_{\ell}$, and $g$ are time-functions;
  $K_{1}$
and $K_{2}$ are   coefficients of  
   ACC adaptation.  
 
It is well-known that there can be 
   string instability of a long enough platoon of ACC-vehicles 
	(\ref{ACC_General})~\cite{IoannouChien2002A,Levine1966A_Aut,Liang1999A_Aut,Liang2000A_Aut,Swaroop1996A_Aut,Swaroop2001A_Aut,Rajamani2012A_Aut,Davis2004B9,Davis2014C}. As found by
	Liang and Peng~\cite{Liang1999A_Aut}, the string instability
	occurs under condition
$K_{2}<(2-K_{1}(\tau^{\rm (ACC)}_{\rm d})^{2})/2\tau^{\rm (ACC)}_{\rm d}$:
If a local speed (or time headway) disturbance appears in the platoon of ACC-vehicles, the disturbance begins to grow in its amplitude. The growth of the disturbance destroys the steadily ACC-vehicle motion in a long enough  platoon
of ACC-vehicles.
  Coefficients $K_{2}$ and $K_{1}$ of classical ACC (\ref{ACC_General}) can be chosen to  satisfy
condition for string stability
	\begin{equation}
K_{2}>(2-K_{1}(\tau^{\rm (ACC)}_{\rm d})^{2})/2\tau^{\rm (ACC)}_{\rm d}.
 \label{ACC_stability}
 \end{equation}   
Under condition (\ref{ACC_stability}), any small 
local speed (or time headway) disturbance that appears in the platoon decays over time.
Therefore,   ACC-vehicles in the platoon remain to move at a time-independent speed at the desired time headway
$\tau^{\rm (ACC)}_{\rm d})$ to each other.
In this review article, we consider {\it only}  ACC-vehicles whose parameters satisfy condition
(\ref{ACC_stability}) for string stability.

\begin{figure}
\begin{center}
\includegraphics[scale=.75]{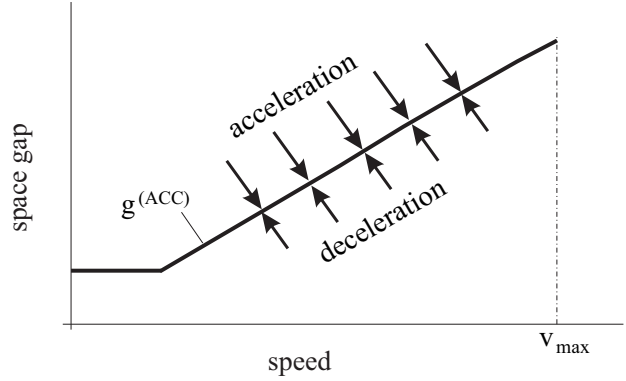}
\end{center}
\caption[]{Operating points of the classical model of   ACC (\ref{ACC_General}) under condition of string stability (\ref{ACC_stability}): Qualitative speed dependence of desired space gap
$g^{\rm (ACC)}=v\tau^{\rm (ACC)}_{\rm d}$ (\ref{g_ACC_F})
(see, e.g.,~\cite{IoannouChien2002A,Levine1966A_Aut,Liang1999A_Aut,Liang2000A_Aut,Swaroop1996A_Aut,Swaroop2001A_Aut,Rajamani2012A_Aut,Davis2004B9,Davis2014C}).
}
\label{ACC_Fig}
\end{figure}

It should be noted that in the works devoted to analysis of the effect of autonomous driving
 on traffic flow~\cite{Treiber_Int1,Dharba1999A,Marsden2001A,VanderWerf2001A,VanderWerf2002A,TreiberH2001A,Shrivastava2002A}, \cite{Kukuchi2003A,BoseIoannou2003A,Suzuki2003A,Zhou2005A,vanArem2006,Kesting2007A,Kesting2008A,Kesting2010A,Shladover2012A}, \cite{Ngoduy2012A,Ngoduy2013A,Papageorgiou2015B,Papageorgiou2015A}, \cite{Papageorgiou2015C,Talebpour2016A,Wang2017A,Mamouei2018A,Perraki2018A,Sharon2017A,Zhou2017B} traffic flow models for human  driving 
vehicles have been used in which  at a given
 time-independent vehicle speed  
there is a single   model solution for a space gap
to the preceding vehicle. Therefore,  
    for such hypothetical steady state
model solutions,
there is a one-dimensional (1D) relationship between a chosen speed and the related desired (or optimal)
space gap to the preceding vehicle.
This  well-known assumption for traffic flow models
of human driving vehicles used, e.g., in~\cite{Treiber_Int1,Dharba1999A,Marsden2001A,VanderWerf2001A,VanderWerf2002A,TreiberH2001A,Shrivastava2002A}, \cite{Kukuchi2003A,BoseIoannou2003A,Suzuki2003A,Zhou2005A,vanArem2006,Kesting2007A,Kesting2008A,Kesting2010A,Shladover2012A}, \cite{Ngoduy2012A,Ngoduy2013A,Papageorgiou2015B,Papageorgiou2015A}, \cite{Papageorgiou2015C,Talebpour2016A,Wang2017A,Mamouei2018A,Perraki2018A,Sharon2017A,Zhou2017B} is qualitatively the same as 
the existence of a desired time headway
$\tau^{\rm (ACC)}_{\rm d}$ of the ACC-vehicle to the 
 preceding 
vehicle: For the classical ACC-rule  (\ref{ACC_General}) that satisfies
condition for string stability   (\ref{ACC_stability}), 
at a given ACC speed $v$
 there is a single operating point for a desired space gap (Fig.~\ref{ACC_Fig}) 
\begin{equation}
g^{\rm (ACC)}=v\tau^{\rm (ACC)}_{\rm d}. 
\label{g_ACC_F}
\end{equation}

 \section{ACC   in Framework of Three-Phase Theory (TPACC) \label{TPACC_St_S}}

 \subsection{Indifferent Zone in Car-Following in Three-Phase Traffic Theory}

A study of real field traffic data~\cite{Kerner1998C,Kerner1998B,Kerner1999B,Kerner1999A,Kerner1999C,Kerner2000,Kerner2001,Kerner2002A,Kerner2002B,KR1996B,KR1997B} shows that
  the existence of a desired time headway
$\tau^{\rm (ACC)}_{\rm d}$ of the ACC-vehicle to the 
 preceding 
vehicle    
is   inconsistent with a basic behavior of
real drivers in car-following: Empirical data shows that
real drivers do not
try to reach a fixed time headway  to the 
 preceding 
vehicle in car-following.

\begin{figure}
\begin{center}
\includegraphics*[scale=.9]{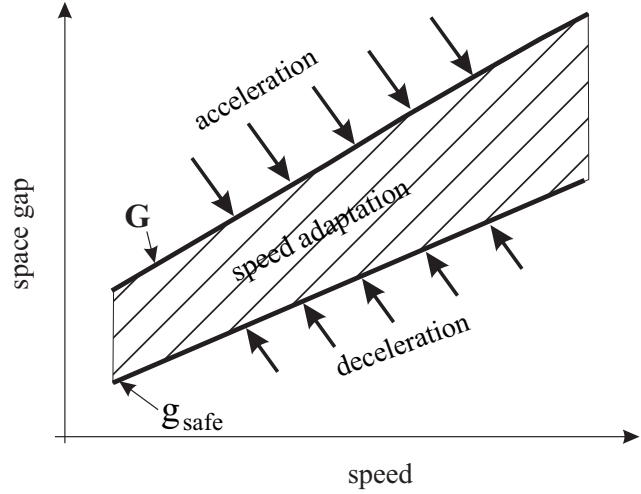}
\end{center}
\caption[]{Qualitative explanation of  indifference zone  in car-following for
TPACC.
 Qualitative presentation of a
  part of 2D-region for operating points of TPACC in
the space-gap--speed plane:   At a given speed of TPACC-vehicle there are the infinity of 
operating points of TPACC~\cite{KernerPat1,KernerPat2,KernerPat3,KernerPat4}.
$G$ is a synchronization space gap, $g_{\rm safe}$ is a safe space gap.
}
\label{Eco_ACC}
\end{figure}

 To explain this empirical fact,  
the author has introduced a hypothesis  about the existence of a two-dimensional (2D)
region of synchronized flow states~\cite{Kerner1998C,Kerner1998B,Kerner1999B,Kerner1999A,Kerner1999C}:
    In the three-phase   theory, it is assumed that
when a driver approaches a slower moving preceding vehicle and the driver cannot pass it, 
the driver  decelerates
within a synchronization space gap $G$. This speed adaptation  
  to the speed of the preceding vehicle occurs
		without caring what the precise  space gap $g$
 to the preceding vehicle is as long as it is not smaller than a safe space 
gap $g_{\rm safe}$~\cite{Kerner1998C,Kerner1998B,Kerner1999B,Kerner1999A,Kerner1999C}.  
The speed adaptation occurring within the synchronization space gap $G$ leads to
 a 2D-region of  synchronized flow states  (dashed region in Fig.~\ref{Eco_ACC})   determined by
  conditions
	\begin{equation}
g_{\rm safe} \leq g \leq G.
\label{G_g_g_s}
\end{equation} 
 The 2D-region of  synchronized flow states  (dashed region in Fig.~\ref{Eco_ACC}) can also be considered  
	$\lq\lq$indifference zone'' in car-following.

	Accordingly to (\ref{G_g_g_s}),  
   a driver  
		does not try to reach a particular (desired or optimal)  time headway
to the preceding vehicle, but adapts the speed while keeping  time headway $\tau^{\rm (net)}=g/v$
in a range $\tau_{\rm safe}\leq \tau^{\rm (net)}\leq \tau_{\rm G}$, where
  $\tau_{\rm G}=G/v$, $\tau_{\rm G}$ is a synchronization time headway,
  $\tau_{\rm safe}=g_{\rm safe}/v$ is  a safe time headway 
	and it is assumed that the speed $v>0$.

In~\cite{Kerner2018C}, we have defined $\lq\lq$autonomous driving in the  framework of   three-phase traffic theory'' (TPACC)\footnote{The  main reason for the use of the word $\lq\lq$three-phase'' in the ACC strategy is as follows.
A 2D-region of operating points of TPACC, i.e., an
indifference zone in car-following shown by dashed region in Fig.~\ref{Eco_ACC}  follows from the driver behavior firstly
  incorporated in   the three-phase   theory
	in which  a 2D-region of steady states of synchronized flow it is assumed~\cite{Kerner1998C,Kerner1998B,Kerner1999B,Kerner1999A,Kerner1999C,Kerner1998D}. 
	The word $\lq\lq$three-phase'' for TPACC  should emphasize both a qualitative difference between the two types of ACCs and the fact that
the idea of TPACC with no fixed  time headway to the preceding vehicle has been taken from
  the three-phase theory.} as follows:
\begin{itemize}
\item An autonomous driving in the  framework of the  three-phase  traffic theory is
the autonomous driving 
  for which
there is {\it no}  fixed  time headway to the preceding vehicle\footnote{A relation of this definition to real autonomous driving vehicles will be discussed in Sec.~\ref{Value}.
}.
\end{itemize}

	In   inventions~\cite{KernerPat1,KernerPat2,KernerPat3,KernerPat4}, 
we have assumed that to satisfy these empirical features of real traffic, at least at some driving conditions in car-following
  acceleration (deceleration) of autonomous driving in the framework of the three-phase theory
	(TPACC)
  should be given by formula~\cite{KernerPat1,KernerPat2,KernerPat3,KernerPat4} 
\begin{eqnarray}
\label{TPACC_Eq1}
a^{\rm (TPACC)}=K_{\rm \Delta v}(v_{\ell}-v) \quad 
 \textrm{at $g_{\rm safe} \leq g \leq G$},  
\end{eqnarray}
where  
$K_{\rm \Delta v}$ is a dynamic coefficient   ($K_{\rm \Delta v}>0$).

 \subsection{Model of TPACC \label{Model_TPACC_S}}

Following~\cite{Kerner2018C,Kerner2019C}, we  study   the following TPACC
    model:
\begin{equation}
a^{\rm (TPACC)}=  
  \left\{
\begin{array}{ll}
K_{\rm \Delta v}(v_{\ell}-v) &  \textrm{at $g \leq G$} \\ 
K_{1}(g-v\tau_{\rm p})+K_{2} (v_{\ell}-v) &  \textrm{at $g> G$}, \\ 
\end{array} \right.
\label{TPACC_main5}  
 \end{equation}
where   it is   assumed   that
\begin{equation}
g\geq g_{\rm safe},
\label{TPACC_main5_1} 
\end{equation}  $\tau_{\rm p}$ is a model parameter that satisfies condition 
\begin{equation}
\tau_{\rm p}<\tau_{\rm G}.
\label{TPACC_main5_5} 
\end{equation}
In comparison with the TPACC-strategy (\ref{TPACC_Eq1}),  
 the TPACC model (\ref{TPACC_main5})  allows us to simulate physical features of 
TPACC-vehicles in mixed traffic flow.

		Under   condition (\ref{TPACC_main5_1}),  
  from the TPACC model (\ref{TPACC_main5}) it follows
 that when the  space gap $g(t)$    
 of the TPACC-vehicle   to the preceding vehicle is within the range
\begin{eqnarray}
g_{\rm safe}(t)\leq g(t) \leq G(t), 
\label{TPACC_main_range_g} 
 \end{eqnarray}
the acceleration (deceleration) of the TPACC-vehicle does not depend on
  the space gap $g(t)$. Conditions (\ref{TPACC_main_range_g}) determine the indifference zone in the space gap for
	TPACC-vehicle in the car-following process. Because time headway of TPACC-vehicle to the preceding vehicle
	is given by formula  
  $\tau^{\rm (net)}=g/v$, conditions (\ref{TPACC_main_range_g}) are equivalent to
	conditions
	\begin{eqnarray}
\tau_{\rm safe}(t)\leq \tau^{\rm (net)}(t)\leq \tau_{\rm G} 
\label{TPACC_main_range} 
 \end{eqnarray}
that determine the indifference zone in time headway
for   
	TPACC-vehicle in the car-following process. 
	In (\ref{TPACC_main_range}),
  it is assumed that $v>0$. 
  
 One of the first traffic flow models  
 in the framework of the three-phase theory   is the Kerner-Klenov microscopic stochastic
model~\cite{KKl,KKl2003A,KKl2009A}.  Because the Kerner-Klenov microscopic stochastic
model~\cite{KKl,KKl2003A,KKl2009A} 
can show the nucleation nature of traffic breakdown 
(F$\rightarrow$S transition) at the bottleneck as observed in empirical data, 
for all simulations of human driving vehicles
in mixed traffic flow we
use   the   Kerner-Klenov traffic flow
 model (see Appendix~\ref{KKl_Model_Ap}).
			In this model,
  a  discrete time $t=n\tau$, where $n=0,1,2,...$; $\tau=$1 s is   time step, is used.
	The model of human driving vehicles of Refs.~\cite{KKl,KKl2003A} is continuous in space.
	We use a version of this model~\cite{KKl2009A} that is discrete in space:
	 A very small   value  of the discretization 
space interval $\delta x=0.01$ m is used in the model. As explained in~\cite{KKl2009A},
this allows us to make  more accurate simulations of traffic breakdown at road bottlenecks.
Because the model for human driving 
vehicles~\cite{KKl,KKl2003A,KKl2009A} is discrete in time, we simulate TPACC-model (\ref{TPACC_main5})
with    discrete time $t=n\tau$. 
For the simplicity of the consideration of the effect of 
either classical ACC-vehicles or
TPACC-vehicles  on traffic breakdown in mixed traffic flow,
discrete    models of   human driving vehicles, classical ACC-vehicles, and
TPACC-vehicles used in simulations 
  have been given in  Appendixes~\ref{KKl_Model_Ap}--\ref{Boundary_Ini_Con_S}.

  \section{Effect of   Single Autonomous Driving Vehicle on   Traffic Breakdown  \label{Prob_S}}

In the near future, we could expect
 mixed traffic flow in which the share of autonomous driving vehicles will be very small.
  Therefore, 
now we consider   mixed traffic flow consisting of 
a small percentage $\gamma=2\%$ of autonomous driving vehicles that are randomly distributed between 
 human driving vehicles. At a
such small share of autonomous driving vehicles in mixed traffic flow, a probability that a platoon of 
several autonomous driving vehicles can occur in mixed traffic flow is negligible.
In other words, almost any autonomous driving vehicle in mixed traffic flow can be considered 
as a single autonomous driving vehicle surrounded by human driving vehicles.
The objective  of our study made in this
section and   Sec.~\ref{ACC_Cl_Param_Prob_S}   is to show that already a
 single autonomous driving vehicle  can effect considerably on
  traffic breakdown in mixed traffic flow
 at the bottleneck.  

\subsection{Probability of Traffic Breakdown  \label{Prob_P_Sub}}

  Traffic breakdown
  occurs in a metastable free flow with respect to the F$\rightarrow$S transition.
  Local speed disturbances caused by vehicle interactions
	in a neighborhood of the bottleneck can randomly initiate traffic breakdown
	in the metastable free flow. Such traffic breakdown has been called {\it spontaneous}
	traffic breakdown (spontaneous F$\rightarrow$S transition).
The larger the amplitude of local
speed disturbances at the bottleneck, the more probable the nucleus occurrence
for the spontaneous breakdown, i.e, the larger 
 the probability of traffic breakdown $P^{\rm (B)}$
at the bottleneck~\cite{KernerBook,Kerner2009,Kerner2017A,Kerner2017B,Kerner_Review3_Int1,Kerner2018B}.

\begin{figure}
\begin{center}
\includegraphics*[scale=.6]{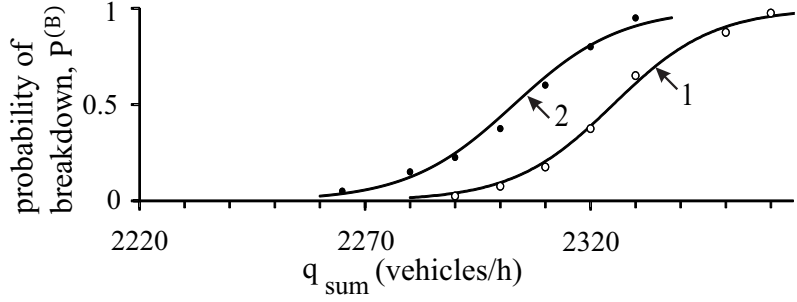}  
\end{center}
\caption[]{Effect of a single  autonomous driving vehicle on the probability $P^{\rm (B)}(q_{\rm sum})$ of traffic breakdown
 at on-ramp bottleneck on single-lane road in mixed traffic flow. Flow-rate functions  of breakdown probability
  $P^{\rm (B)}(q_{\rm sum})$ (curves 1 and2) are calculated  through the change in
the on-ramp inflow rate $q_{\rm on}$ at a given flow rate $q_{\rm in}=$ 2000 vehicles/h.
  Curve 1 is related to
traffic flow without autonomous driving vehicles as well as to
mixed traffic
 flow with 2$\%$ of TPACC-vehicles. Curve 2 is related to mixed traffic flow with 2$\%$ of ACC-vehicles 
with $\tau^{\rm (ACC)}_{\rm d}=$1.3 s. 
Simulation parameters of ACC   and TPACC   are identical ones:
  $\tau^{\rm (ACC)}_{\rm d}=\tau_{\rm p}=$1.3 s,
 $\tau_{\rm G}=$1.4 s, $K_{1}=0.3 \ s^{-2}$ and $K_{2}=K_{\rm \Delta v}=0.6 \ s^{-1}$;
 maximum speed in free flow $v_{\rm free}=30$m/s (108 km/h),
vehicle length $d=$7.5 m (see Appendix~\ref{KKl_Model_Ap}); maximum acceleration (deceleration) of ACC- and TPACC-vehicle
  (under condition that vehicle speed is smaller than some safe speed)
		$a_{\rm max}=b_{\rm max}=3 \ {\rm m}/{\rm s}^2$ (see Appendix~\ref{Cla_ACC_S}).
}
\label{Prob2_ACC}
\end{figure}

\begin{figure}
\begin{center}
\includegraphics*[scale=.6]{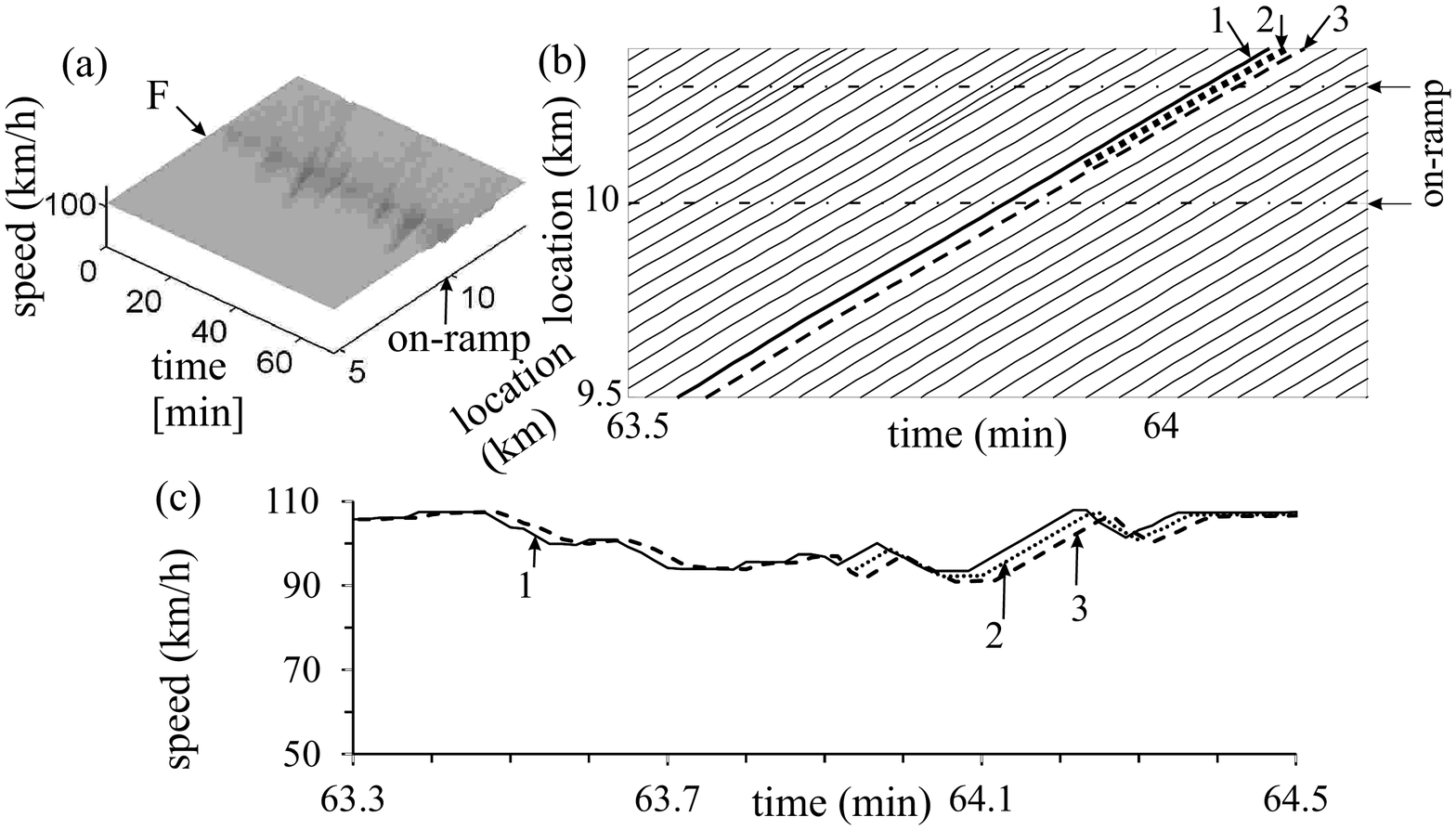}  
\end{center}
\caption[]{Explanation of the result shown by curve 1 of Fig.~\ref{Prob2_ACC} that  
 a single TPACC-vehicle does not effect on the probability of traffic breakdown
 at on-ramp bottleneck in mixed traffic flow.
 Speed disturbances occurring at on-ramp bottleneck through 
a single TPACC-vehicle:
(a)  Speed in space and time.   (b) Fragment of vehicle trajectories.
 (c) Microscopic speeds along vehicle trajectories
shown by  the same numbers in (b), respectively. In (b, c),
vehicles 1 and 2 are human driving vehicles whereas    vehicle 3 is TPACC-vehicle.    
Mixed traffic flow   with   2$\%$ of TPACC-vehicles;
  $q_{\rm in}=$ 2000 vehicles/h, $q_{\rm on}=$ 280 vehicles/h.
Other
  model parameters   are 
	the same as those in Fig.~\ref{Prob2_ACC}.  
	In (a, b) the on-ramp merging region that is within road locations 
$x_{\rm on} \leq x \leq x_{\rm on}^{\rm (e)}$ (see Appendix~\ref{Models_Bott_Sec})
 is labeled by $\lq\lq$on-ramp''.
}
\label{Break_TPACC}
\end{figure}

For calculation of the probability $P^{\rm (B)}(q_{\rm sum})$
of traffic breakdown in free flow at the bottleneck, 
at each given value of the flow rate downstream of the bottleneck
 $q_{\rm sum}=q_{\rm in}+q_{\rm on}$
	different simulation realizations (runs) $N_{\rm r}=$ 40 during the same time interval for
	the observing of traffic flow $T_{\rm ob}=$ 30 min have been made (Fig.~\ref{Prob2_ACC}).
	The different  realizations have been performed
	at the same set of model parameters, however, at different values 
	of the initial values of random function $rand()$
	in the traffic flow model (see Appendix~\ref{Fluc_KKl} and Appendix~\ref{Stoch_Time_Del_A}). Then, $P^{\rm (B)}(q_{\rm sum})=n_{\rm r}/N_{\rm r}$,
	where $n_{\rm r}$ is the number of realizations in which traffic breakdown  has occurred during the time interval $T_{\rm ob}$ (more detailed explanations of the calculation of the flow-rate function
	$P^{\rm (B)}(q_{\rm sum})$ have been given in
	  the book~\cite{Kerner2017A}).  
		
		\subsection{Speed Disturbances caused by Single TPACC and Single Classical ACC at Bottleneck}
		
\begin{figure}
\begin{center}
\includegraphics*[scale=.6]{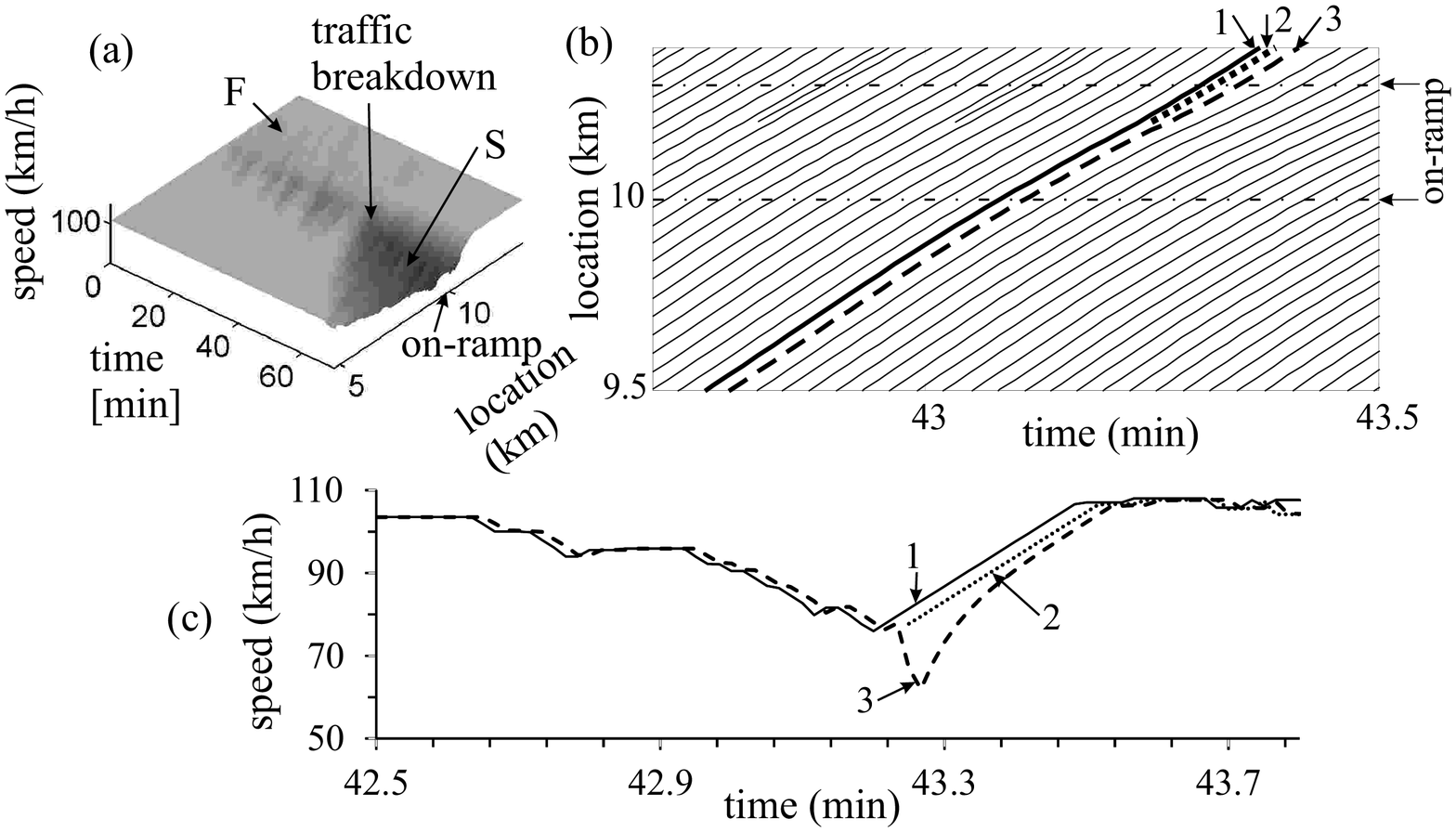}  
\end{center}
\caption[]{Explanation of the effect of a single classical ACC-vehicle on the probability of traffic breakdown
 at on-ramp bottleneck in mixed traffic flow   associated with curve 2 of
  Fig.~\ref{Prob2_ACC}.
 Speed disturbances occurring at on-ramp bottleneck through 
a single ACC-vehicle (a--c):
(a)  Speed in space and time.   (b) Fragment of vehicle trajectories.
 (c) Microscopic speeds along vehicle trajectories
shown by  the same numbers in (b), respectively. In (b, c),
vehicles 1 and 2 are human driving vehicles whereas    vehicle 3 is ACC-vehicle.  
Mixed traffic flow   with   2$\%$ of ACC-vehicles;
  $q_{\rm in}=$ 2000 vehicles/h, $q_{\rm on}=$ 280 vehicles/h.
Other
  model parameters for ACC-vehicles   are 
	the same as those in Fig.~\ref{Prob2_ACC}. In (a), F -- free flow, S -- synchronized flow.
	In (a, b) the on-ramp merging region that is within road locations 
$x_{\rm on} \leq x \leq x_{\rm on}^{\rm (e)}$ (see Appendix~\ref{Models_Bott_Sec})
 is labeled by $\lq\lq$on-ramp''.
}
\label{Break_ACC}
\end{figure}

		 Single  
	TPACC-vehicles moving in  such mixed traffic flow
	cause  very small speed disturbances at the bottleneck   (Fig.~\ref{Break_ACC} (a--c)). 
	Indeed, we have found that probability of traffic breakdown remains
	in this mixed flow the same as that in traffic flow consisting of human drivers only
	(curve 1 in Fig.~\ref{Prob2_ACC}). Thus, single TPACC-vehicles do not effect on
	the probability of traffic breakdown at the bottleneck.

	Contrarily to TPACC-vehicles, single classical ACC-vehicles effect considerably  on
	the probability of traffic breakdown at the bottleneck: The probability of traffic breakdown
	can increase even when $\gamma=2\%$ of classical ACC-vehicles is in 
	mixed traffic flow
	(curves 2 and 3  in Fig.~\ref{Prob2_ACC}). This is because 
	already a single  ACC-vehicle
can initiate traffic breakdown at the bottleneck   (Fig.~\ref{Break_ACC} (a, b)).
This deterioration of traffic flow
	through classical autonomous driving is explained by
  the occurrence of a large amplitude speed disturbance caused by  
a classical ACC-vehicle at the bottleneck (dashed 
vehicle trajectory 3 in Fig.~\ref{Break_ACC} (b, c)). 

\subsection{Explanation of Effect of Single Autonomous Driving Vehicle
 on Speed Disturbance at Bottleneck   \label{Exp_Prob_Sub}}

The occurrence of very different amplitudes of
local speed disturbances caused by the propagation of  single classical ACC
and TPACC-vehicles through the bottleneck (compare
dashed vehicle  trajectories  3 in Fig.~\ref{Break_TPACC} (b, c)
and Fig.~\ref{Break_ACC} (b, c)) can be understood if we consider time functions of  the space gap
$g(t)$   and the acceleration (deceleration) of the ACC- and TPACC-vehicles
(Fig.~\ref{Accel_ACC_TPACC}). We can see that after the space gap $g(t)$ due to   merging of vehicle 2
decreases abruptly (labeled by $\lq\lq$merging of vehicle 2" in Figs.~\ref{Accel_ACC_TPACC} (a, c)),
both ACC-vehicle and TPACC-vehicle decelerate to   increase the space gap $g(t)$.
However, the deceleration of ACC-vehicle
(Fig.~\ref{Accel_ACC_TPACC} (b)) is considerably stronger than that of TPACC-vehicle
(Fig.~\ref{Accel_ACC_TPACC} (d)). 
  The strong deceleration of the ACC-vehicle explains
the strong speed reduction of the ACC-vehicle shown in trajectory  3 in Fig.~\ref{Break_ACC} (c).

\begin{figure}
\begin{center}
\includegraphics*[scale=.5]{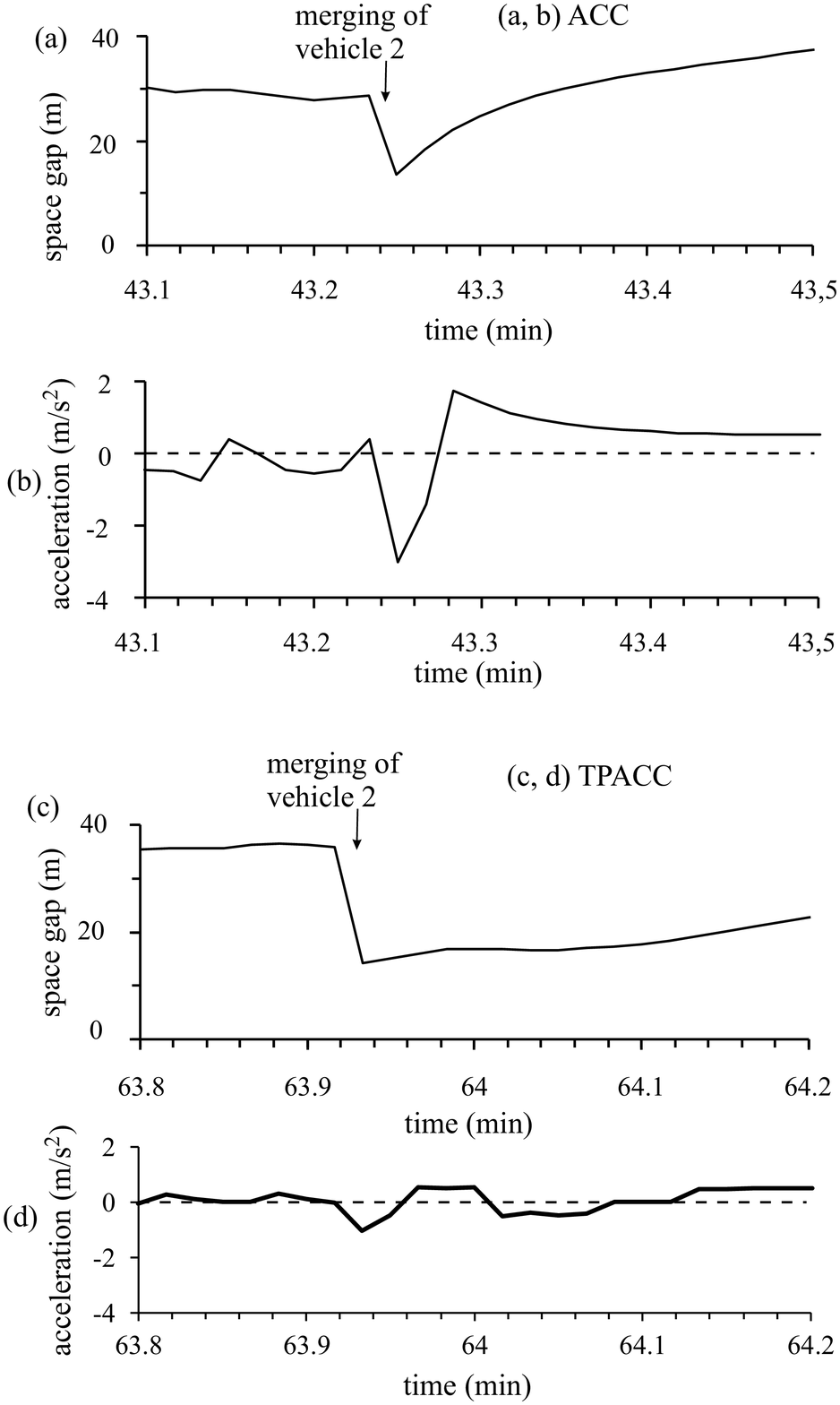}  
\end{center}
\caption[]{Explanations of speed disturbances caused by   propagation of a single autonomous driving vehicle through bottleneck. (a, b)
Time functions of  the space gap $g(t)$ (a) and the acceleration (deceleration) of the ACC-vehicle
(b) related to vehicle  trajectory  3 in Fig.~\ref{Break_ACC} (b, c).
(c, d)
Time functions of  the space gap $g(t)$ (c) and the acceleration (deceleration) of the TPACC-vehicle
(d) related to vehicle  trajectory  3 in Fig.~\ref{Break_TPACC} (b, c).
}
\label{Accel_ACC_TPACC}
\end{figure}

\begin{figure}
\begin{center}
\includegraphics*[scale=.6]{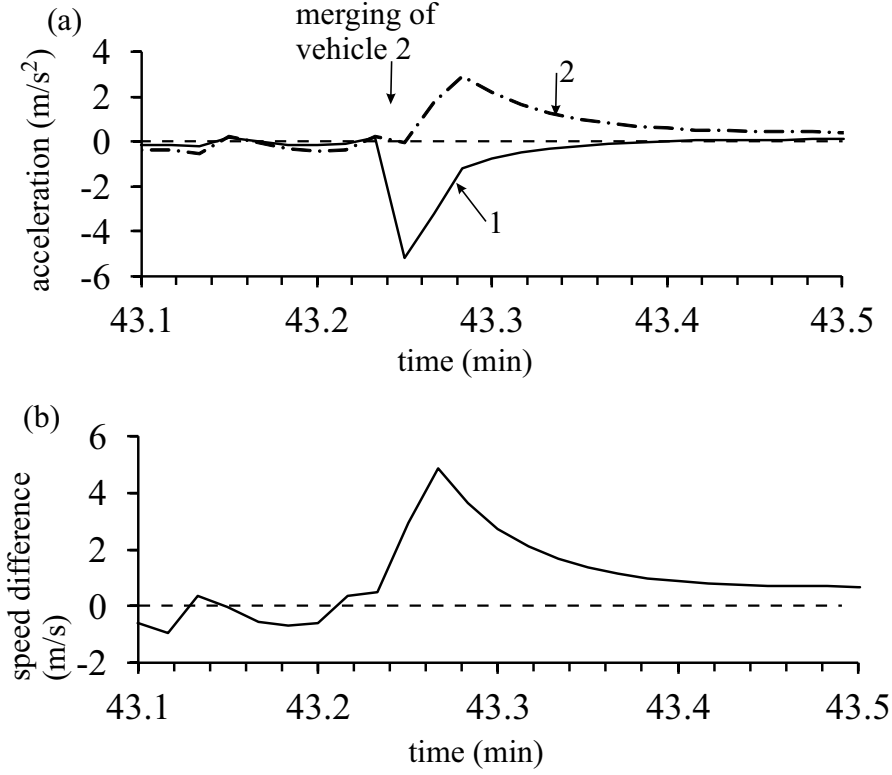}  
\end{center}
\caption[]{Continuation of Fig.~\ref{Accel_ACC_TPACC} (a, b). (a) Components of formula 
(\ref{ACC_General}) as time functions; solid curve 1 is time dependence of the term
  $K_{1}(g-v\tau^{\rm (ACC)}_{\rm d})$, 
	dashed-dotted curve 2 is   time dependence of the term $K_{2} (v_{\ell}-v)$.
(b) Relative speed between the preceding vehicle and ACC-vehicle.
 Note that the   deceleration
calculated through from $K_{1}(g-v\tau^{\rm (ACC)}_{\rm d})$ in (\ref{ACC_General})  
reaches a large negative value about $- 5.2 \ {\rm m/s^{2}}$.
It should be noted that as long as condition $v(t)\leq v_{\rm s}(t)$ is satisfied,
in the ACC- and TPACC-models there is a limitation of the deceleration
by the value $- 3 \ {\rm m/s^{2}}$. This explains   the maximum deceleration of ACC-vehicle
$- 3 \ {\rm m/s^{2}}$ shown in Fig.~\ref{Accel_ACC_TPACC} (b).
}
\label{Accel2_ACC}
\end{figure}

\begin{figure}
\begin{center}
\includegraphics*[scale=.6]{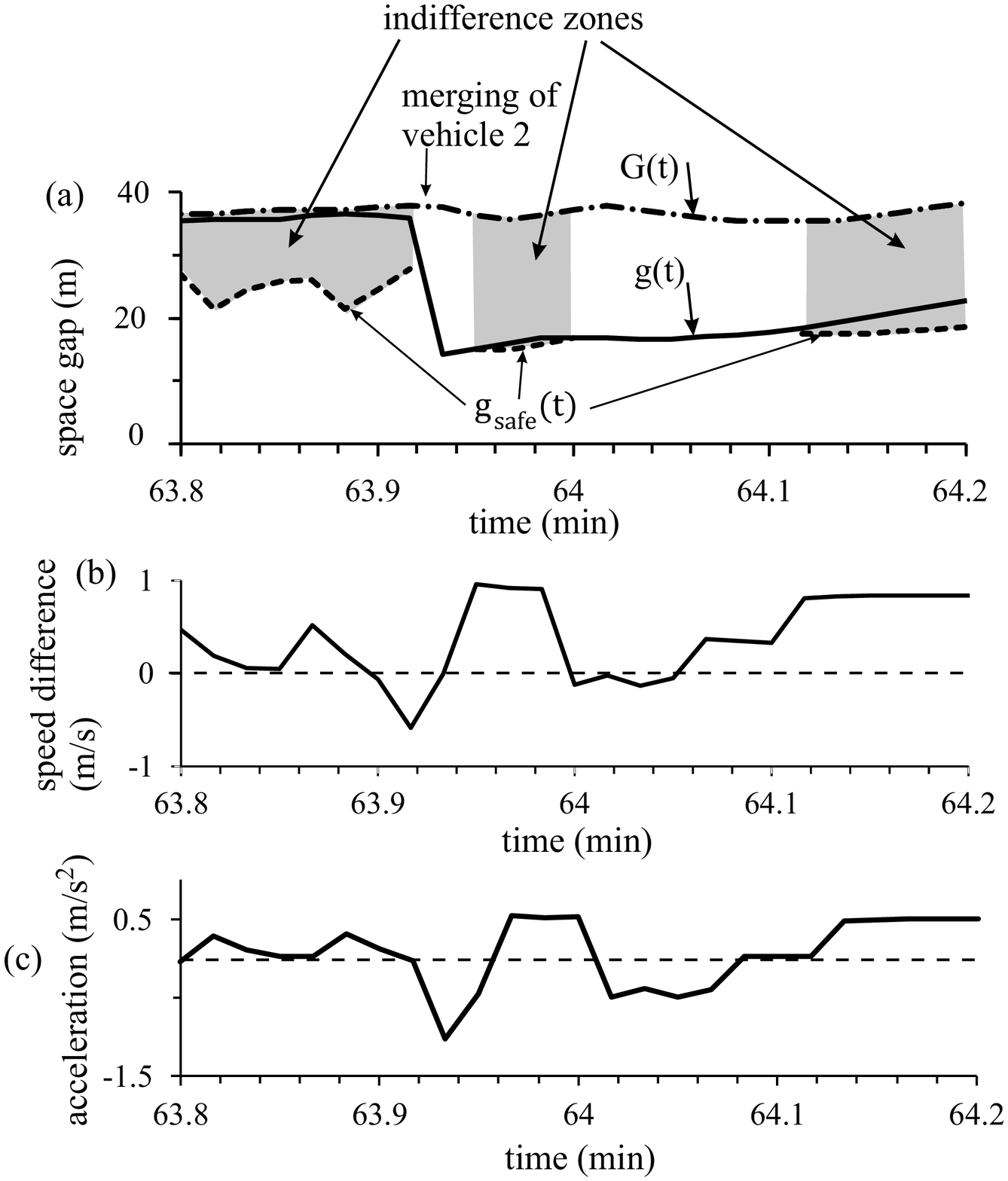}  
\end{center}
\caption[]{Continuation of Fig.~\ref{Accel_ACC_TPACC} (c, d). (a) Time function
 of  the space gap $g(t)$ of TPACC-vehicle together with indifference zones
(gray regions);
$G(t)$ is synchronization space gap;   $g_{\rm safe}(t)$ (dashed curves)
is  safe space gap; gray regions show time intervals in which condition
 (\ref{TPACC_main_range_g})  for indifference zone is satisfied.
(b) Relative speed $\Delta v=v_{\ell}-v$ between the speed of the preceding vehicle and
the speed of  TPACC-vehicle.
(c) Acceleration (deceleration) of TPACC-vehicle given in a larger scale 
as that in Fig.~\ref{Accel_ACC_TPACC} (d).
}
\label{Accel2_TPACC}
\end{figure}

 To understand this very different microscopic dynamic behavior of ACC- and TPACC-vehicles,
we should mention that  in the models of ACC- and TPACC-vehicles 
(see Appendixes~\ref{Cla_ACC_S} and~\ref{App_TPACC_Model})
there is a safe speed $v_{\rm s}$ that depends on  
 the speed of the preceding vehicle $v_{\ell}$
and on the space gap $g(t)$ between vehicles:
When the vehicle speed of an autonomous driving  vehicle
 satisfies condition $v(t)\leq v_{\rm s}(t)$,
then the acceleration (deceleration) of autonomous driving  vehicles
is determined by formula (\ref{ACC_General}) for ACC and formulas
(\ref{TPACC_main5}) for TPACC. It should be noted that  
condition $v(t)\leq v_{\rm s}(t)$ is equivalent to condition 
$g(t)\geq g_{\rm safe}(t)$ for the space gap. For this reason, when 
the space gap  
$g(t)< g_{\rm safe}(t)$, then
the deceleration of an autonomous driving  vehicle should be at least as strong as
it follows from the formulation for the safe speed
(see Appendix~\ref{KKl_Model_Ap}). The formulation for the safe speed
  is the same for human driving vehicles, ACC-vehicles, and TPACC-vehicles.
The formulation for the safe speed $v_{\rm s}(v_{\ell},g)$ 
 guaranties a collision-less motion of  autonomous driving  vehicles.

It turns out that although  the space gap $g(t)$ due to   merging of vehicle 2
decreases abruptly (labeled by $\lq\lq$merging of vehicle 2" in Fig.~\ref{Accel_ACC_TPACC} (a)),
  the deceleration of  ACC-vehicle   determined by 
formula (\ref{ACC_General}) leads to a stronger vehicle deceleration  (Fig.~\ref{Accel2_ACC} (a)) than
it follows from the formulation for the safe speed $v_{\rm s}(v_{\ell},g)$.
Therefore, the stronger vehicle deceleration determined by 
formula (\ref{ACC_General}) is applied. 
Due to the merging of vehicle 2 from the on-ramp,
 there two effects that follow each other:
(i) the abrupt decrease in the space gap $g(t)$
and (ii) the subsequent increase in the speed difference $\Delta v=v_{\ell}-v$.
 Because the effect (i) occurs
earlier, firstly
the deceleration of ACC-vehicle
is determined by the term
 $K_{1}(g-v\tau^{\rm (ACC)}_{\rm d})$ in (\ref{ACC_General}) 
(curve 1 in Fig.~\ref{Accel2_ACC} (a)).
The  subsequent increase in the speed difference $\Delta v=v_{\ell}-v$
that is responsible for the positive value of the term $K_{2} (v_{\ell}-v)$
 in (\ref{ACC_General}) (curve 2 in Fig.~\ref{Accel2_ACC} (a)) cannot 
prevent the strong deceleration of ACC-vehicle mentioned above 
(Fig.~\ref{Accel_ACC_TPACC} (b)).  
 
Microscopic behavior of TPACC-vehicle  
(Fig.~\ref{Accel2_TPACC}) is qualitatively different from that of ACC-vehicle
(Fig.~\ref{Accel2_ACC}) discussed above.
Before vehicle 2 merges onto the main road, the space gap $g(t)$ 
of TPACC-vehicle has satisfied condition (\ref{TPACC_main_range_g}) of the indifference zone
(first gray region in Fig.~\ref{Accel2_TPACC} (a) that is related to time interval before labeling
$\lq\lq$merging of vehicle 2").
After the space gap $g(t)$ due to   merging of vehicle 2
decreases abruptly (labeled by $\lq\lq$merging of vehicle 2" in Fig.~\ref{Accel2_TPACC} (a)), 
  condition $g(t)< g_{\rm safe}(t)$ is   satisfied for TPACC-vehicle. In contrast with ACC-vehicle,
	 the deceleration of  TPACC-vehicle   determined by 
formula (\ref{TPACC_main5}) leads to a weaker vehicle deceleration than
it follows from the formulation for the safe speed $v_{\rm s}(v_{\ell},g)$.
	As a result, 
	TPACC-vehicle decelerates in accordance with the formulation of the safe speed
	$v_{\rm s}(v_{\ell},g)$.
	
	As mentioned,   the formulation of the safe speed
	$v_{\rm s}(v_{\ell},g)$ is related to the behavior of a human driving vehicle
	(Appendix~\ref{Safe_speed_kkl}).
		The safe speed
	$v_{\rm s}(v_{\ell},g)$ depends considerably stronger
	on the speed difference $\Delta v=v_{\ell}-v$ than on the space gap $g(t)$.  
	When under
	condition $g(t)< g_{\rm safe}(t)$ the speed difference $\Delta v=v_{\ell}-v\geq 0$,
	the   deceleration of the TPACC-vehicle associated with the safety conditions
		(Figs.~\ref{Accel_ACC_TPACC} (d) and~\ref{Accel2_TPACC} (c))
	is considerably smaller than the deceleration of the ACC-vehicle given by
  formula (\ref{ACC_General})
 	(Figs.~\ref{Accel_ACC_TPACC} (b) and~\ref{Accel2_ACC} (a)). 
	Moreover, when  the value $\Delta v>0$ 
is large enough, 
condition $g(t)> g_{\rm safe}(t)$ for the indifference zone is satisfied
(second gray region in Fig.~\ref{Accel2_TPACC} (a)). 
 The time evolution of the speed difference
$\Delta v(t)$ (Fig.~\ref{Accel2_TPACC} (b))
determines in the large degree the deceleration (acceleration)
of TPACC-vehicle (Fig.~\ref{Accel2_TPACC} (c)). Over time, there is  an alternation of
 indifference zones (\ref{TPACC_main_range_g}) (in which formulas
(\ref{TPACC_main5}) are valid) and safety deceleration regions
(in which the TPACC-vehicle moves in accordance with the formulation for the safe speed)
 (Fig.~\ref{Accel2_TPACC} (a)). This alternation
determines a slowly increase in the space gap $g(t)$ of TPACC-vehicle (Fig.~\ref{Accel2_TPACC} (a)).
Finally, TPACC-vehicle moves in the indifference zone only.
 
Both safety conditions of TPACC-vehicle and car-following behavior (\ref{TPACC_main5})
   at the bottleneck (trajectory 3 in  Fig.~\ref{Break_TPACC} (c)) are similar as those
 for  human driving vehicles (trajectories 1 and 2 in  Fig.~\ref{Break_TPACC} (c)).
\begin{itemize}
\item For this reason, we can consider autonomous driving in the framework of the three-phase theory (TPACC) as
$\lq\lq$autonomous driving learning from real driving behavior".
\end{itemize}

In particular, when due to the vehicle merging  the space gap $g(t)$
 becomes shorter  than the safe space gap
$g_{\rm safe}$ and the speed difference  $\Delta v=v_{\ell}-v\approx 0$,
a TPACC vehicle decelerates as slowly as a human driving vehicle does.
This explains why in simulations of mixed traffic flow with  TPACC vehicles 
presented in Fig.~\ref{Break_TPACC} (b, c) no large speed disturbances occur at the bottleneck.
This explains also why in contrast with classical autonomous driving
(curve 2 in Fig.~\ref{Prob2_ACC})
single TPACC-vehicles do not effect on the probability of traffic breakdown
at the bottleneck in mixed traffic flow (curve 1 in Fig.~\ref{Prob2_ACC}).

  \section{Effect of Dynamic Parameters of Classical ACC and TPACC on 
	Traffic Breakdown    at Bottleneck \label{ACC_Cl_Param_Prob_S}}

  \subsection{Time Headway of ACC and Breakdown Probability    
	\label{ACC_Cl_Param_Prob_Sub}} 
	
\begin{figure}
\begin{center}
\includegraphics*[scale=.65]{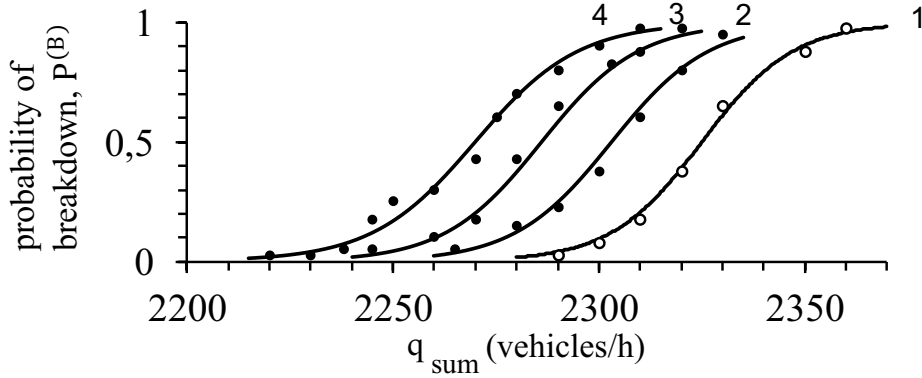}
\end{center}
\caption[]{Dependence of the probability $P^{\rm (B)}(q_{\rm sum})$ of traffic breakdown
 at on-ramp bottleneck on single-lane road on ACC and TPACC parameters.   Flow-rate functions  of breakdown probability
  $P^{\rm (B)}(q_{\rm sum})$ (curves 1--4) are calculated  through the change in
the on-ramp inflow rate $q_{\rm on}$ at a given flow rate $q_{\rm in}=$ 2000 vehicles/h.
  Curve 1 is related to
traffic flow without autonomous driving vehicles as well as to
mixed traffic
 flow with 2$\%$ of TPACC-vehicles. Curves 2--4 are related to mixed traffic flow with 2$\%$ of ACC-vehicles. 
Simulation parameters of ACC   are: curve 2 is the same as curve 2 in Fig~\ref{Prob2_ACC};
curve 3 is related to $\tau^{\rm (ACC)}_{\rm d}=$1.5 s;
curve 4 is related to $\tau^{\rm (ACC)}_{\rm d}=$2 s.
For curve 1 with 2$\%$ of TPACC-vehicles in the mixed traffic flow
 the following sets of TPACC-parameters ($\tau_{\rm p}, \ \tau_{\rm G}$) have been
used: (1.3, \ 1.4) s, (1.5, \ 1.6) s, (2, \ 2.2) s.
Other model parameters are the same as those in Fig~\ref{Prob2_ACC}.
}
\label{Probability_ACC}
\end{figure}

	In Sec.~\ref{Prob_S} we have shown that
	traffic breakdown  can be caused by a single classical ACC-vehicle at the on-ramp bottleneck
	with the breakdown probability $P^{\rm (B)}>0$, whereas the probability of traffic breakdown through TPACC-vehicles at the same parameters is equal to zero. However, this result is related to a chosen
	value of  a desired time headway
$\tau^{\rm (ACC)}_{\rm d}=$ 1.3 s in Eq.~\ref{ACC_General}
	(curve 2 in Figs.~\ref{Prob2_ACC} and~\ref{Probability_ACC}).

Therefore, a question arises whether this result
remains for a broad range of the desired time headway of ACC 
and the related  range of time headway of TPACC. A study of this question for a   range of the desired time headway of ACC 
\begin{equation}
1.3 \ \leq \tau^{\rm (ACC)}_{\rm d}\leq \ {\rm 2 \ s}
\label{range_time}
\end{equation}
shows that at any desired time headway of ACC (\ref{range_time}) the probability of
traffic breakdown caused by single ACC-vehicles is larger than in traffic flow of human drivers.
Moreover, the longer the desired time headway of ACC $\tau^{\rm (ACC)}_{\rm d}$ is,
the larger the probability of traffic breakdown at the same
other model parameters (curves 2--4 in Fig.~\ref{Probability_ACC}).

Contrarily to classical ACC-vehicles, when, in accordance with (\ref{range_time}), we choose
parameters of TPACC-model (\ref{TPACC_main5}) in the ranges 
\begin{equation}
1.3 \ \leq \tau_{\rm p}\leq \ {\rm 2 \ s},
\label{range_time_TPACC}
\end{equation}
\begin{equation}
1.4 \ \leq \tau_{\rm G}\leq \ {\rm 2.2 \ s},
\label{range_time_TPACC_G}
\end{equation}
then no noticeable change in the probability of traffic breakdown
through TPACC-vehicles can be found (curve 1 in Fig.~\ref{Probability_ACC}).

\subsection{Analysis of Disturbances caused by Classical ACC at Bottleneck \label{ACC_Cl_Stat_Sub}} 

To understand   the dependence of the ACC parameters   on traffic breakdown (curves 2--4 in Fig.~\ref{Probability_ACC}),
we have  chosen the on-ramp inflow rate $q_{\rm on}=$ 280 vehicles/h
 and the flow rate on the main road $q_{\rm in}=$ 2000 vehicles/h   at which  
the probability of traffic breakdown
  $P^{\rm (B)}$  at the bottleneck is equal to zero for
	mixed traffic flow with $2 \%$  TPACC-vehicles (curve 1 in Fig.~\ref{Probability_ACC}), whereas
	the probability of traffic breakdown in mixed traffic flow with $2\%$  ACC-vehicles satisfies conditions
	\begin{equation}
0<P^{\rm (B)}<1
\label{Prob_ACC_formula}
\end{equation}
 for any chosen desired time headway of ACC in (\ref{range_time})  (curves 2--4 in Fig.~\ref{Probability_ACC}).

  \begin{figure}
\begin{center}
\includegraphics*[scale=.5]{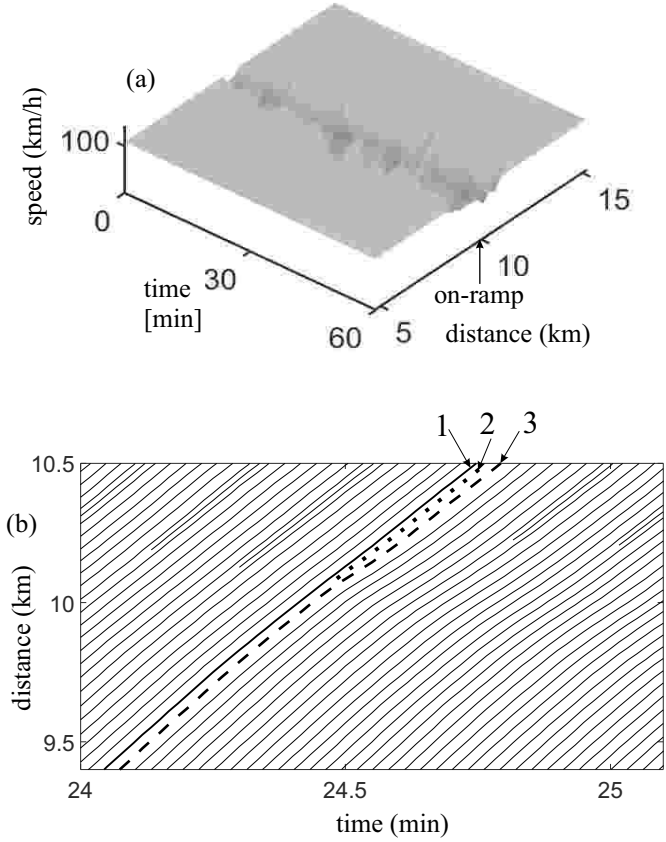}
\end{center}
\caption[]{One of the realizations of simulation
of mixed traffic flow in which under conditions (\ref{Prob_ACC_formula})
 no traffic breakdown is observed during
the observation time interval $T_{\rm ob}=$ 60 min in mixed traffic flow
with $2 \%$ ACC-vehicles with $\tau^{\rm (ACC)}_{\rm d}$1.3 s:
(a) Vehicle speed in space and time. (b) Fragment of vehicle trajectories.
The on-ramp inflow rate $q_{\rm on}=$ 280 vehicles/h
 and the flow rate on the main road $q_{\rm in}=$ 2000 vehicles/h.
Other model parameters are the same as those in Fig.~\ref{Probability_ACC}.
}
\label{Realization_ACC_1-3}
\end{figure}

  \begin{figure}
\begin{center}
\includegraphics*[scale=.5]{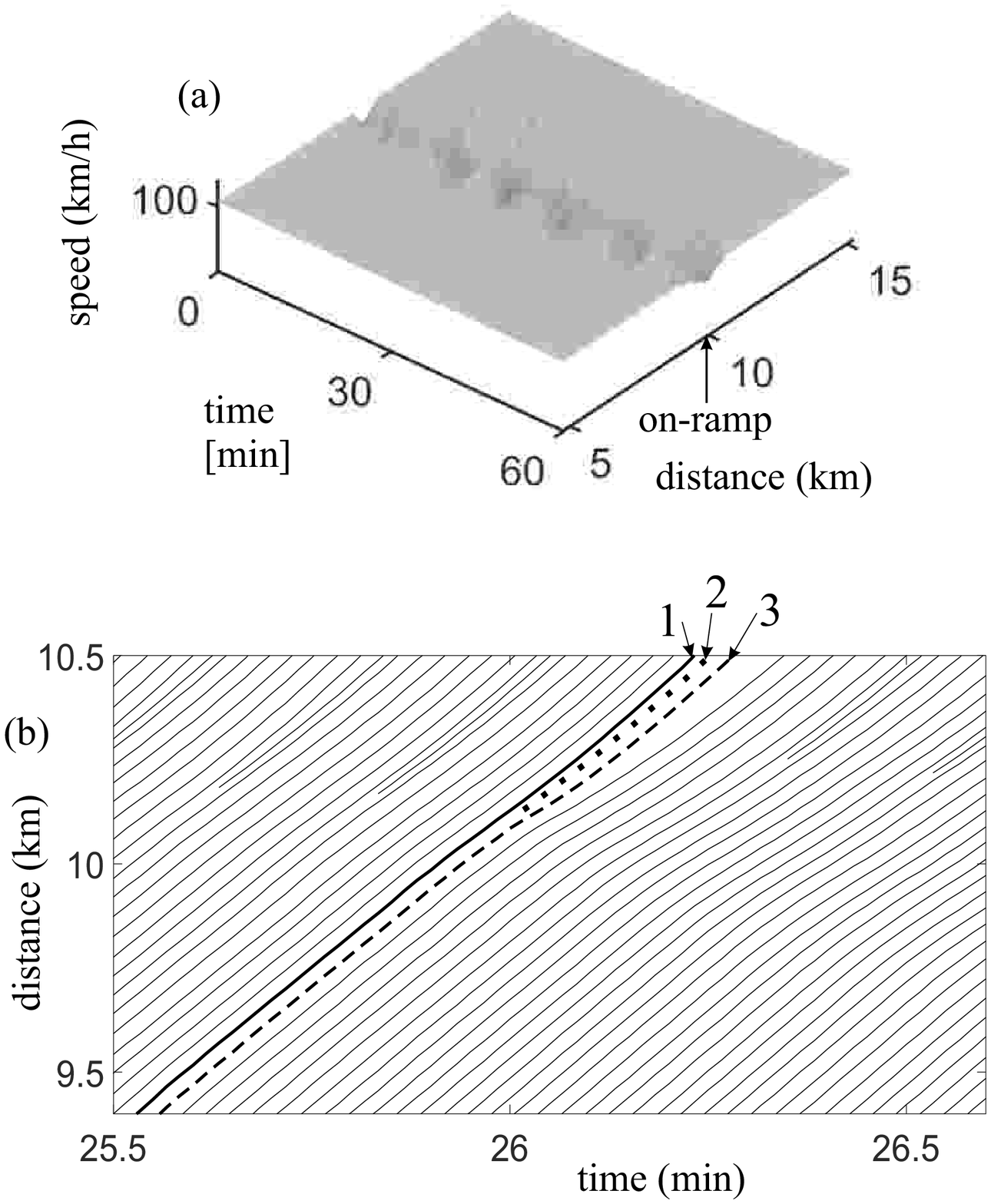}
\end{center}
\caption[]{One of the realizations of simulation
of mixed traffic flow in which under conditions (\ref{Prob_ACC_formula})
 no traffic breakdown is observed during
the observation time interval $T_{\rm ob}=$ 60 min in mixed traffic flow
with $2 \%$ ACC-vehicles with $\tau^{\rm (ACC)}_{\rm d}=$1.5 s:
(a) Vehicle speed in space and time. (b) Fragment of vehicle trajectories.
The on-ramp inflow rate $q_{\rm on}=$ 280 vehicles/h
 and the flow rate on the main road $q_{\rm in}=$ 2000 vehicles/h.
Other model parameters are the same as those in Fig.~\ref{Probability_ACC}.
}
\label{Realization_ACC_1-5}
\end{figure}

  \begin{figure}
\begin{center}
\includegraphics*[scale=.55]{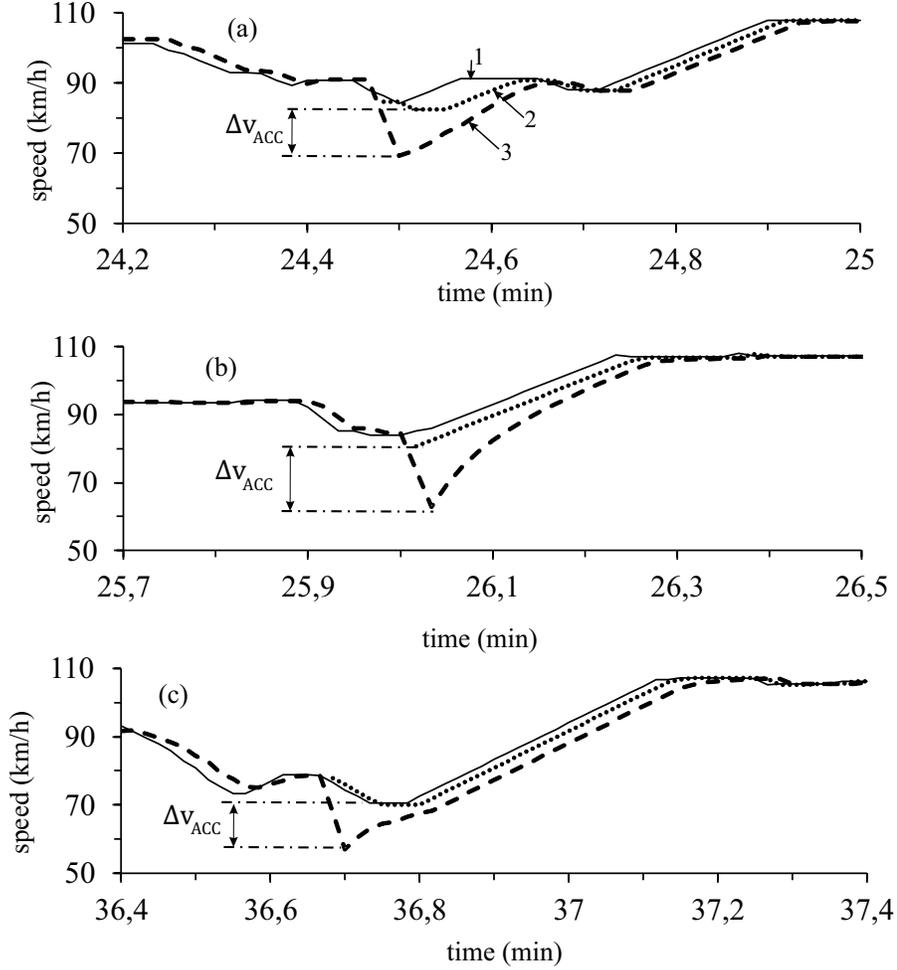}
\end{center}
\caption[]{Microscopic speed of ACC-vehicles following a human driving vehicle that merges from the on-ramp
on the main road for different values
$\tau^{\rm (ACC)}_{\rm d}$:
(a) The simulation realization shown in Fig.~\ref{Realization_ACC_1-3} for
$\tau^{\rm (ACC)}_{\rm d}=$1.3 s.
(b) The simulation realization shown in Fig.~\ref{Realization_ACC_1-5} for
$\tau^{\rm (ACC)}_{\rm d}=$1.5 s.
(c) One of the simulation realization for
$\tau^{\rm (ACC)}_{\rm d}=$2 s.
The on-ramp inflow rate $q_{\rm on}=$ 280 vehicles/h
 and the flow rate on the main road $q_{\rm in}=$ 2000 vehicles/h.
Other model parameters are the same as those in Fig.~\ref{Probability_ACC}.
}
\label{Disturbances_ACC_1-3_1-5_2}
\end{figure}

\begin{figure}
\begin{center}
\includegraphics*[scale=.65]{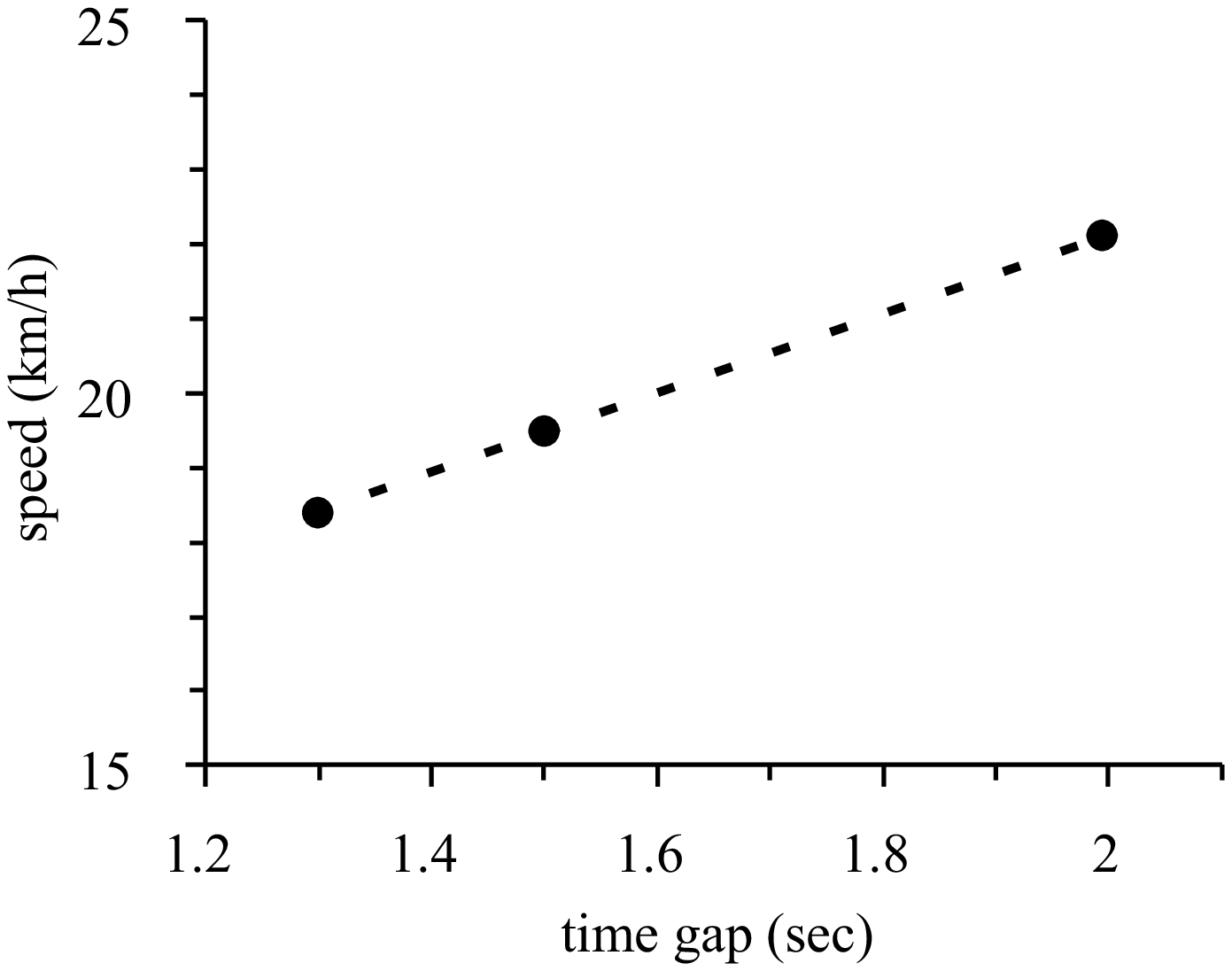}
\end{center}
\caption[]{Mean values of the amplitude
of local speed reduction of ACC-vehicles following a human driving vehicle that merges from the on-ramp
on the main road for different values
$\tau^{\rm (ACC)}_{\rm d}$  (black circles) that  are the same as those in Fig.~\ref{Disturbances_ACC_1-3_1-5_2}.
}
\label{DifferenceDisturbances_ACC_1-3_1-5_2}
\end{figure}

	Conditions (\ref{Prob_ACC_formula}) allows us to choose those simulation realizations
	for
	mixed traffic flow with $2 \%$  ACC-vehicles in which no traffic breakdown is observed during the observation time interval
	$T_{\rm ob}$ for any chosen desired time headway of ACC within the range (\ref{range_time}).
	For these simulation realizations, we make a statistical analysis of the amplitudes of  
	speed disturbances caused by ACC-vehicles for the case, when
	an ACC-vehicle follows a human driving vehicle that merges from the on-ramp onto the main road.
	Examples of a simulation realization for $\tau^{\rm (ACC)}_{\rm d}=$1.3 s
	and a simulation realization for $\tau^{\rm (ACC)}_{\rm d}=$1.5 s
	are shown, respectively, in Figs.~\ref{Realization_ACC_1-3} and~\ref{Realization_ACC_1-5}.
	
	As the amplitude (denoted by $\Delta v_{\rm ACC}$) of a speed disturbance caused by an ACC-vehicle, we consider a difference between the  speed of vehicle 2   that merges from the on-ramp onto the main road 
	and the minimum speed of the ACC-vehicle
	(vehicle 3 in Fig.~\ref{Disturbances_ACC_1-3_1-5_2}) following this merging vehicle 2:
\begin{equation}
\Delta v_{\rm ACC}=v_{\ell, m}-v^{\rm (ACC)}_{\rm min},
\label{Combain08}
\end{equation}
where $v_{\ell, m}$ is the speed of   vehicle 2 at time step $n=m$ when vehicle 2    becomes the preceding vehicle for the
  ACC-vehicle, $v^{\rm (ACC)}_{\rm min}$ is the minimum speed of the ACC-vehicle.

	As in Fig.~\ref{Break_ACC}, in Fig.~\ref{Realization_ACC_1-3} (b)  
vehicles 1 and 2 are human driving vehicles, whereas    vehicle 3 is ACC-vehicle.
The microscopic ACC-vehicle speed (vehicle 3 in Fig.~\ref{Realization_ACC_1-3} (b))
as a time-function exhibits a relatively large amplitude of a local speed reduction
 (Fig.~\ref{Disturbances_ACC_1-3_1-5_2} (a)).
A large amplitude of a local speed reduction exhibits also
an ACC-vehicle with $\tau^{\rm (ACC)}_{\rm d}=$1.5 s
as it is shown in Fig.~\ref{Disturbances_ACC_1-3_1-5_2} (b).

Qualitative the same simulation results as that in Fig.~\ref{Realization_ACC_1-3} 
 have been found 
 for simulation realizations in which no traffic breakdown is observed during
the observation time interval $T_{\rm ob}=$ 60 min in mixed traffic flow with $2\%$  ACC-vehicles
 satisfying conditions (\ref{Prob_ACC_formula})
for any  chosen desired time headway of ACC within the range (\ref{range_time}):
When a human driving vehicle merges from the on-ramp onto the main road, then an ACC-vehicle that follows this vehicle
exhibits always a large amplitude of a local speed reduction.
An example for a large amplitude of a local speed reduction
 that shows an ACC-vehicle with $\tau^{\rm (ACC)}_{\rm d}=$ 2 s
  is shown in Fig.~\ref{Disturbances_ACC_1-3_1-5_2} (c).
	
	The amplitude of a local speed reduction
 that shows an ACC-vehicle following a human driving vehicle that merges from the on-ramp onto the main road
is a random value that depends on a simulation realization. Random amplitudes
of the local speed disturbances are realized for each of the
chosen desired time headway of ACC   $\tau^{\rm (ACC)}_{\rm d}$.
This means that
each of the   amplitudes of local speed disturbances of the ACC-vehicles shown in Fig.~\ref{Disturbances_ACC_1-3_1-5_2}
 are related to the associated simulation realization {\it only}.

For this reason,  we have made a statistical analysis of the amplitude
of the local speed disturbances: For each of the
chosen desired time headway of ACC   $\tau^{\rm (ACC)}_{\rm d}$ shown in Fig.~\ref{Disturbances_ACC_1-3_1-5_2}, we
have studied $N_{\rm r}=$ 40 different random simulation realizations.
The mean values of the amplitude of local speed disturbance found from these random realizations
are presented by black circles in  Fig.~\ref{DifferenceDisturbances_ACC_1-3_1-5_2}.

We can see from dash line in Fig.~\ref{DifferenceDisturbances_ACC_1-3_1-5_2} 
 that the longer the chosen desired time headway of ACC   $\tau^{\rm (ACC)}_{\rm d}$ is, the
larger the mean amplitude of the local speed disturbance caused by the ACC-vehicle at the on-ramp bottleneck.
On the other hand, the larger the mean amplitude of the local speed disturbance (local speed reduction) at the bottleneck is, the larger the probability of traffic breakdown in the metastable free flow at the bottleneck.
This result explains the increase in the probability of traffic breakdown  at the bottleneck  with the increase in
the value of the  desired time headway of ACC   $\tau^{\rm (ACC)}_{\rm d}$ shown in Fig.~\ref{Probability_ACC}.

  \section{Effect of Dynamic Rules of Autonomous Driving on Disturbances   at Bottleneck \label{Dyn_Rules_S}}
	
	We have shown that the longer the chosen desired time headway of ACC   $\tau^{\rm (ACC)}_{\rm d}$ is, the
larger the mean amplitude of the local speed disturbance caused by the ACC-vehicle at the on-ramp bottleneck
(Fig.~\ref{DifferenceDisturbances_ACC_1-3_1-5_2}). 

Contrarily to this feature
of the classical ACC, we have found that the mean amplitude of
 the local speed disturbance caused by the TPACC-vehicle at the on-ramp bottleneck
 does not almost depend on the parameters of TPACC
within the parameter ranges  (\ref{range_time_TPACC}), (\ref{range_time_TPACC_G}).
For this reason, some effect
of single TPACC-vehicles on
the probability of traffic breakdown at the on-ramp bottleneck cannot be found (curve 1 in Fig.~\ref{Probability_ACC}).

	  \begin{figure}
\begin{center}
\includegraphics*[scale=.6]{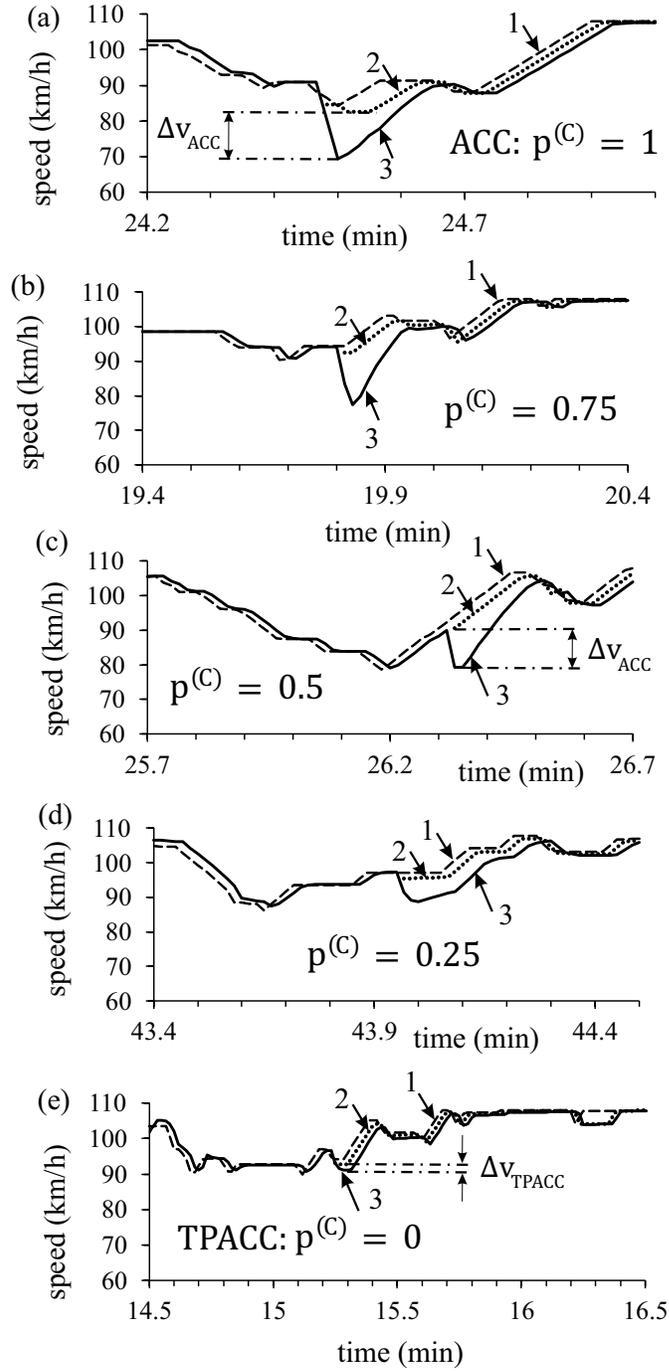}
\end{center}
\caption[]{Local speed reduction 
caused by ACC-vehicle at on-ramp bottleneck related to ACC-model
(\ref{Combain01})--(\ref{Combain05}) in one of simulation realizations
for five different values  $p^{\rm(C)}=$ 0 (a), 0.25 (b), 0.5 (c), 0.75 (d), and 1 (e).
Vehicles 1 and 2 are human driving vehicles, vehicle 3 is the ACC-vehicle.
Model parameters are:  $\tau_{\rm G}=$1.4 s, $\tau_{\rm p}=$1.3 s, 
$K_{1}=$0.3 $s^{-2}$, $K_{2}=K_{\rm \Delta v}=$0.6$s^{-1}$.
Other model parameters are the same as those in Fig.~\ref{Probability_ACC}. 
}
\label{Disturbances_TPACC_ACC_1-3}
\end{figure}

	  \begin{figure}
\begin{center}
\includegraphics*[scale=.65]{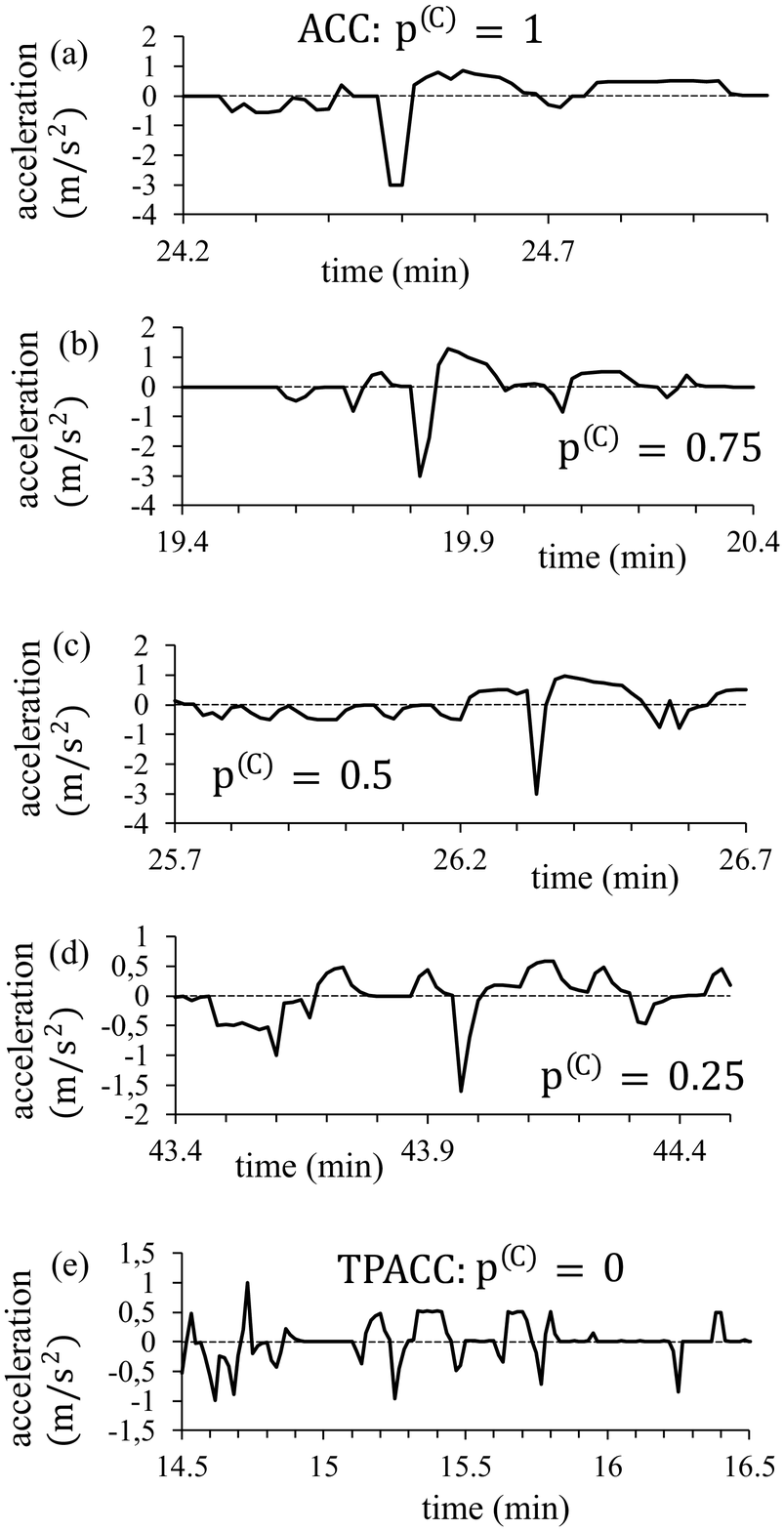}
\end{center}
\caption[]{Deceleration (acceleration) of  
  ACC-vehicle   in the same simulation realizations as those in Fig.~\ref{Disturbances_TPACC_ACC_1-3}, respectively,
for five different values  $p^{\rm(C)}=$ 0 (a), 0.25 (b), 0.5 (c), 0.75 (d), and 1 (e).
}
\label{Deceleration_TPACC_ACC_1-3}
\end{figure}

\begin{figure}
\begin{center}
\includegraphics*[scale=.65]{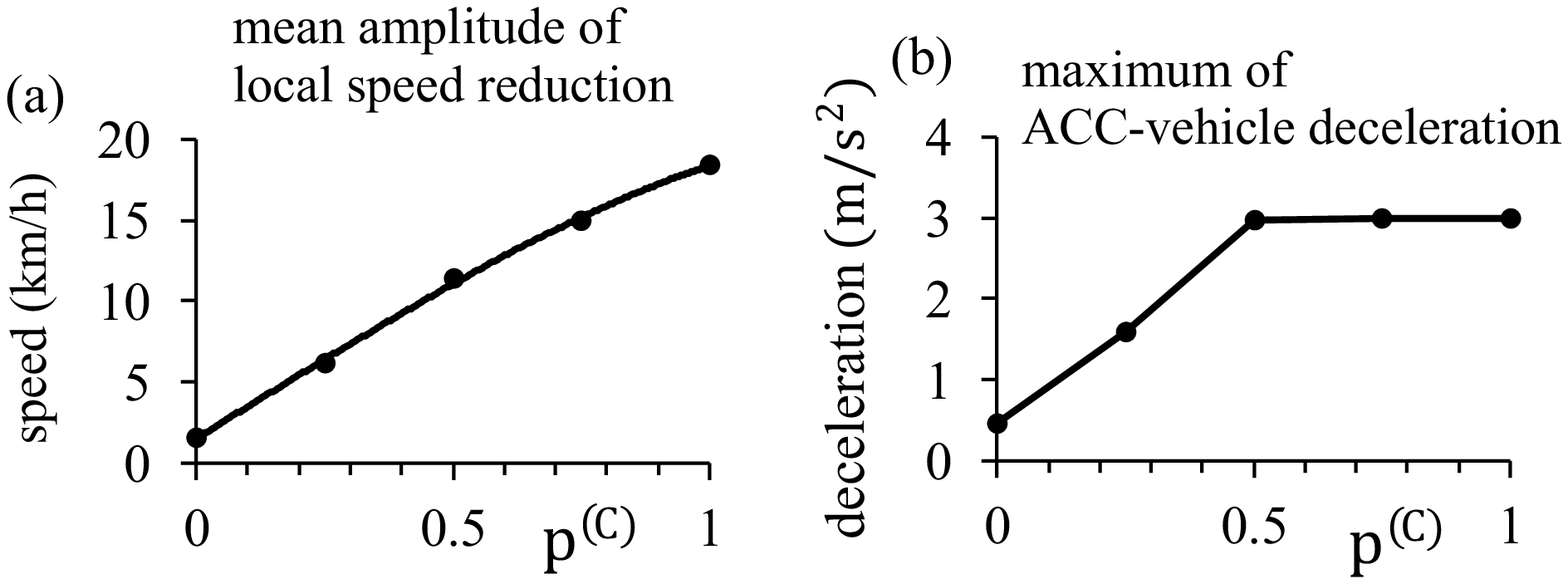}
\end{center}
\caption[]{Characteristics of local speed reductions 
caused by ACC-vehicles at on-ramp bottleneck as a function of   value  $p^{\rm(C)}$
for model parameters of Fig.~\ref{Disturbances_TPACC_ACC_1-3}: 
(a) Mean amplitude of local speed reduction. (b)  Mean value of maximum ACC-vehicle deceleration within local speed reduction.
}
\label{DifferenceDisturbances_TPACC_ACC_1-3}
\end{figure}

	  \begin{figure}
\begin{center}
\includegraphics*[scale=.65]{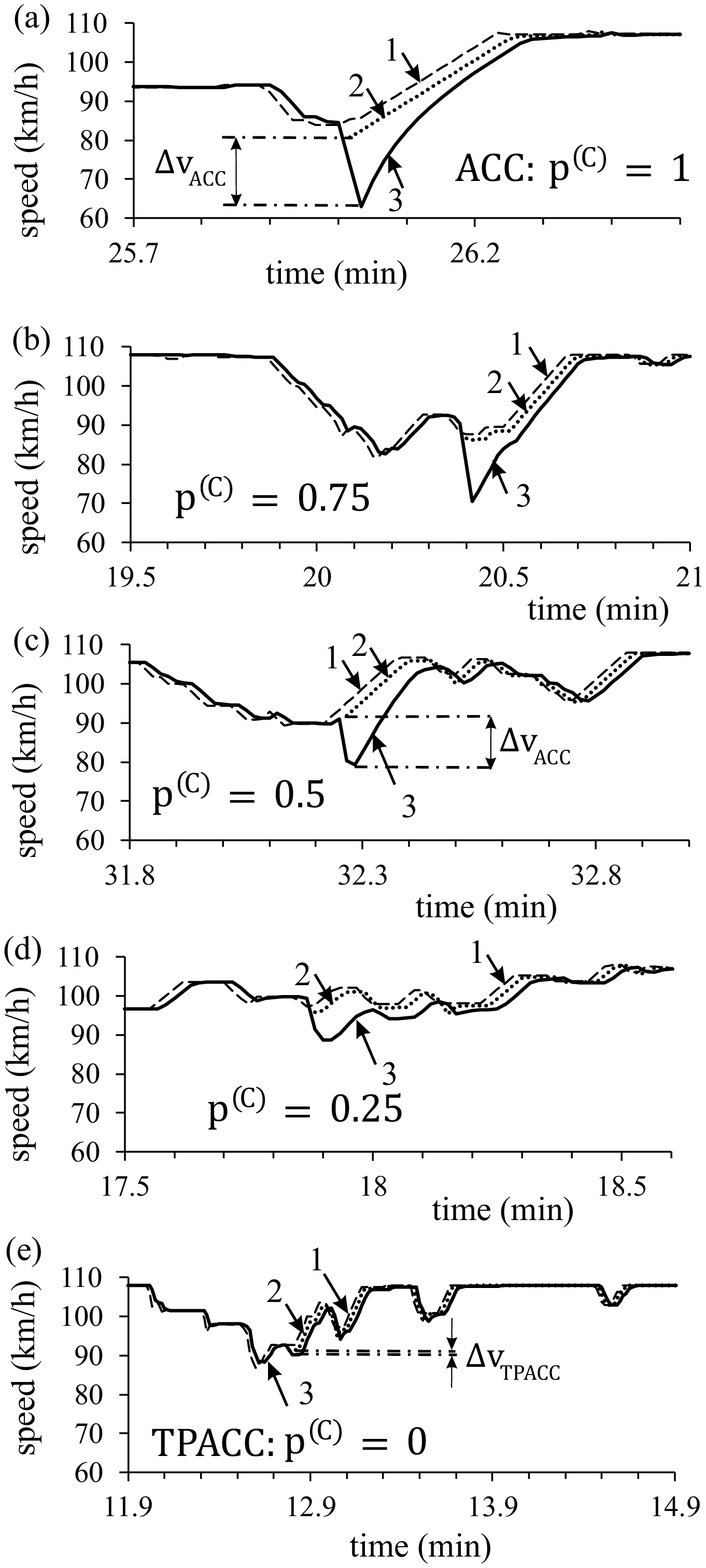}
\end{center}
\caption[]{Local speed reduction 
caused by ACC-vehicle at on-ramp bottleneck  related to ACC-model
(\ref{Combain01})--(\ref{Combain05}) in one of simulation realizations
for five different values  $p^{\rm(C)}=$ 0 (a), 0.25 (b), 0.5 (c), 0.75 (d), and 1 (e).
Vehicles 1 and 2 are human driving vehicles, vehicle 3 is the ACC-vehicle.
  $\tau_{\rm G}=$1.6 s, $\tau_{\rm p}=$1.5 s. 
Other model parameters are the same as those in Fig.~\ref{Disturbances_TPACC_ACC_1-3}.
}
\label{Disturbances_TPACC_ACC_1-5}
\end{figure}

	  \begin{figure}
\begin{center}
\includegraphics*[scale=.6]{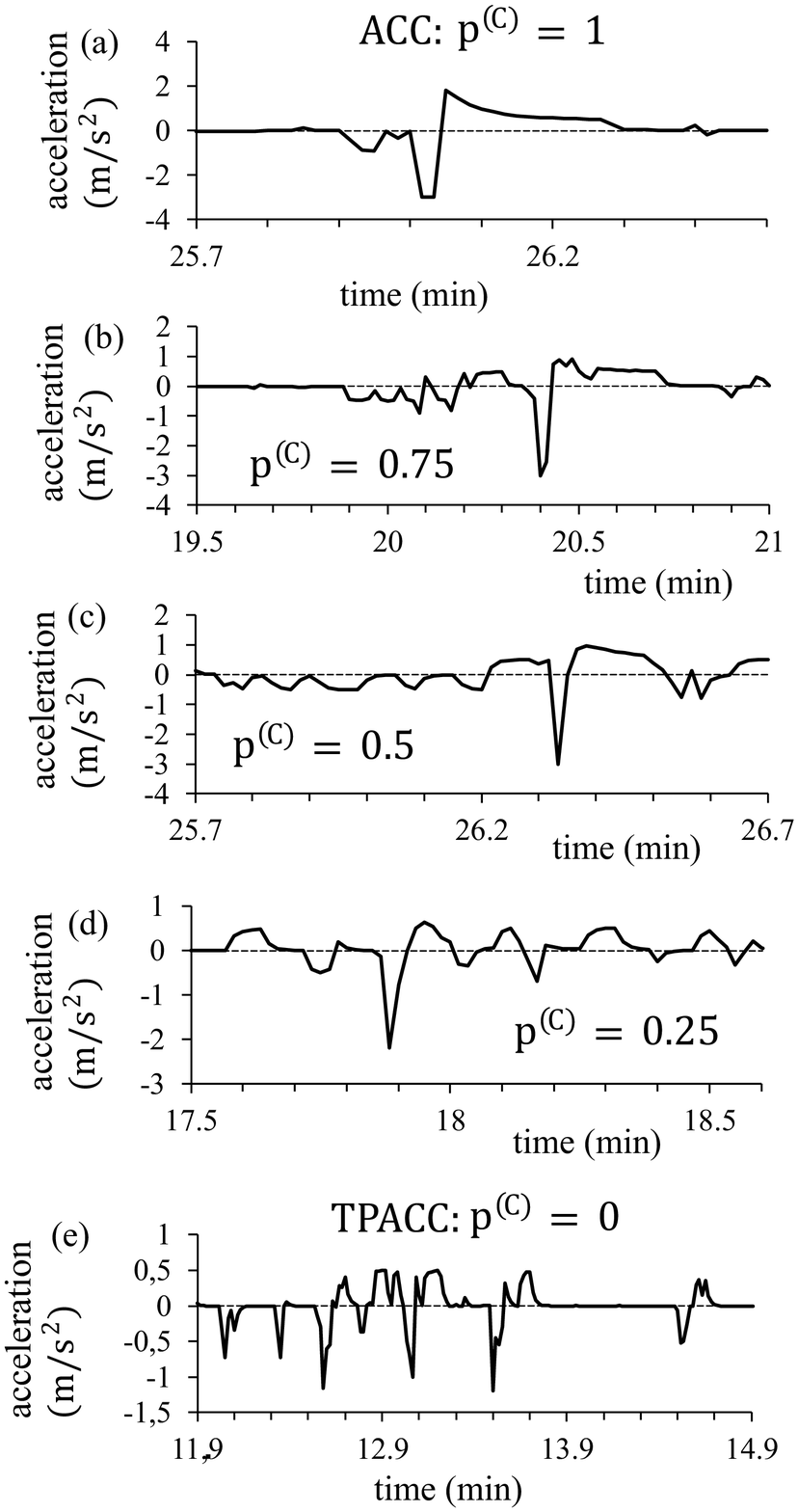}
\end{center}
\caption[]{Deceleration (acceleration) of  
  ACC-vehicle   in the same simulation realizations as those in Fig.~\ref{Disturbances_TPACC_ACC_1-5}, respectively,
for five different values   $p^{\rm(C)}=$ 0 (a), 0.25 (b), 0.5 (c), 0.75 (d), and 1 (e).
}
\label{Deceleration_TPACC_ACC_1-5}
\end{figure}

\begin{figure}
\begin{center}
\includegraphics*[scale=.6]{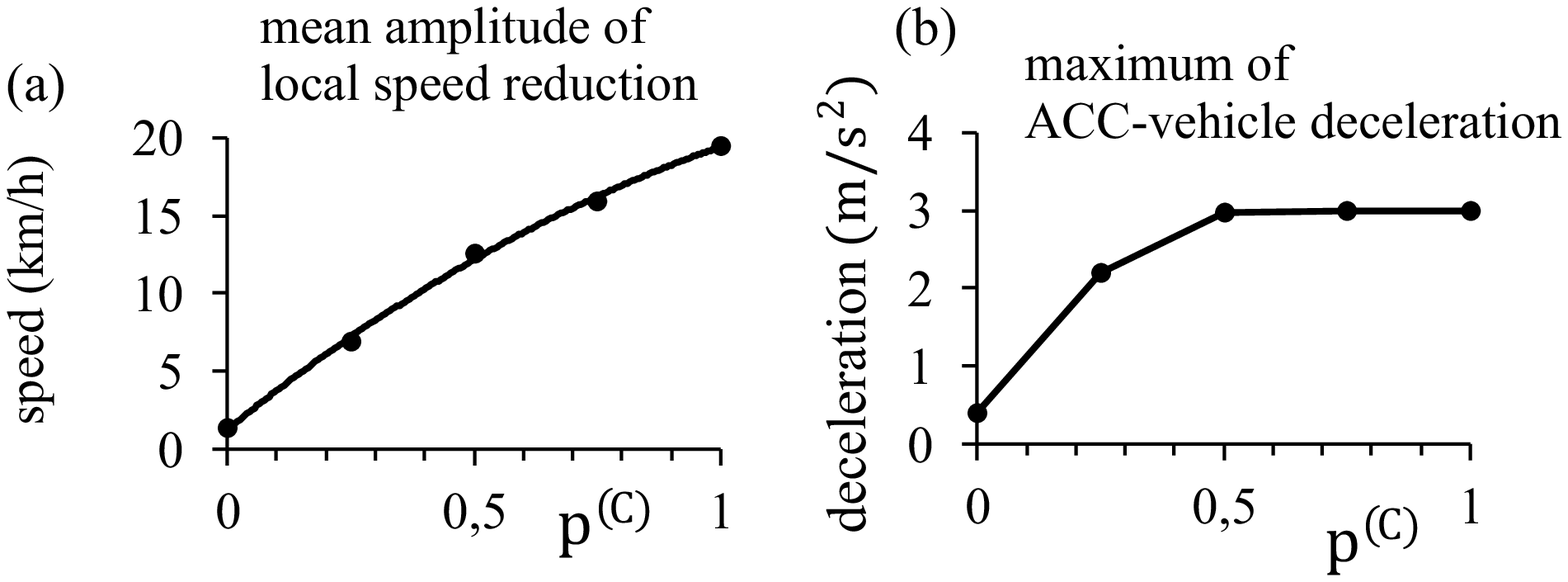}
\end{center}
\caption[]{Characteristics of local speed reductions 
caused by ACC-vehicles at on-ramp bottleneck as a function of   value  $p^{\rm(C)}$
for model parameters of Fig.~\ref{Disturbances_TPACC_ACC_1-5}: 
(a) Mean amplitude of local speed reduction. (b)   Mean value of maximum ACC-vehicle deceleration within local speed reduction.
}
\label{DifferenceDisturbances_TPACC_ACC_1-5}
\end{figure}

	  \begin{figure}
\begin{center}
\includegraphics*[scale=.65]{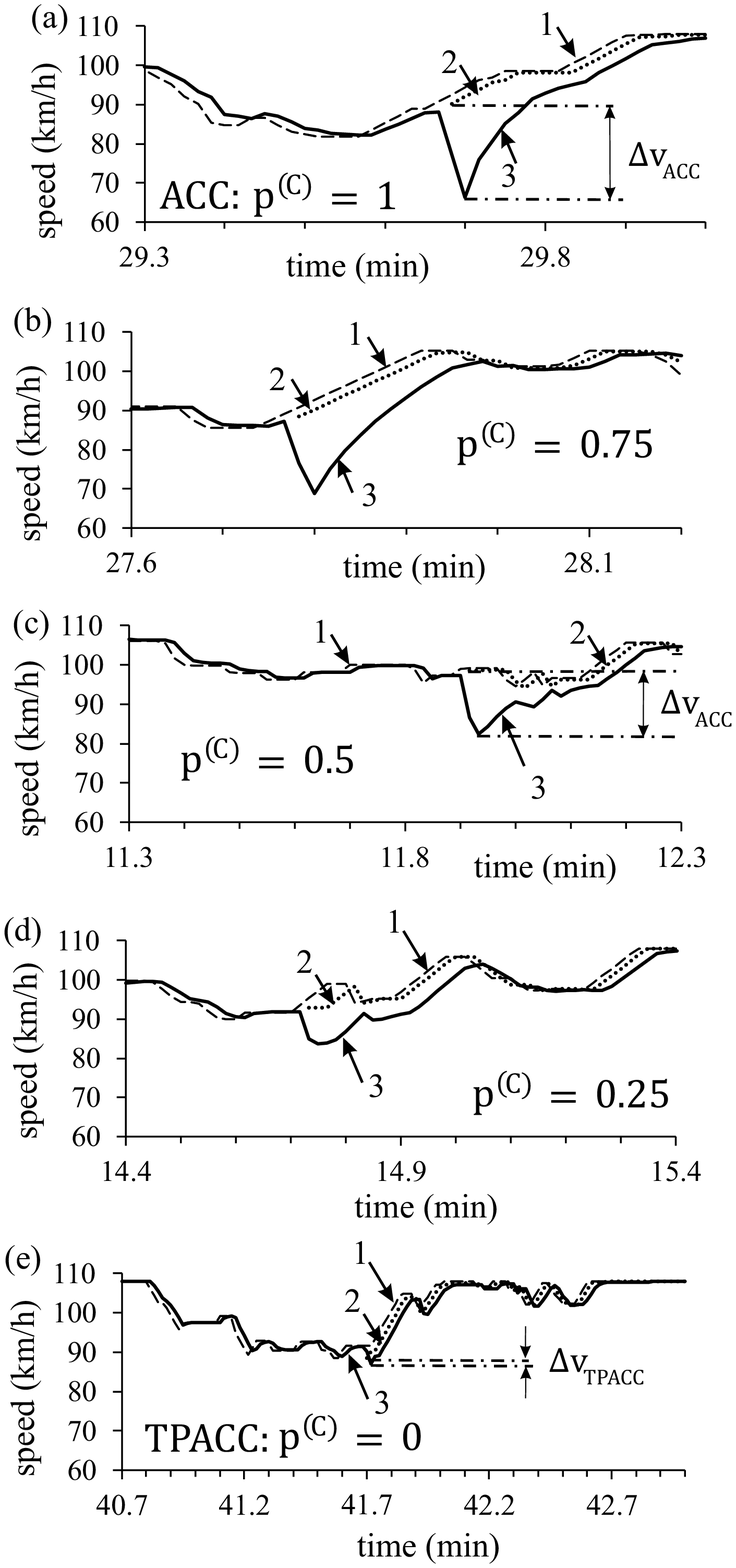}
\end{center}
\caption[]{Local speed reduction 
caused by ACC-vehicle at on-ramp bottleneck  related to ACC-model
(\ref{Combain01})--(\ref{Combain05}) in one of simulation realizations
for five different values  $p^{\rm(C)}=$ 0 (a), 0.25 (b), 0.5 (c), 0.75 (d), and 1 (e).
Vehicles 1 and 2 are human driving vehicles, vehicle 3 is the ACC-vehicle.
   $\tau_{\rm G}=$2.2 s, $\tau_{\rm p}=$2 s.
	Other model parameters are the same as those in Fig.~\ref{Disturbances_TPACC_ACC_1-3}.
}
\label{Disturbances_TPACC_ACC_2}
\end{figure}

	  \begin{figure}
\begin{center}
\includegraphics*[scale=.6]{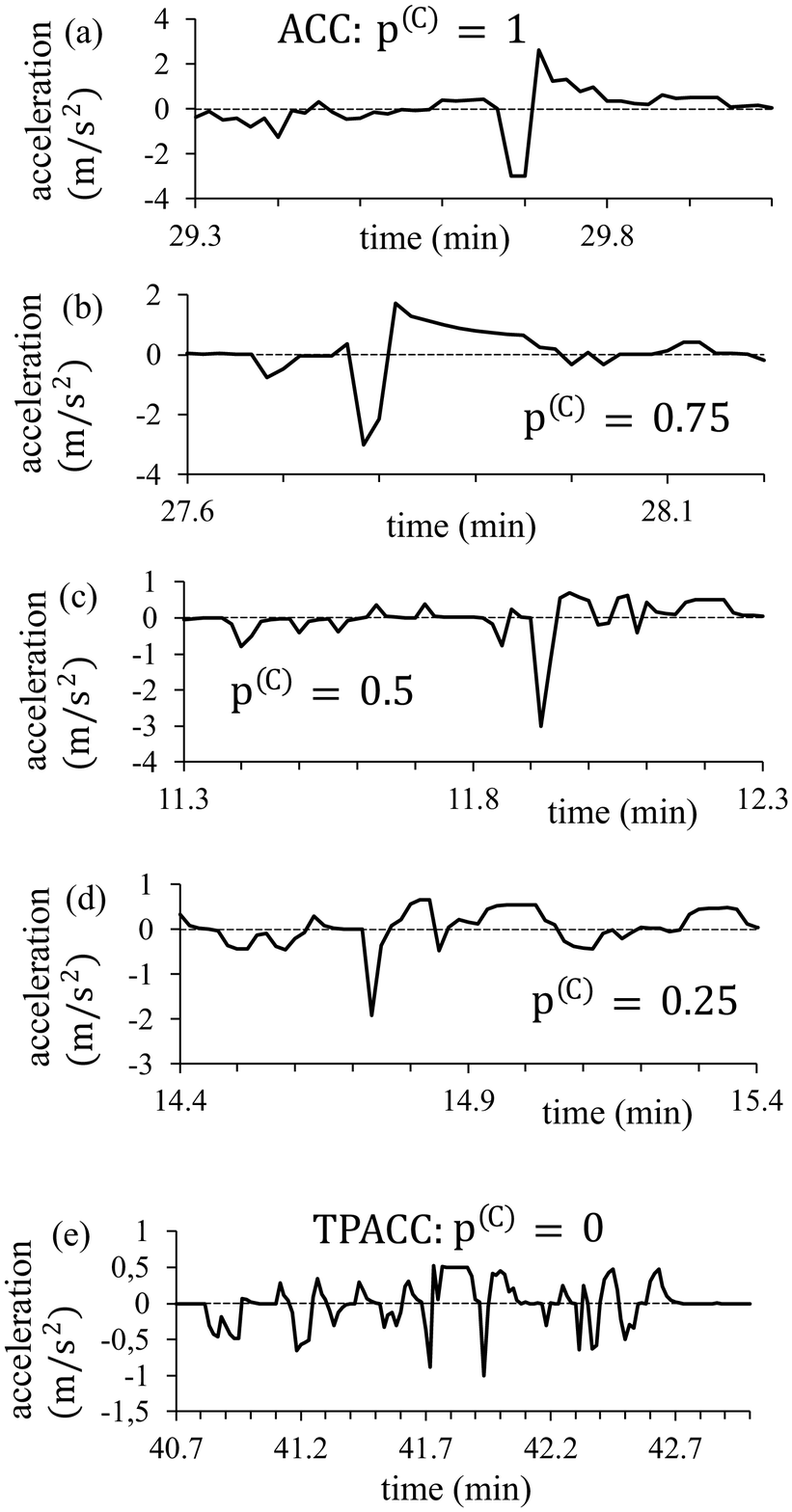}
\end{center}
\caption[]{Deceleration (acceleration) of  
  ACC-vehicle   in the same simulation realizations as those in Fig.~\ref{Disturbances_TPACC_ACC_2}, respectively,
for five different values  $p^{\rm(C)}=$ 0 (a), 0.25 (b), 0.5 (c), 0.75 (d), and 1 (e).
}
\label{Deceleration_TPACC_ACC_2}
\end{figure}

\begin{figure}
\begin{center}
\includegraphics*[scale=.6]{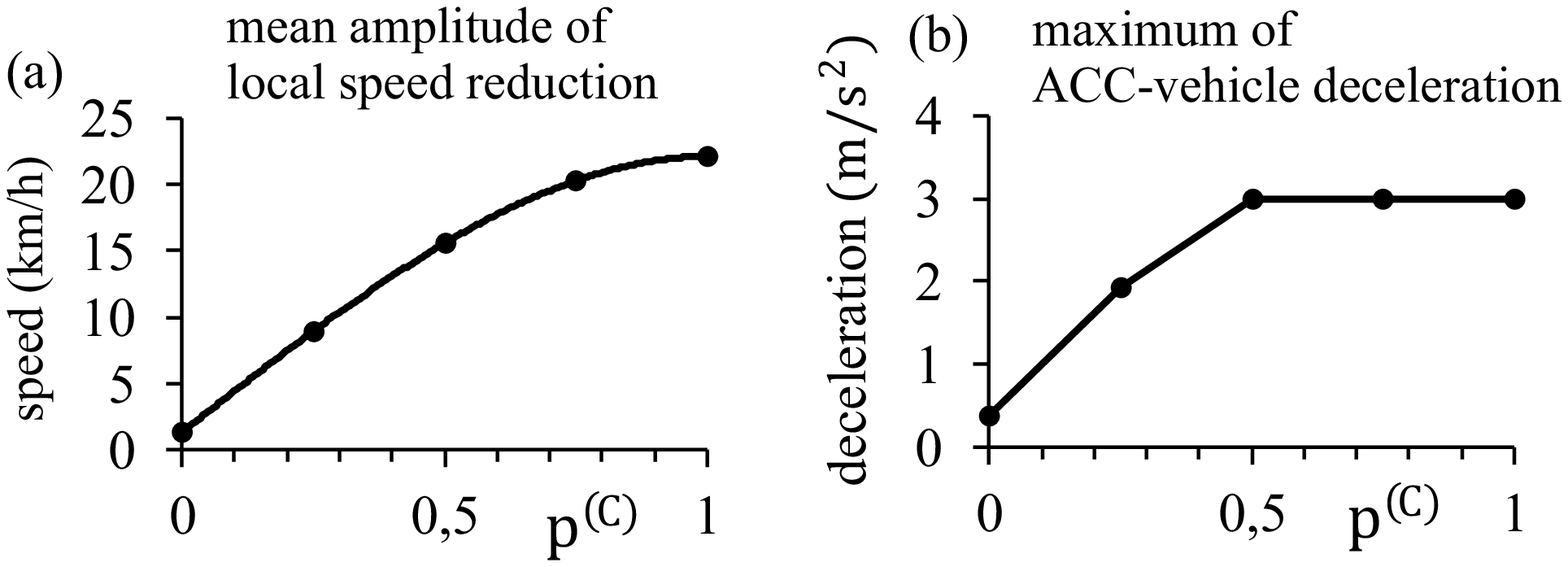}
\end{center}
\caption[]{Characteristics of local speed reductions 
caused by ACC-vehicles at on-ramp bottleneck as a function of   value  $p^{\rm(C)}$
for model parameters of Fig.~\ref{Disturbances_TPACC_ACC_2}: 
(a) Mean amplitude of local speed reduction. (b)   Mean value of maximum ACC-vehicle deceleration within local speed reduction.
}
\label{DifferenceDisturbances_TPACC_ACC_2}
\end{figure}
  
The  mathematical difference between dynamic rules of
classical ACC  (\ref{ACC_General}) and TPACC (\ref{TPACC_main5}) seems to be small.
Therefore, the following question arises: How can
such seeming small mathematical difference between dynamic rules of
classical ACC and TPACC lead to such a large effect on the
local speed disturbances in free flow and, respectively, to a large effect on the 
probability of traffic breakdown at the   bottleneck?

To answer this question, we introduce a model of ACC,
in which rather than a classical formula (\ref{ACC_General}) for ACC-acceleration,
the ACC-acceleration denoted by $a^{\rm(C)}$ is given by the following formula:
\begin{eqnarray}
a^{\rm(C)}=
\left\{\begin{array}{ll}
\tilde a^{\rm(C)} &  \textrm{at $g \leq G^{\rm(C)}$,} \\
a^{\rm(ACC)} &  \textrm{at $g > G^{\rm(C)}$}, \\
\end{array} \right.
\label{Combain01}
\end{eqnarray}
where 
\begin{equation}
\tilde a^{\rm(C)}=a^{\rm(2D)}(1-p^{\rm(C)})+a^{\rm(ACC)}p^{\rm(C)},
\label{Combain02}
\end{equation}
\begin{equation}
a^{\rm(2D)}=K_{\rm \Delta v}\Delta v,
\label{Combain03}
\end{equation}
\begin{equation}
a^{\rm(ACC)}=K_{1}(g -v \tau_{\rm p})+ K_{2}\Delta v,
\label{Combain04}
\end{equation}
\begin{equation}
G^{\rm(C)} =G (1-p^{\rm(C)})+v \tau_{\rm p}p^{\rm(C)},
\label{Combain05}
\end{equation}
a constant parameter $p^{\rm(C)}$ satisfies condition 
\begin{equation}
0 \leq p^{\rm(C)} \leq 1,
\label{Combain06}
\end{equation}
it is also assumed that conditions  (\ref{TPACC_main5_1}) and (\ref{TPACC_main5_5}) are satisfied. 
 A discrete in time version of the model of ACC
(\ref{Combain01})--(\ref{Combain05})  used in simulations is presented in Appendix~\ref{Models_ACC_TP_Sec}.

It should be emphasized that   at
$p^{\rm(C)}=1$ and  $\tau^{\rm (ACC)}_{\rm d}=\tau_{\rm p}$ the ACC-model (\ref{Combain01})--(\ref{Combain05}) 
transforms to the classical model for ACC  (\ref{ACC_General}). 
Contrarily, at $p^{\rm(C)}=0$
 the ACC-model (\ref{Combain01})--(\ref{Combain05}) transforms to the TPACC-model  (\ref{TPACC_main5}).
In other words,  in the ACC model (\ref{Combain01})--(\ref{Combain05}) by the increasing
 in parameter $p^{\rm(C)}$ (\ref{Combain06})  
  the rules for ACC-vehicle motion (\ref{Combain01})--(\ref{Combain05})
 are continuously changed from the rules for the TPACC-model  (\ref{TPACC_main5}) at $p^{\rm(C)}=0$
 to the rules for the classical ACC-model (\ref{ACC_General}) at  
 $p^{\rm(C)}=1$.

  We have found that
the larger the value   $p^{\rm (C)}$ in the ACC-model (\ref{Combain01})--(\ref{Combain05})
 is, the stronger on average the reaction of the ACC-vehicle (vehicle 3 in 
Figs.~\ref{Disturbances_TPACC_ACC_1-3},
~\ref{Disturbances_TPACC_ACC_1-5}, and~\ref{Disturbances_TPACC_ACC_2}) on
the difference between
the desired time headway of ACC   $\tau^{\rm (ACC)}_{\rm d}=\tau_{\rm p}$ and a current time headway to the preceding vehicle
(vehicle 2) is. This result remains for   different values $\tau_{\rm p}$ related to time headway ranges
(\ref{range_time}), (\ref{range_time_TPACC})  
and (\ref{range_time_TPACC_G}), respectively 
(Figs.~\ref{Disturbances_TPACC_ACC_1-3}--\ref{DifferenceDisturbances_TPACC_ACC_2}).
 In particular,
  we have found the following general results:

(i) When the parameter $p^{\rm(C)}$ increases from $p^{\rm(C)}=0$ (TPACC)  to $p^{\rm(C)}=1$ (classical ACC),
  the  amplitude of a local speed reduction   (Figs.~\ref{Disturbances_TPACC_ACC_1-3},
~\ref{Disturbances_TPACC_ACC_1-5}, and~\ref{Disturbances_TPACC_ACC_2})
caused by the ACC-vehicle (\ref{Combain01})--(\ref{Combain05})
at the on-ramp bottleneck
increases continuously. Therefore,
the increase in the mean amplitude of a local speed reduction  caused by the ACC-vehicle (\ref{Combain01})--(\ref{Combain05})
at the on-ramp bottleneck  is almost linear
function of the increase in  parameter $p^{\rm (C)}$ (Figs.~\ref{DifferenceDisturbances_TPACC_ACC_1-3} (a),
~\ref{DifferenceDisturbances_TPACC_ACC_1-5} (a), and~\ref{DifferenceDisturbances_TPACC_ACC_2} (a)).

(ii) When the parameter $p^{\rm(C)}$ increases from   $p^{\rm(C)}=0$ (TPACC) to
$p^{\rm(C)}=1$ (classical ACC),  the   deceleration of the ACC-vehicle (\ref{Combain01})--(\ref{Combain05})
within a local speed reduction caused by the ACC-vehicle (\ref{Combain01})--(\ref{Combain05}) at the bottleneck increases strongly. 
For values of $p^{\rm(C)}$ that are close to 1,
the   deceleration of the ACC-vehicle reaches  
the maximum value $- 3 \ \rm m s^{-2}$ (Figs.~\ref{DifferenceDisturbances_TPACC_ACC_1-3} (b),
~\ref{DifferenceDisturbances_TPACC_ACC_1-5} (b), and~\ref{DifferenceDisturbances_TPACC_ACC_2} (b)) chosen in the model
of the ACC-vehicle (\ref{Combain01})--(\ref{Combain05}) for usual driving conditions\footnote{Under model
 parameters  used in simulations there have been found no cases when
a security deceleration caused by safety conditions
are realized: Safety conditions can lead to a considerably stronger ACC deceleration than the value
 $- 3 \ \rm m s^{-2}$. Safety conditions in the model 
the ACC-vehicle (\ref{Combain01})--(\ref{Combain05}) are mathematically described by safe speed
$v_{{\rm s},n}$ as shown in Appendix~\ref{Models_ACC_TP_Sec}.
}.
When for $p^{\rm (C)}=1$ (classical ACC) the desired time headway of ACC   $\tau^{\rm (ACC)}_{\rm d}=\tau_{\rm p}$  
 increases from 1.3 s to 2 s, the duration of the time interval within which
the   deceleration of the ACC-vehicle reaches  
the maximum value $- 3 \ \rm m s^{-2}$ increases considerably 
(Figs.~\ref{Deceleration_TPACC_ACC_1-3} (b),
~\ref{Deceleration_TPACC_ACC_1-5} (b), and~\ref{Deceleration_TPACC_ACC_2} (b)).

This   comparison of classical ACC (\ref{ACC_General}) with TPACC (\ref{TPACC_main5}) confirms that   
the stronger the reaction of ACC-vehicle on the difference between
the desired time headway of ACC   $\tau^{\rm (ACC)}_{\rm d}$ and a current time headway to the preceding vehicle,
the larger on average  the amplitude of local speed reduction caused by the ACC-vehicle
at the bottleneck and, therefore, the larger the probability of traffic breakdown at the bottleneck.
Indeed, within the time headway range (\ref{TPACC_main_range}) TPACC-vehicle ($p^{\rm (C)}=0$)
does no react on a current time headway to the preceding vehicle. Due to the existence of such an indifferent zone
in car-following the TPACC-vehicle either does not produce
a local speed reduction at the bottleneck at all or the local speed reduction caused by the TPACC-vehicle
 is of a very small amplitude. 

Contrarily to TPACC ($p^{\rm(C)}=0$), when $p^{\rm (C)}>0$ and it 
 increases  continuously, the  deceleration of
the ACC-vehicle (\ref{Combain01})--(\ref{Combain05}) at the bottleneck caused by 
  the difference between
the desired time headway of ACC-vehicle   $\tau^{\rm (ACC)}_{\rm d}=\tau_{\rm p}$ and a current time headway to the preceding vehicle
increases on average continuously
(Figs.~\ref{Deceleration_TPACC_ACC_1-3} (a--d),
~\ref{Deceleration_TPACC_ACC_1-5} (a--d), and~\ref{Deceleration_TPACC_ACC_2} (a--d)).
This leads to  the largest  mean  amplitude of a local speed reduction caused by the ACC-vehicle (\ref{Combain01})--(\ref{Combain05}) at the bottleneck at   $p^{\rm (C)}=1$ (Figs.~\ref{DifferenceDisturbances_TPACC_ACC_1-3} (a),
~\ref{DifferenceDisturbances_TPACC_ACC_1-5} (a), and~\ref{DifferenceDisturbances_TPACC_ACC_2} (a)).
This explains the following results found in simulations:
 For a given time headway $\tau^{\rm (ACC)}_{\rm d}$, the larger the value $p^{\rm (C)}$ in the model of
the ACC-vehicle (\ref{Combain01})--(\ref{Combain05}), the larger  
 the probability of traffic breakdown at the bottleneck for the same other model parameters.
This increase in the probability of traffic breakdown becomes the stronger, 
the longer time headway  $\tau^{\rm (ACC)}_{\rm d}=\tau_{\rm p}$ is   in the ACC-model
 (\ref{Combain01})--(\ref{Combain05}). This result correlates with the conclusion of
  Sec.~\ref{ACC_Cl_Param_Prob_Sub} that  
  the longer the desired time headway    $\tau^{\rm (ACC)}_{\rm d}$ of classical ACC
	($p^{\rm (C)}=1$ in (\ref{Combain01})--(\ref{Combain05})), the stronger the increase in the
probability of traffic breakdown caused by a single classical ACC-vehicle at the bottleneck
(curves 2--4 in Fig.~\ref{Probability_ACC}).

\section{Platoons of Autonomous Driving Vehicles and
  Probability of Traffic Breakdown in Mixed Traffic Flow \label{Prob_Platoon_S}}  

\begin{figure}
\begin{center}
\includegraphics*[scale=.55]{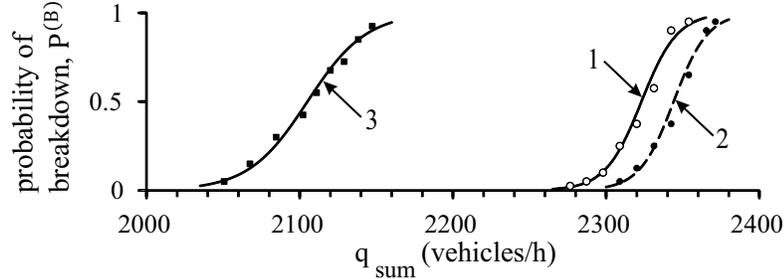}
\end{center}
\caption[]{Probability of traffic breakdown at on-ramp bottleneck
as a function of the flow rate $q_{\rm sum}=q_{\rm in}+q_{\rm on}$
at a given on-ramp inflow rate $q_{\rm on}=$ 320 vehicles/h  
   in mixed traffic flow
with 20$\%$ autonomous driving vehicles:
Curve 1 is related to
traffic flow without autonomous driving vehicles. Curves 2 and 3 are related to mixed traffic flow with
TPACC-vehicles (curve 2) and ACC-vehicles (curve 3). 
Simulation parameters of ACC   and TPACC   
   are, respectively, the same as those in Fig.~\ref{Prob2_ACC}.
}
\label{Break2_ACC}
\end{figure}

 If the share of autonomous driving vehicles in mixed traffic flow increases,
 (Fig.~\ref{Break2_ACC}), the probability of traffic breakdown caused by ACC-vehicles
that deteriorate traffic  
can increase   considerably 
(curve 3 in Fig.~\ref{Break2_ACC}). 

Contrarily to classical ACC-vehicles, long enough platoons of TPACC-vehicles in mixed traffic flow
  decrease the breakdown probability (curve 2 in Fig.~\ref{Break2_ACC}).
	This is explained by   small local speed disturbances
	 caused by TPACC-vehicles in mixed free flow at the bottleneck in comparison with large
	local speed disturbances 
	 caused by classical ACC-vehicles, as already discussed in Sec.~\ref{Dyn_Rules_S}.

In particular, the  reduction of
  the probability of traffic breakdown in mixed traffic flow
at the bottleneck though already small platoons of TPACC-vehicles
(curve 2 in Fig.~\ref{Break2_ACC}) is   explained by the speed adaptation effect 
of the three-phase theory that is the basis of TPACC  (\ref{TPACC_main5}):
 At each vehicle speed, the TPACC-vehicle   makes an arbitrary choice 
in   time headway that satisfies conditions (\ref{TPACC_main_range}). In other words,  
 the TPACC-vehicle 
accepts different values of time headway  at different times and does not control
  a fixed time headway to the preceding vehicle.
	This dynamic behavior of a platoon of TPACC-vehicles decreases 
	the amplitude of local speed disturbances at the bottleneck~\cite{Kerner2018C}.
 This explains why, in contrast to classical ACC-vehicles, TPACC-vehicles  decrease   the probability of traffic breakdown in  mixed traffic flow.

We note that in mixed traffic flow with $\gamma= 20 \%$ of classical
 ACC-vehicles at the on-ramp inflow rate
 $q_{\rm on}=$ 320 vehicles/h the probability of traffic breakdown
reaches the maximum value 
$P^{\rm (B)}=1$ already at $q_{\rm in}=$ 1830 vehicles/h  (curve 3 in Fig.~\ref{Break2_ACC}).
However, it should be noted that at parameters of ACC-vehicles used for simulations of
 curve 3 in Fig.~\ref{Break2_ACC} any platoon of the ACC-vehicles satisfies condition
(\ref{ACC_stability}) for string stability. This means that
if, rather than mixed traffic flow (curve 3 in Fig.~\ref{Break2_ACC}), we consider 
   free  flow consisting of $\gamma= 100 \%$ of classical
 ACC-vehicles, then in accordance with results of Ref.~\cite{Kerner2018C}
 no traffic breakdown at $q_{\rm on}=$ 320 vehicles/h and $q_{\rm in}=$ 1830 vehicles/h
is realized at the bottleneck. The latter result remains true,
 even when
 the flow rate upstream of the bottleneck increases to $q_{\rm in}=$ 2000 vehicles/h\footnote{
As found in~\cite{Kerner_Review3_Int1}, if the percentage of the ACC-vehicles in mixed traffic flow
increases continuously above the value $\gamma= 20 \%$ used in Fig.~\ref{Break2_ACC} (curve 3), then
 there should be
some critical percentage of the ACC-vehicles in mixed traffic flow
$\gamma_{\rm cr, \ increase}$: At
$\gamma=\gamma_{\rm cr, \ increase}$,
  the shift
of the function $P^{\rm (B)}(q_{\rm sum})$ to the left in the flow rate axis  
  reaches its maximum. When the percentage of ACC-vehicles increases subsequently, i.e,
 $\gamma>\gamma_{\rm cr, \ increase}$, then 
the function $P^{\rm (B)}(q_{\rm sum})$    shifts to the right
 in the flow rate axis in comparison with the case 
$\gamma=\gamma_{\rm cr, \ increase}$.
This behavior of the function $P^{\rm (B)}(q_{\rm sum})$ in mixed traffic flow can be explained 
as follows~\cite{Kerner_Review3_Int1}.
At ACC-parameters under consideration (Fig.~\ref{Break2_ACC}), any platoon of ACC-vehicle is stable.
Therefore, when long enough platoons of stable ACC-vehicles begin to build in mixed traffic flow,
these platoons can suppress growing speed disturbances in free flow at the bottleneck.
We should emphasize that the value $\gamma_{\rm cr, \ increase}$ is a large one (in simulations
presented in~\cite{Kerner_Review3_Int1} it is $\gamma_{\rm cr, \ increase} \approx 35 \%$).
Therefore, it is related to a non-realistic case (at least in the near future). In this
non-realistic case traffic breakdown in mixed traffic flow
is mostly determined by behavior of platoons of ACC-vehicles,
rather than dynamic interactions between human driving vehicles and ACC-vehicles
discussed above (curve 3 in Fig.~\ref{Break2_ACC}).}.

			\section{Traffic Stream  Characteristics of Mixed Traffic   Flow    \label{Stream_S}}
	
	In traffic engineering,   the   flow--density 
	(the fundamental diagram) and speed--flow relationships are often used to study the effect of
	traffic control and management on  
	macroscopic traffic stream  characteristics~\cite{May,Manual2000,Manual2010,Gartner,Gartner2,ElefteriadouBook2014_Int1}. 
	To  answer a question of how   traffic flow is affected when the TPACC strategy versus the ACC strategy is considered~\cite{Kerner2018C,Kerner2019C}, in this section 
	  we make   a   study of 
	traffic stream flow characteristics related
	to simulations of mixed traffic flow 
	presented in Figs.~\ref{Prob2_ACC} and~\ref{Break2_ACC}.
For simplicity, we consider  below macroscopic traffic stream  characteristics
		with the use of speed--flow relationships. The associated  flow--density 
	 flow--density  relationships (the fundamental diagram) can be found in~\cite{Kerner2018C}.
	
\subsection{Mixed Traffic   Flow with 2$\%$ Autonomous Driving Vehicles \label{Stream2_S}}

\begin{figure}
\begin{center}
\includegraphics*[scale=.7]{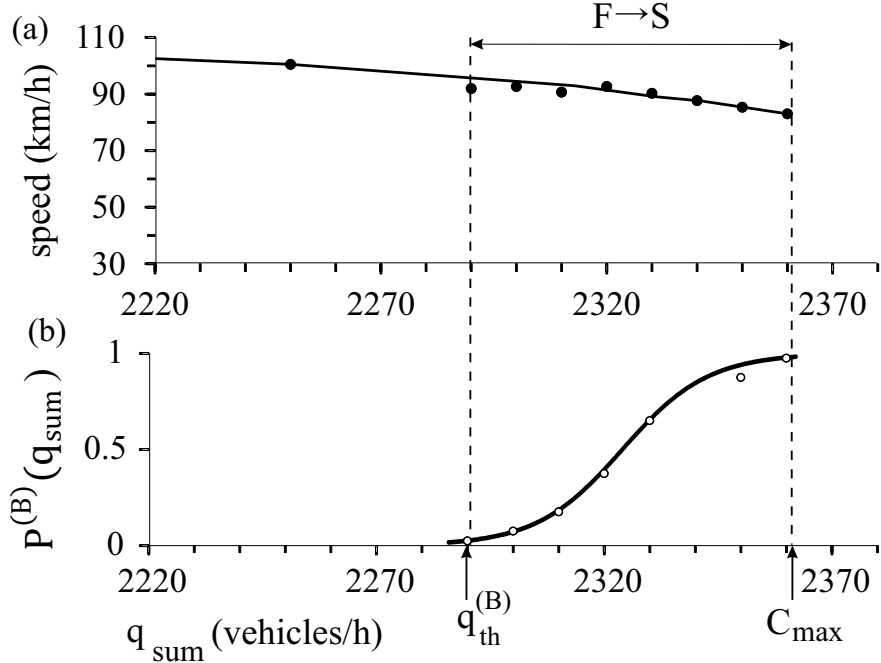}
\end{center}
\caption[]{Stream flow characteristics
of free flow on single-lane road with
on-ramp bottleneck in
traffic   without autonomous driving vehicles and in mixed traffic with 2$\%$ of TPACC-vehicles:
(a) A part of speed--flow relationship for larger flow rates. (b) The
breakdown probability $P^{\rm (B)}(q_{\rm sum})$; function $P^{\rm (B)}(q_{\rm sum})$
 is curve 1
 from Fig.~\ref{Prob2_ACC}.
	Traffic stream flow characteristics have been calculated as follows. At each given flow rate $q_{\rm sum}$ 
	(black points on the characteristics),
	5-min averaged data for the speed, density, and
	flow rate have been measured with the use of a virtual road detector installed at the end of the on-ramp merging region $x=$ 10.3 km. The data has been measured {\it only} during time interval  within which
	free flow has been observed in a simulation realization.  Then, as by the calculation of $P^{\rm (B)}(q_{\rm sum})$
	in Fig.~\ref{Prob2_ACC}, 
 $N_{\rm r}=$ 
40 different realizations have been simulated  for each of  the chosen flow rates   $q_{\rm sum}$.  
	This allows us to make a statistical analysis of  the average
	speed and density in the traffic stream.
	Black points on the speed--flow     relationship are related to 	 
	 the average values of the speed and density derived from this statistical analysis.
	Other model parameters are the same as those in Fig.~\ref{Prob2_ACC}.
	 Calculated values:
	$q^{\rm (B)}_{\rm th, \ TPACC}=q^{\rm (B)}_{\rm th}=2290$
	and $C_{\rm max, \ TPACC}=C_{\rm max}=2360$ vehicles/h.
}
\label{Stream}
\end{figure}

	In Fig.~\ref{Stream} (a), we show a part of the  speed--flow relationship for larger flow rates in free flow
 without autonomous driving vehicles.  It turns out that    traffic stream flow characteristics are  
  identical for
traffic without autonomous driving vehicles and for mixed traffic  with 2$\%$ of TPACC-vehicles: 
Single TPACC-vehicles  do not affect the stream flow characteristics in free  flow  (Fig.~\ref{Stream} (a)).
This   corresponds results of Sec.~\ref{Prob_S}: 2$\%$ of TPACC-vehicles do not affect
 the probability of
traffic breakdown $P^{\rm (B)}(q_{\rm sum})$ (curve 1 shown in Fig.~\ref{Prob2_ACC}).

\begin{figure} 
 \begin{center}
\includegraphics[scale=.65]{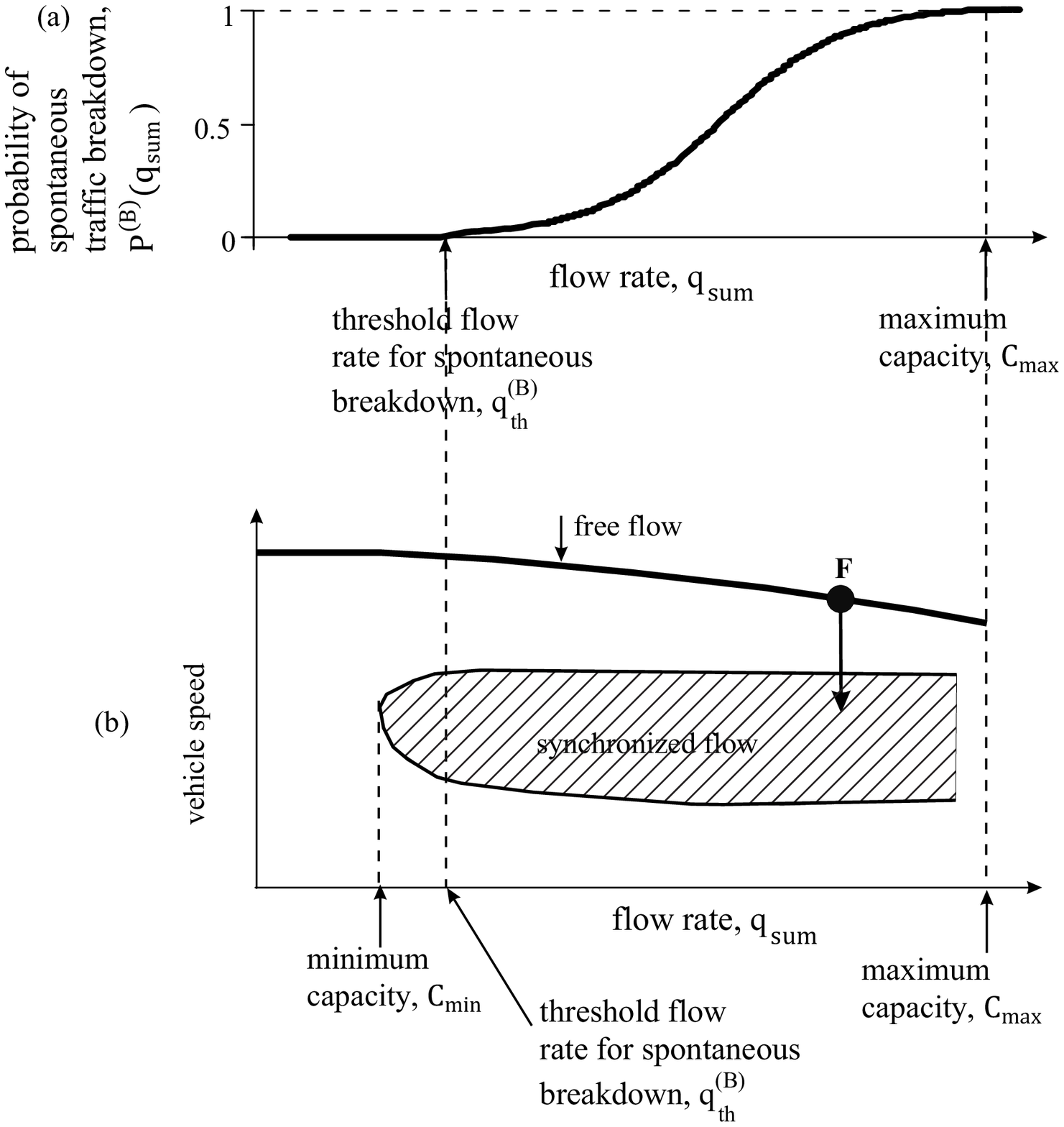}
 \end{center}
\caption{Qualitative explanation  of
   condition (\ref{th_max}): (a)  Qualitative flow-rate function of the breakdown probability  $P^{\rm (B)}(q_{\rm sum})$.
		(b) Z-Characteristic of traffic breakdown with
		states for free flow and synchronized flow taken from Fig.~\ref{Breakdown_nuc_simple2}.
	Adapted from~\cite{Kerner2017A}.
}
\label{Breakdown_nuc3}
\end{figure}

In the three-phase theory
(see  books~\cite{KernerBook,Kerner2009,Kerner2017A} and~\cite{Kerner2017B,Kerner2018A}), there is a deep connection between the flow-rate dependence of the probability of
traffic breakdown $P^{\rm (B)}(q_{\rm sum})$ (Fig.~\ref{Prob2_ACC}) and
the overall flow as well as other  
traffic stream flow characteristics
  (Fig.~\ref{Stream}).
	In particular, on traffic stream flow characteristics   one should distinguish
	a flow rate range (Figs.~\ref{Stream} and~\ref{Breakdown_nuc3})
	\begin{equation}
	q^{\rm (B)}_{\rm th} \leq q_{\rm sum} \leq C_{\rm max}.
	\label{th_max}
	\end{equation}
	
		Within the flow rate range (\ref{Cap_q_F_Int_Met}), free flow is in a metastable state with respect to
	traffic breakdown (F$\rightarrow$S transition) at the bottleneck 
	(Fig.~\ref{Breakdown_nuc3}).
	A characteristic flow rate $q_{\rm sum}=q^{\rm (B)}_{\rm th}$ in (\ref{th_max})
	is a threshold flow rate for spontaneous traffic breakdown at the bottleneck:
	At $q_{\rm sum}<q^{\rm (B)}_{\rm th}$ the breakdown probability 
	$P^{\rm (B)}=0$ (Fig.~\ref{Breakdown_nuc3} (a)), i.e., no 
	spontaneous
	traffic breakdown can
	occur 
	during a time interval of the observation of traffic flow $T_{\rm ob}$. 
	
	A characteristic flow rate
	$q_{\rm sum}=C_{\rm max}$ in (\ref{Cap_q_F_Int_Met}) and (\ref{th_max}) is the 
	maximum highway capacity (Figs.~\ref{Breakdown_nuc_simple2} and~\ref{Breakdown_nuc3}):
	At $q_{\rm sum}\geq C_{\rm max}$ the breakdown probability 
	$P^{\rm (B)}=1$ (Fig.~\ref{Breakdown_nuc3} (a)), i.e.,  
	spontaneous
	traffic breakdown does
	occur at the bottleneck
	during the time interval $T_{\rm ob}$.
	
The larger the values $q^{\rm (B)}_{\rm th}$ and $C_{\rm max}$ for the traffic stream, the larger
	is on average the overall flow.   Therefore, the characteristic flow rates $q^{\rm (B)}_{\rm th}$
	and $C_{\rm max}$, which determine the boundaries 
	of 
	the flow rate range (\ref{th_max}) (Fig.~\ref{Breakdown_nuc3}),
	are  basic statistical characteristics of
	the overall flow in the  
	traffic stream in the framework of the three-phase theory\footnote{A  more detailed discussion  of the application of the three-phase theory for
	a statistical analysis of 
	traffic stream characteristics (like the definition  
	of $\lq\lq$stochastic highway capacity of free flow at a bottleneck'' and its theoretical justification
	made in the three-phase theory) 
	as well as a critical consideration of the three-phase 
	theory~\cite{KernerBook,Kerner2009,Kerner2017A,Kerner1999B,Kerner1999A,Kerner1999C,Kerner_Review_Int1} versus
	the classical traffic flow theories and models reviewed 
	in~\cite{Helbing2001,Haight1963A,Gazis2002,Gartner,Gartner2,ElefteriadouBook2014_Int1,Da,Sch,Brockfeld2003,Bellomo,Ferrara2018A,Leu,Mahnke,MahnkeKLub2009A,Wid,Wh2,Treiber_Int1,Schadschneider2011,Saifuzzaman2015A,Pa1983,New,Nagel2003A,Nagatani_R}  are out of scope of this article:
	Such a detailed analysis
	has already been made in the book~\cite{Kerner2017A} as well as in~\cite{Kerner2018A}.}.

\begin{figure}
\begin{center}
\includegraphics*[scale=.7]{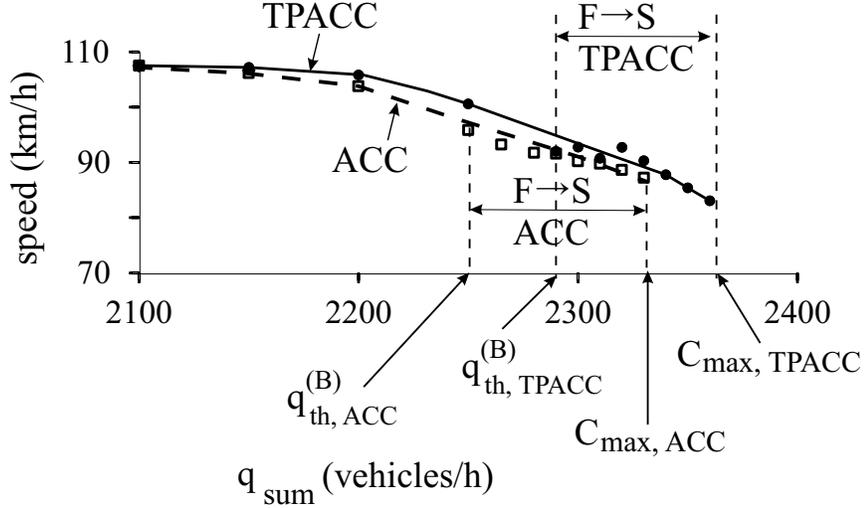}
\end{center}
\caption[]{Comparison of traffic stream flow characteristics  
for free flow on single-lane road with
on-ramp bottleneck in  mixed traffic with 2$\%$ of autonomous driving vehicles.
A parts of  
 speed--flow  relationship for larger flow rates.
Solid curves  $\lq\lq$TPACC''  
 are related to TPACC-vehicles.
Dashed curves $\lq\lq$ACC''   are related to  classical ACC-vehicles.
Stream flow characteristics have been calculated as explained in caption to Fig.~\ref{Stream}.
  Other model parameters are the same as those in Fig.~\ref{Prob2_ACC}. Calculated values:
$q^{\rm (B)}_{\rm th, \ TPACC}=2290$
	and $C_{\rm max, \ TPACC}=2360$ vehicles/h; $q^{\rm (B)}_{\rm th, \ ACC}=2265$
	and $C_{\rm max, \ ACC}=2330$ vehicles/h.
}
\label{Stream2}
\end{figure}

	For the further analysis, we denote
	the flow rate range (\ref{th_max}) on traffic stream characteristics
	by the arrow $\lq\lq$F$\rightarrow$S'' (Figs.~\ref{Stream},~\ref{Stream2}, and~\ref{Stream3}).

	We denote the statistical characteristics of the overall flow $q^{\rm (B)}_{\rm th}$,
  $C_{\rm max}$ in (\ref{th_max}) for mixed traffic flow 
	by  $q^{\rm (B)}_{\rm th, \ TPACC}$, $C_{\rm max, \ TPACC}$, when autonomous driving vehicles are TPACC-vehicles,
	and by $q^{\rm (B)}_{\rm th, \ ACC}$, $C_{\rm max, \ ACC}$ for classical ACC-vehicles, respectively.
	We have found that   the overall flow characteristics  
	do not change on average  
	in mixed traffic with 2$\%$ of  TPACC-vehicles (Fig.~\ref{Stream}):
 $q^{\rm (B)}_{\rm th, \ TPACC}=q^{\rm (B)}_{\rm th}$ and
	$C_{\rm max, \ TPACC}=C_{\rm max}$.

	In contrast with 
	mixed traffic flow with 2$\%$ of TPACC-vehicles, we have found that both values $q^{\rm (B)}_{\rm th, \ ACC}$
	and $C_{\rm max, \ ACC}$ decrease   in mixed traffic flow with 2$\%$ of classical ACC-vehicles
	(Fig.~\ref{Stream2}, curve $\lq\lq$ACC''). This means that already 2$\%$ of classical ACC-vehicles
	reduce on average the overall flow in the
	traffic stream. As explained in Sec.~\ref{Dyn_Rules_S},
	this result is associated with a large  local  speed disturbance
	  caused by  a   classical ACC-vehicle at the bottleneck: Within the flow rate range
	(\ref{th_max}),
	the large local speed disturbance can
	initiate a nucleus for spontaneous traffic breakdown at the bottleneck.

		\subsection{Mixed Traffic   Flow with 20$\%$ Autonomous Driving Vehicles}

	\begin{figure}
\begin{center}
\includegraphics*[scale=.7]{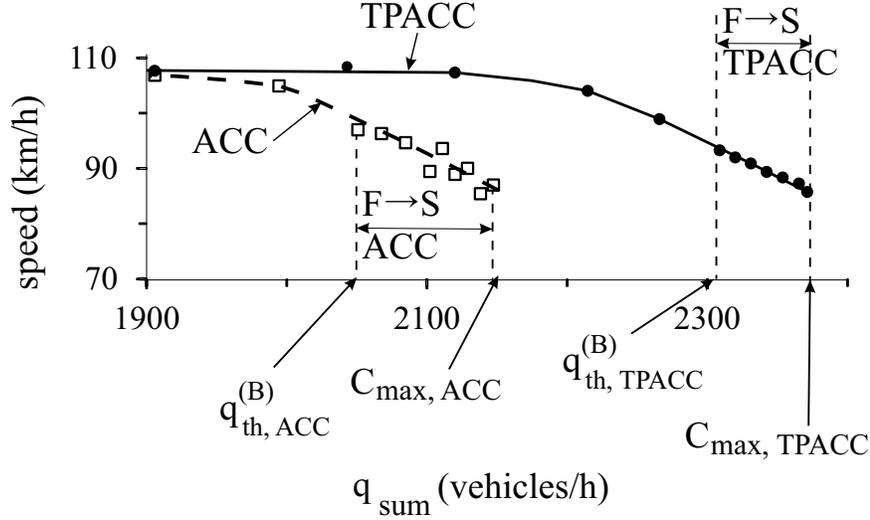}
\end{center}
\caption[]{Comparison of traffic stream flow characteristics  
for free flow on single-lane road with
on-ramp bottleneck in  mixed traffic with 20$\%$ of autonomous driving vehicles.
A part of  
 speed--flow   relationship for larger flow rates.
Solid curve $\lq\lq$TPACC''   is related to TPACC-vehicles.
Dashed curve $\lq\lq$ACC''   is related to  classical ACC-vehicles.
Stream flow characteristics have been calculated as explained in caption to Fig.~\ref{Stream}.
In simulations,   at $q_{\rm sum}\leq$ 2000 vehicles/h
	there is no on-ramp inflow ($q_{\rm on}=0$);    at
	$q_{\rm sum}=q_{\rm in}+q_{\rm on}>$ 2000 vehicles/h,
	the increase in $q_{\rm sum}$ has been achieved through increase in $q_{\rm on}$
	at constant $q_{\rm in}=$ 2000 vehicles/h. Simulation parameters of ACC   and TPACC   
   are, respectively, the same as those in Fig.~\ref{Prob2_ACC}. Calculated values:
$q^{\rm (B)}_{\rm th, \ TPACC}=2308$
	and $C_{\rm max, \ TPACC}=2371$ vehicles/h; $q^{\rm (B)}_{\rm th, \ ACC}=2050$
	and $C_{\rm max, \ ACC}=2147$ vehicles/h.
}
\label{Stream3} 
\end{figure}

	We might assume that vehicles implementing TPACC strategy
(\ref{TPACC_main5})  can reduce the overall flow for the traffic stream due to their different use of available space on the road. However, simulations presented in~\cite{Kerner2018C} 
show that 
no such adverse effect  for the traffic stream occurs. 
	Contrarily, rather than reduction of the overall flow  through the use of TPACC-vehicles,
	we have found that  TPACC-vehicles increase on average the  overall flow
	  (compare values $q^{\rm (B)}_{\rm th, \ TPACC}$, $C_{\rm max, \ TPACC}$ in Fig.~\ref{Stream}
		with, respectively, these values given in caption to Fig.~\ref{Stream3}).
	
	In contrast with 
	 TPACC-vehicles, we have found that 
	  classical ACC-vehicles reduce  on average the overall flow in mixed traffic flow
		(Fig.~\ref{Stream3}). As explained in Secs.~\ref{Prob_S}--\ref{Dyn_Rules_S}, 
		this effect of the overall flow reduction  
			caused by classical ACC-vehicles is explained by 
			the occurrence of large local  speed disturbances at the bottleneck in mixed traffic
			flow. The local speed disturbances
			initiate traffic breakdown in the mixed traffic flow at considerably smaller  flow rates
			in comparison with  
			traffic flow  without
			classical ACC-vehicles. 
		 Thus, through the strong effect of autonomous driving vehicles
			on traffic breakdown at the bottleneck, in simulations we cannot resolve  
		the  effect of the different use
		 of available space on the road
		by TPACC-vehicles and ACC-vehicles
		on the overall flow.

	Thus, we have found that    autonomous driving based
		on the TPACC strategy can   
increase on average the overall flow   for   mixed traffic flow. Contrarily to the TPACC strategy,
autonomous driving based
		on the classical ACC strategy decreases on average the overall flow
		(Figs.~\ref{Stream2} and~\ref{Stream3})\footnote{Simulations presented in~\cite{Kerner2018C} show that even in a non-realistic case of  traffic consisting of 100$\%$ of automated driving vehicles
	the effect  of large local speed disturbances at the bottleneck on the overall flow caused by the classical ACC
	 is also stronger than the   effect of the different use
		 of available space on the road by the ACC-vehicles and TPACC-vehicles.
Indeed, under simulation parameters of ACC and TPACC 
 used in Fig.~\ref{Prob2_ACC}, the maximum highway capacity for TPACC-vehicles
$C_{\rm max, \ TPACC}=$ 
2353 vehicles/h is slightly larger than that for ACC-vehicles $C_{\rm max, \ ACC}=$ 2322.6  vehicles/h.}.

  \section{Discussion \label{Dis_S}}
	
		\subsection{Conclusions 
\label{Conl_S}}
	
	In the  article, we have reviewed results of   studies of the ACC in the framework of three-phase theory (TPACC)
	as well as the effect of autonomous driving on traffic breakdown. Additionally, we have presented novel results
	of the article that have been achieved through a 
  mathematical model for rules of ACC  (\ref{Combain01})--(\ref{Combain05}) introduced in this   paper.
Through the use of this ACC model  (\ref{Combain01})--(\ref{Combain05}), we have found the following results:
	\begin{itemize}
\item In a wide range of the desired time headway to the preceding vehicle classical
	ACC-vehicle causes a large local speed disturbance at a highway bottleneck. This large disturbance increases
		the probability of traffic breakdown:  
		Even a single   autonomous driving
vehicle based on the classical ACC-strategy   can   provoke traffic breakdown at the bottleneck
in   mixed traffic flow.
Respectively, classical ACC-vehicles decreases   the maximum highway capacity.		
\item At the same chosen flow rate,
		the longer the desired time headway of the
		classical ACC-vehicle to the preceding vehicle is, the larger the probability of traffic breakdown
		caused by a single classical ACC-vehicle at the bottleneck.  
		\item
		Contrarily to classical ACC-vehicles, in  a wide range of headway times TPACC-vehicles can even reduce local speed disturbances at the bottleneck.
		For this reason, single TPACC-vehicles does not change the probability of traffic breakdown.
		\end{itemize}

We can make the following general conclusions:
	\begin{description}
	\item 1. For the enhancing of future mixed traffic flow, 
	the dynamics of   autonomous driving vehicles   should learn from
	some common features of human driving vehicles\footnote{There is at least one exclusion
 from this   $\lq\lq$learning''  of common features of human driving for autonomous driving vehicles:
A human  exhibits a finite
  driver reaction time. In accordance with the three-phase theory, the driver reaction time
	is responsible for moving jam emergence in synchronized flow  (Sec.~\ref{Simp_Ex_Sec}).
	To avoid moving jams  in synchronized flow, the reaction time
	of    autonomous driving vehicles on unexpected deceleration of the preceding vehicle
	or on a sudden reduction of the spacing  to the preceding vehicle
	between 
	should be made
	as short as zero.  }. Indeed, if the dynamics of the autonomous driving vehicle is qualitatively different from the common  
	dynamic behavior of human driving vehicles, an autonomous driving vehicle  might be considered an 
	$\lq\lq$obstacle'' for
	human driving vehicles in mixed traffic flow. This can reduce traffic safety as well as decrease highway capacity.
	This explains the
 necessity of the use of    autonomous driving vehicles those dynamics in car-following is familiar
for humans. One of the important features of the common dynamic behavior of human driving vehicles
	found out in empirical data and firstly incorporated in the three-phase traffic theory is an indifference zone in car-following in synchronized flow leading to a two-dimensional (2D) region
of synchronized flow states.
	\item 2.
	To find   results of the effect of
autonomous driving vehicles on   mixed traffic flow  that are valid for the reality, the dynamics of human driving vehicles in mixed traffic flow must be simulated with a microscopic traffic flow model in the framework of the three-phase theory. 
	Indeed, we have explained that and why if one of the standard traffic flow models for
simulations of human driving vehicles is used, then  results of a study of the effect of  autonomous driving vehicles are invalid   for the real world.
We   use the term $\lq\lq$standard" for
  traffic flow
	models     that
 are related to the state-of-the-art in traffic and transportation research.
Examples of standard (classical) traffic flow
	models can be found in papers, reviews, and books~\cite{Helbing2001,Haight1963A,Gazis2002,Gartner,Gartner2,ElefteriadouBook2014_Int1,Da,Sch,Brockfeld2003,Bellomo,Ferrara2018A,Leu,Mahnke,MahnkeKLub2009A,Wid,Wh2,Treiber_Int1,Schadschneider2011,Saifuzzaman2015A,Pa1983,Newell1963,New,Nagel2003A,Nagatani_R}. This   criticism on the 
	standard (classical) traffic flow
	models has been explained in  reviews~\cite{Kerner2018D,Kerner2018A,Kerner2017B,Kerner_Review_Int1,Kerner_Review2_Int1,Kerner_Review3_Int1,Kerner2018B}.
	\end{description}

The  advantages of TPACC are associated with the absence of a fixed desired time headway to the preceding vehicle in the TPACC strategy: An TPACC-vehicle exhibits a large indifference zone within 
the time headway range (\ref{TPACC_main_range}) within which the TPACC-vehicle does not
control time headway to the preceding vehicle. 
As we have explained in this paper,  due to the large
  indifference zone  within  the time headway range (\ref{TPACC_main_range}),
	the TPACC-vehicle should not necessarily decelerate as strong as the preceding vehicle when a local short-time speed disturbance appears at a road bottleneck.  This dynamic behavior of TPACC-vehicles decreases local
	speed disturbances in free flow at a road bottleneck. In its turn, the decrease in the amplitude of
	local  speed disturbances at road bottlenecks results in a
	decrease in the   probability of traffic breakdown in the traffic stream.

		\subsection{About Applicability of     Model Results for     Future Autonomous Driving in Mixed Traffic Flow   \label{Value}}
	
	\subsubsection{Simple Models of Autonomous Driving Reproducing  Features of  Mixed Traffic Flow}
 
 Although the 
	TPACC model (\ref{TPACC_main5})~\cite{Kerner2018C} as well as the ACC model  (\ref{Combain01})--(\ref{Combain05})
	introduced in this paper
	have allowed us to understand the physics  
  of  TPACC-vehicles in mixed traffic flow,  
  the following question can arise: Whether can   the simple 
	TPACC model (\ref{TPACC_main5}) and the ACC model  (\ref{Combain01})--(\ref{Combain05})  be applicable for  
	reliable statements about physical features of real mixed traffic flow?
	To answer this question, we consider firstly some features of the 
	TPACC model (\ref{TPACC_main5})  that might be seemed at the first glance 
	as non-realistic for real traffic
	flow.

  It seems that TPACC model
	(\ref{TPACC_main5})
is mathematically of small incremental value from the pre-existing classical ACC model (\ref{ACC_General}).
However,   we have above shown that
 this $\lq\lq$small incremental mathematical value'' exhibits
 a large physical effect on traffic flow. This large physical effect on traffic flow through the TPACC strategy
(\ref{TPACC_main5}) is associated with the TPACC physical feature mentioned above: Through the indifference zone of TPACC (\ref{TPACC_main5}), a TPACC-vehicle does not react on the time headway change within the
   time headway range (\ref{TPACC_main_range}). This dynamic behavior of TPACC-vehicles decreases local speed disturbances in free flow at the bottleneck. The reduction of the local speed disturbances
	results in a
	decrease in the breakdown probability in the traffic stream.

  This review deals with a  subset of the functionality required for  autonomous driving, namely longitudinal following a given leader. Other   challenges for autonomous driving such as lateral dynamics  or sensor-related problems, which are important to
	satisfy a safety motion of   autonomous driving vehicles on multi-lane highways
	and urban areas (see, e.g.,~\cite{Ioannou,Ioannou1996,Ioannou2006,Meyer2014A,Bengler2014A,Maurer2015A,Rajamani2012A_Aut}), are not tackled in this review paper.
	Therefore, a question can arise in what degree   results
	derived for TPACC are  related to future autonomous driving.
	
	As mentioned in Sec.~\ref{Introduction}, in empirical data the qualitative flow-rate dependence
	of the probability of traffic breakdown at a road bottleneck does not depend on the number of highway lanes
	(on  features of lateral dynamics of vehicles), on the bottleneck type, and on  real vehicle technology  
	(during last 30 years vehicle technology was  
	changed considerably, however, qualitative empirical features of traffic breakdown did not change). 
In accordance with the three-phase theory  that explains
all known empirical features of 
traffic breakdown~\cite{KernerBook,Kerner2009,Kerner2017A}, the simple ACC-model and TPACC-model    
used in the paper reflect dynamic vehicle features  that are responsible for traffic breakdown.
  For this reason, the result of the paper that at the same model parameters
 classic ACC-vehicles (\ref{ACC_General})   increase 
  the breakdown probability, whereas  
 TPACC-vehicles   (\ref{TPACC_main5})   decrease   the breakdown probability
	proves that  the use of indifference zones of the three-phase
		theory can have benefits for future autonomous driving.
 This is because contrarily with
  (\ref{ACC_General}), human driving vehicles do not control
	time headway within the time headway range  
	(\ref{TPACC_main_range})~\cite{KernerBook,Kerner2009,Kerner2017A}.	Thus,	the TPACC vehicles, which can be considered autonomous driving $\lq\lq$learning''  from empirical human driving behavior, can
	decrease the breakdown probability.
 
 \subsubsection{TPACC as ACC Learning from  Driver Behavior}
  
	Results of this review allow us to assume that
	 future systems for autonomous driving should be developed whose rules are consistent with
	those of human driving vehicles. 
Otherwise, we could expect that  autonomous driving vehicles
 can be considered as $\lq\lq$obstacles'' for drivers.
The  physics of autonomous driving in the  framework of  the three-phase   theory studied in this paper
  emphasizes that future
		autonomous driving should be developed in which both the longitudinal dynamics  
		(TPACC) and lateral dynamics  
		should learn from  driver behavior. In particular,  the longitudinal and 
	  lateral dynamics of autonomous driving vehicles
		should be consistent with the existence of indifference zones of the three-phase
		theory~\cite{KernerBook,Kerner2009,Kerner2017A,Kerner1999B,Kerner1999C}.
		
		\subsubsection{About Simulations of Autonomous Driving in Dangerous Traffic Situations}
		
		A question can arise from the choice of   the model time step  $\tau=$ 1 s 
	in Eqs.~(\ref{ACC_dynamics_Eq})--(\ref{next2_ACC})  of Appendix~\ref{Cla_ACC_S}  
		and  Eqs.~(\ref{TPACC_main})--(\ref{next2_TPACC}) of Appendix~\ref{App_TPACC_Model} that have been used for numerical simulations
	of ACC model (\ref{ACC_General}) and TPACC model (\ref{TPACC_main5}), respectively.
	In these models of ACC and TPACC 
	the time step  $\tau=$ 1 s determines 
	the safe space gap $g_{\rm safe}=v \tau$ under hypothetical steady state conditions in which
	all vehicles move at time-independent speed $v$.
 Contrarily to the ACC model (\ref{ACC_dynamics_Eq})--(\ref{next2_ACC}) and the 
	TPACC model (\ref{TPACC_main})--(\ref{next2_TPACC}),
	typical ACC controllers in vehicles that on the market have update time intervals $\tau$ of 100 ms or less.  
	Indeed, there may be some very 
	dangerous traffic situations in real traffic in which the safe time headway for an ACC-vehicle is quickly
	reached and, therefore, the ACC-vehicle must decelerate strongly
	already after a 
	time interval that is a much shorter than 1 s to avoid the
	collision with the preceding vehicle.
	Therefore, to avoid collisions, {\it real} ACC-controllers must
	have update time intervals $\tau$ of 100 ms or less. However, at model time step $\tau=$ 1 s	through the choice in the mathematical formulation of the safe speed in  Eqs.~(\ref{ACC_dynamics_Eq})--(\ref{next2_ACC}) of Appendix~\ref{Cla_ACC_S}  
		and  Eqs.~(\ref{TPACC_main})--(\ref{next2_TPACC}) of Appendix~\ref{App_TPACC_Model}
	  as well as in
	the model of human driving vehicles (see Secs.~\ref{Safe_speed_kkl} and~\ref{Cla_ACC_S} of Appendix),
{\it collision-less}   traffic flow is guaranteed
	 in {\it any} dangerous traffic situation that can occur in simulations of traffic flow.
	
	In other words, to disclose the physics of 
	TPACC it is   sufficient the choice of  
	the update time $\tau=$ 1 s in {\it simulations} of   TPACC behavior.
	To explain this, we should note that Eqs.~(\ref{next1_TPACC}), (\ref{next2_TPACC}) of Appendix~\ref{App_TPACC_Model} effect on
		TPACC dynamics {\it only} under condition $g_{n}< g_{\rm safe, n}$, i.e., when
		the space gap becomes smaller than the safe one. 
	This is because the physics of TPACC  disclosed in this paper
	is {\it solely} determined by Eq.~(\ref{TPACC_main}):
	Under condition
	$g_{n}\geq g_{\rm safe, n}$,    Eqs.~(\ref{next1_TPACC}), (\ref{next2_TPACC})
do not change TPACC acceleration (deceleration)
calculated through Eq.~(\ref{TPACC_main}).
The same conclusion is also valid for the ACC model Eqs.~(\ref{ACC_dynamics_Eq})--(\ref{next2_ACC}) of Appendix~\ref{Cla_ACC_S}.
	
	\subsection{Can Vehicular Traffic consisting of  100$\%$ Autonomous Vehicles be Real Option in The Future?  \label{Mixed_S}}
	
	In this review paper, we have focused on  mixed traffic flow with a small percentage   of 
autonomous driving vehicles. There are the following reasons for this limitation of the study made in this paper.
First, in the near future only a very small
percentage   of 
autonomous driving vehicles in mixed traffic flow can be expected.
Second, we would like to study whether already
 a single autonomous driving vehicle surrounded by human driving vehicles
can effect on traffic breakdown at the bottleneck.
Indeed, we have found that
 already a single autonomous driving vehicle whose dynamics is
based on the classical (standard) approach  
can provoke traffic breakdown at the bottleneck. We have also found  
  that such a deterioration of traffic system
does not occur when a autonomous driving vehicle learns from driver behavior in car-following as introduced in the three-phase theory.

  Dynamic rules of autonomous driving vehicles can be developed that are
 totally different from the dynamic behavior of human driving vehicles.    In other words,
for traffic flow consisting of 100$\%$ of autonomous driving vehicles
the dynamic rules of autonomous driving vehicle
should not necessarily be consistent with the dynamic behavior of human driving vehicles. 
One of the consequences is that in   traffic flow consisting of 100$\%$ of autonomous driving vehicles
 highway capacity can be considerably larger in comparison with highway capacity of  mixed traffic flow studied in the paper.
Therefore, a question can arise: Why does the effect of autonomous driving vehicle on traffic breakdown and capacity of
 mixed traffic flow studied in this paper is important for future vehicular traffic?

To answer this question, we should discuss whether and when traffic flow consisting of 100$\%$ of autonomous driving vehicles
in a traffic network without human driving vehicles    could be possible to expect. First, we assume that in the future all vehicles are autonomous driving vehicles only. 
 We could expect that there is at least one reason that might prevent the realization of this case:
\begin{itemize}
\item   {\it Service costs for autonomous vehicles} could be many times larger than those for conventional vehicles driving by humans. Indeed,    the check of systems for autonomous driving should be made much frequently than it is needed for the case for a manual driving vehicle.  This is because a sudden failure of a system for autonomous driving in a vehicle can lead to an accident with  catastrophic consequences for both passengers of the autonomous driving vehicle and passengers of several other following vehicles in traffic flow. Such a frequent check  of systems for autonomous driving could lead to enormous service costs that could   be paid by only a (small) part of  vehicle owners\footnote{To explain this statement, we should note that mean time headway in vehicular traffic can reach values 1--2 seconds or   less.  Even for an autonomous driving vehicle in which there is a possibility for 
  driver control of the vehicle, none of passengers is able to take 
	the vehicle under control during the short time interval 1--2 seconds.}.
\end{itemize}

If in the future   some of the vehicles moving in traffic networks
 are manual driving ones, then we can assume that two independent  from each 
other traffic networks might be developed: (i) One network is dedicated  to autonomous driving vehicles only. (ii) Another network  is dedicated  to   human driving vehicles only. We could expect that there is at least one reason that might prevent the realization of this case:
\begin{itemize}
\item    {\it Extremely high   costs} for the development of two independent  from each 
other traffic networks\footnote{We should recall that traffic congestion might be prevented, if either  highways with much enough lanes
 were build   or many enough parallel   highways for some 
travel routes in a network were build.
However, due to the extremely high road repairing costs this well-known idea could not be realized.
  Indeed, after 20--30 years almost each highway should be repaired. The result is 
the extremely high road repairing costs and/or very many road-works that act as highway bottlenecks for traffic breakdown. }.
\end{itemize}

It is clear that the organization of high-speed highway lanes, or   some separated roads,  or else 
 {\it a part} of a traffic network dedicated to connected
autonomous vehicles    is possible as, for example, suggested 
in~\cite{Davis2018A,Xiaonian_Shan2018A,Jing-Peng-Wang2019A,Stewart2019A,Lanhang_Ye2018A,Richards2018A}. In this case, in
another network part in which human driving vehicles can move, 
mixed traffic flow is realized. 
Therefore, it could be expected that   mixed traffic flow will remain the reality   also for future vehicular traffic.

 \section{Future Directions}

	In this review paper, we have made   a comparison of the effect of classical ACC-vehicles and TPACC-vehicles on the probability of traffic breakdown at a road bottleneck in mixed traffic flow. In this scenario of the application of autonomous driving vehicles, 
  rather than only some specific values of TPACC and ACC parameters as made in~\cite{Kerner2018C,Kerner2019C}
	we have considered a wide range of headway times   to demonstrate that the TPACC strategy   can exhibit advantages in comparison with the classical ACC. Moreover, through the use of ACC model  (\ref{Combain01})--(\ref{Combain05}) we could understand 
	the physical difference between classical ACC strategy and TPACC  in more details.

However, a change in other parameters of ACC     might give   different results.    Additionally, incorporating cooperative merging between ACC vehicles could reduce the tendency to initiate breakdown at highway bottlenecks.   We believe that related detailed studies of the TPACC model, which are out of the scope of this review,  will be a very interesting task of future investigations of the physics of autonomous driving.

 \appendix 

\section{Kerner-Klenov Microscopic Stochastic Traffic Flow Model \label{KKl_Model_Ap}}

  In this Appendix, we make   explanations 
 of the Kerner-Klenov stochastic microscopic three-phase model   for  
  human driving vehicles~\cite{KKl,KKl2003A,KKl2009A}
	and model parameters used for simulations of  mixed traffic flow presented in the main text.

\subsection{Update Rules of Vehicle Motion  
  \label{Identical_KKl_Up}}

In a discrete model version of the Kerner-Klenov stochastic microscopic three-phase model used in all simulations presentred in the main text,  rather than the continuum space co-ordinate~\cite{KKl},   
a discretized space co-ordinate with a small enough value  of the discretization 
space interval $\delta x$   is used~\cite{KKl2009A}. 
Consequently,  
  the vehicle speed and acceleration (deceleration) discretization intervals are $\delta v=\delta x/\tau$
 and   $\delta a=\delta v/\tau$, respectively, where $\tau$ is time step. Because in 
  the discrete model version   discrete (and dimensionless) values of space coordinate, speed and acceleration 
 are used, which are measured respectively in  values $\delta x$, $\delta v$ and  $\delta a$, 
and time is  measured in values of $\tau$,
   value $\tau$ in all formulas  is assumed below to be the dimensionless value $\tau=1$.
	 In the discrete model version 
used for all simulations,
 the discretization cell $\delta x=$ 0.01 m   is used.

    A choice of $\delta x=0.01$  m  made in the   model  
     determines the accuracy of vehicle speed calculations {\it in comparison} with the initial continuum in space stochastic model  of~\cite{KKl}.
 We have found that    the discrete   model   exhibits
similar characteristics of phase transitions and
resulting congested patterns at highway bottlenecks as those
in the
 continuum  model                     
 at $\delta x$  that satisfies the conditions
\begin{equation}
\delta x/\tau^{2} \ll b, \ a, \ a^{\rm (a)}, \ a^{\rm (b)}, \ a^{\rm (0)},
\label{cond_Stoch}
\end{equation}
where model parameters for driver deceleration and acceleration $b$,
$a$, $a^{\rm (a)}$, $a^{\rm (b)}$, $a^{\rm (0)}$
will be explained below.

Update rules of vehicle motion  
in the discrete model for identical drivers and
identical vehicles moving in a road lane  are as follows~\cite{KKl2009A}:
\begin{equation}
v_{n+1}=\max(0, \min({v_{\rm free}, \tilde v_{n+1}+\xi_{n}, v_{n}+a \tau, v_{{\rm s},n} })),
\label{final}   
\end{equation}
\begin{equation}
\label{next_x}
x_{n+1}= x_{n}+v_{n+1}\tau,
\end{equation}
where the index $n$ corresponds 
to the discrete time $t_{\rm n}=\tau n, \ n=0,1,...$; 
$v_{n}$ is the vehicle speed at time step $n$, $a$ is the maximum acceleration,
$\tilde v_{n}$ is the vehicle speed  without  speed fluctuations $\xi_{n}$:
\begin{equation}
 \tilde v_{n+1}= \min(v_{\rm free},  v_{{\rm s},n}, v_{{\rm c},n}),
 \label{final2}
 \end{equation}
\begin{equation}
v_{{\rm c},n}=\left\{
\begin{array}{ll}
v_{ n}+\Delta_{ n} &  \textrm{at $g_{n} \leq G_{ n}$} \\
v_{ n}+a_{ n}\tau &  \textrm{at $g_{n}> G_{ n}$}, \\
\end{array} \right.  
\label{delta} 
\end{equation}
\begin{equation}
\Delta_{n}=\max(-b_{ n}\tau, \min(a_{ n}\tau, \ v_{ \ell,n}-v_{ n})),
 \label{final3}
 \end{equation} 
\begin{equation}
 g_{n}=x_{\ell, n}-x_{n}-d,
 \label{gap_formula}
\end{equation}
the subscript $\ell$  
marks variables related to the preceding vehicle,
$v_{{\rm s}, n}$ is a safe speed at time step $n$,
$v_{\rm free}$ is the free flow speed in free flow,
  $\xi_{n}$ describes   speed fluctuations;
	$g_{n}$ is a space gap between two vehicles following each other;
  $G_{n}$ is the synchronization space gap;
all vehicles have the same length $d$. The vehicle length $d$ includes
the mean space gap between vehicles that are in a standstill within a wide moving jam.
Values $a_{n}\geq 0$ and $b_{n}\geq 0$ in (\ref{delta}), (\ref{final3}) restrict changes in speed per time step
when the vehicle accelerates or adjusts the speed to that of the preceding vehicle.
 
 \subsection{Synchronization Space Gap  and Hypothetical Steady States of Synchronized Flow \label{Syn_Gap_kkl}}
 
 Equations (\ref{delta}), (\ref{final3}) 
describe the adaptation of the vehicle speed to the speed of the preceding vehicle, i.e.,
the speed adaptation effect in synchronized flow. This
vehicle speed adaptation takes place within  the synchronization gap  $G_{n}$: 
At 
\begin{equation}
g_{n}\leq G_{n}
\label{x_x_G_n}
\end{equation}  
the vehicle tends to
adjust its speed to  the speed of the preceding vehicle. This means that the vehicle
decelerates if $v_{n}> v_{\ell,n}$, and accelerates if $v_{n}< v_{\ell,n}$.

In   (\ref{delta}),
the synchronization gap $G_{n}$ depends on the
vehicle speed $v_{n}$ and on the speed of the preceding vehicle $v_{\ell, n}$:
 \begin{equation}
 G_{n}=G(v_{n}, v_{\ell,n}),
  \label{Syn_Gap}
\end{equation}
 \begin{equation}
G(u, w)=\max(0,  \lfloor k\tau u+  a^{-1}u(u-w) \rfloor),
  \label{Syn_Gap2}
\end{equation}
where  $k>1$ is constant; $\lfloor{z}\rfloor$ denotes the integer part of  $z$.

 \begin{figure}
\begin{center}
\includegraphics*[scale=.6]{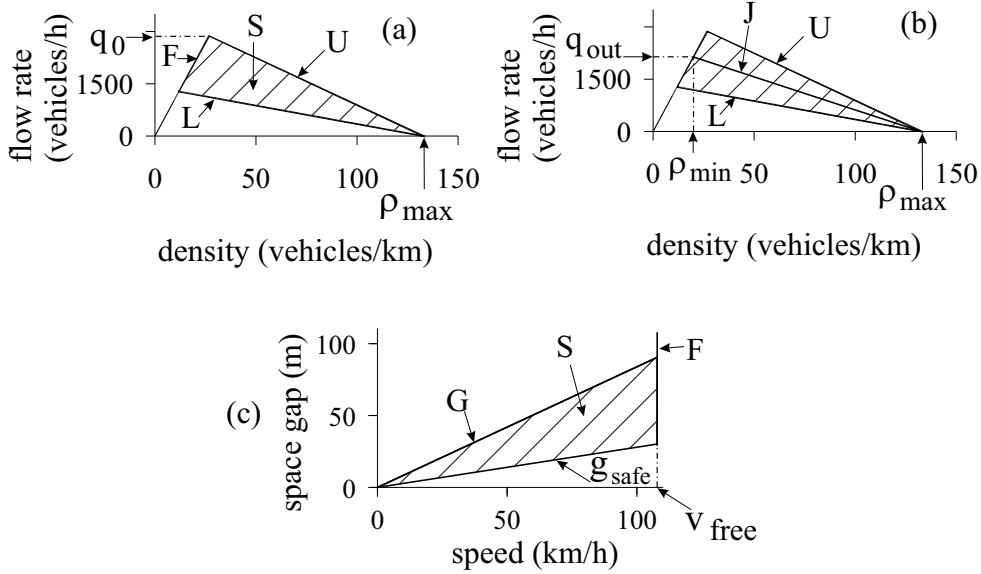}
\end{center}
\caption[]{Steady speed states for the Kerner-Klenov
traffic flow model in the flow--density   (a, b) and in the space-gap--speed planes (c).
 In (a, b),  $L$ and $U$ are, respectively,   lower and upper boundaries of 2D-regions of steady states of synchronized flow. In (b),
 J is the line $J$ whose slope is equal to the characteristic mean velocity $v_{\rm g}$ of a wide moving jam;  in the flow--density plane,
 the line $J$
  represents  the propagation of the downstream front of the wide moving jam with time-independent  
	velocity $v_{\rm g}$.  F -- free flow, S -- synchronized flow.
}
\label{KKl_Steady}
\end{figure}

The speed adaptation effect within the synchronization distance
is related to the  hypothesis of the three-phase 
theory:  
 Hypothetical steady states of synchronized flow 
cover  
 a 2D region in the flow--density  (Fig.~\ref{KKl_Steady} (a)).
  Boundaries $F$, $L$, and $U$ of this 2D-region shown in Fig.~\ref{KKl_Steady} (a)
   are, respectively, associated with the free flow speed in
	free flow, a synchronization space gap $G$, and a safe space gap   $g_{\rm safe}$.
	A speed-function of the safe space gap   $g_{\rm safe}(v)$ is found from the equation
	\begin{equation}
v=v_{\rm s}(g_{\rm safe}, \ v).
\label{g_safe}
\end{equation}
	
Respectively, as for the continuum model (see Sec.~16.3 of the book~\cite{KernerBook}), for the discrete model   hypothetical steady   states of synchronized flow 
   cover a 2D-region in the flow--density plane (Fig.~\ref{KKl_Steady} (a, b)).
  However, because the speed $v$ and space gap $g$
   are integer in the discrete model,
   the steady states do not form a continuum in the flow--density plane
   as they do in the continuum model. The inequalities
   \begin{equation}
  v\leq  v_{\rm free}, \quad g\leq G(v, \ v), \quad    g \geq   g_{\rm safe}(v),
\label{2D_d}
\end{equation}
 define a 2D-region in  
the space-gap--speed plane (Fig.~\ref{KKl_Steady} (c))
 in which the hypothetical steady states exist for the discrete model, when all model fluctuations are neglected.
  
	In (\ref{2D_d}), we have taken into account that in the hypothetical steady states of synchronized flow
 vehicle speeds and space gaps  
are assumed to be time-independent and the speed of each of the vehicles is equal to the speed of the associated preceding vehicle:
 $v=v_{\ell}$.
 However, due to model fluctuations, steady   states of synchronized flow are destroyed, i.e.,
 they do not exist in simulations; this explains the term $\lq\lq$hypothetical" steady states of synchronized flow. 
Therefore, rather than steady states some non-homogeneous in space and time traffic states occur.
In other words, steady states are related to a hypothetical model fluctuation-less limit of
homogeneous in space and time vehicle motion that does not realized in real simulations. Driver time delays
are described through model fluctuations. Therefore, any application of
  the Kerner-Klenov stochastic microscopic three-phase traffic flow
	model without model fluctuations has {\it no sense}.
	In other words, for the description of real
	spatiotemporal traffic flow phenomena, model speed fluctuations
	incorporated in this model are needed.

  \subsection{Model Speed Fluctuations \label{Fluc_KKl}}

 In the   model, 
   random vehicle deceleration
 and acceleration are applied depending on  
whether the vehicle decelerates or accelerates, or else maintains its speed:
 \begin{equation}
  \xi_{ n}=\left\{
\begin{array}{ll}
\xi_{\rm a} &  \textrm{if  $S_{ n+1}=1$}   \\
- \xi_{\rm b} &  \textrm{if $S_{ n+1}=-1$} \\
\xi^{(0)} &  \textrm{if  $S_{ n+1}=0$}.
\end{array} \right.
\label{noise_CA}
\end{equation}
State of vehicle motion $S_{ n+1}$ in (\ref{noise_CA})  
 is determined by   formula
 \begin{equation}
 S_{ n+1}=\left\{
\begin{array}{ll}
-1 &  \textrm{if $\tilde v_{ n+1}< v_{ n}$}   \\
1 &  \textrm{if $\tilde v_{ n+1}> v_{ n}$} \\
0 &  \textrm{if $\tilde v_{ n+1}= v_{ n}$}.
\end{array} \right.
\label{state_CA}
\end{equation}

In  (\ref{noise_CA}), $\xi_{\rm b}$, $\xi^{(0)}$, and $\xi_{\rm a}$ are random sources for deceleration and acceleration that are as follows:
 \begin{equation} 
 \xi_{\rm b}=a^{(\rm b)} \tau \Theta (p_{\rm b}-r),
 \label{xi_dec} 
 \end{equation}
\begin{equation}
\xi^{(0)}=a^{(0)}\tau \left\{
\begin{array}{ll}
-1 &  \textrm{if $r<p^{(0)}$} \\
1 &  \textrm{if $p^{(0)} \leq r<2p^{(0)}$} \quad {\rm and}  \  v_{n}>0 \\
0 &  \textrm{otherwise},
\end{array} \right.
\label{noise_CA_}
\end{equation}
 \begin{equation} 
 \xi_{\rm a}=a^{(\rm a)} \tau \Theta (p_{\rm a}-r),
 \label{xi_acc} 
 \end{equation}
  $p_{\rm b}$ is probability of   
random vehicle deceleration, $p_{\rm a}$ is probability of   
random vehicle acceleration, $p^{(0)}$  
and $a^{(0)}\leq a$ are constants,  
$r={\rm rand (0,1)}$,
 $\Theta (z) =0$ at $z<0$ and $\Theta (z) =1$ at $z\geq 0$,
$a^{(\rm a)}$ and $a^{(\rm b)}$ are model parameters (see Table~\ref{table_parameters}),
which in some applications can be chosen as speed functions
$a^{(\rm a)}=a^{(\rm a)} (v_{n})$ and $a^{(\rm b)}=a^{(\rm b)} (v_{n})$.

\subsection{Stochastic Time Delays of Acceleration and
Deceleration \label{Stoch_Time_Del_A}}

  To simulate     time delays either in  vehicle
acceleration or in  vehicle
deceleration,  $a_{n}$ and  $b_{n}$ in (\ref{final3}) 
are taken as the following stochastic functions
\begin{equation}
a_{n}=a  \Theta (P_{\rm 0}-r_{\rm 1}),
\label{final_a}
 \end{equation}
 \begin{equation}  
b_{n}=a  \Theta (P_{\rm 1}-r_{\rm 1}),
\label{final_b}
 \end{equation}
\begin{equation}
P_{\rm 0}=\left\{
\begin{array}{ll}
p_{\rm 0} & \textrm{if $S_{ n} \neq 1$} \\
1 &  \textrm{if $S_{ n}= 1$},
\end{array} \right.
\label{final_P_0}  
 \end{equation}
 \begin{equation}
P_{\rm 1}=\left\{
\begin{array}{ll}
p_{\rm 1} & \textrm{if $S_{ n}\neq -1$} \\
p_{\rm 2} &  \textrm{if $S_{ n}= -1$},
\end{array} \right.
\label{final_P_1}
 \end{equation}
$r_{1}={\rm rand}(0,1)$, $p_{\rm 1}$ is constant,
$p_{\rm 0}=p_{\rm 0}(v_{n})$ and $p_{\rm 2}=p_{\rm 2}(v_{n})$ are speed functions
 (see Table~\ref{table_parameters}).

\subsection{Simulations of Slow-to-Start Rule \label{SlowToStart_S}}

 In the model, simulations of
the well-known effect of the driver time delay in acceleration  
at the downstream front of synchronized flow or a wide moving jam 
known as a slow-to-start rule~\cite{kkl_Tak,B1998} are made 
 as a collective effect through the use of Eqs. (\ref{delta}), (\ref{final3}), and  a
random value of vehicle acceleration (\ref{final_a}). Eq.~(\ref{final_a})
with $P_{0}=p_{0}<1$ is
applied 
 only   if the vehicle
did not accelerate at the former time step ($S_{n}\neq 1$); in the latter case,  
a vehicle accelerates with some probability $p_{0}$ that depends on the speed
$v_{n}$; otherwise $P_{\rm 0}=1$
(see formula (\ref{final_P_0})).

  The mean time delay in vehicle acceleration is equal to 
\begin{equation}
\tau^{\rm (acc)}_{\rm del}(v_{n})=\frac{\tau}{p_{0}(v_{n})}.
\label{KKl_ac_ac}
\end{equation}
From formula (\ref{KKl_ac_ac}), it follows that     
 the mean time delay in vehicle acceleration from 
  a standstill within a wide moving jam (i.e., when in formula (\ref{KKl_ac_ac})
	the speed $v_{n}=0$) is equal to
  \begin{equation}
\tau^{\rm (acc)}_{\rm del}(0)=\frac{\tau}{p_{0}(0)}.
\label{KKl_ac_ac_jam}
\end{equation}
The mean time delay in vehicle acceleration from a standstill within a wide moving 
jam  
  determines the parameters of the line $J$  
	in the flow--density plane (Fig.~\ref{KKl_Steady} (b)).

Probability $p_{\rm 0}(v_{n})$ in (\ref{final_P_0})   is chosen to be an increasing 
speed function (see  Table~\ref{table_parameters}).
Because the speed within synchronized flow is larger than zero,
 the mean time delay in vehicle acceleration at the downstream front of   synchronized flow  that we denote by 
    \begin{equation}
\tau^{\rm (acc)}_{\rm del, \ syn}=\tau^{\rm (acc)}_{\rm del}(v_{n}), \quad v_{n}>0
\label{KKl_ac_ac_syn}
\end{equation}
 is shorter  
  than the mean time delay in vehicle acceleration at the downstream front of   the wide moving jam  
 $\tau^{\rm (acc)}_{\rm del}(0)$:
     \begin{equation}
\tau^{\rm (acc)}_{\rm del, \ syn} < \tau^{\rm (acc)}_{\rm del}(0).
\label{KKl_ac_ac_syn_jam}
\end{equation}

 \subsection{Safe Speed  \label{Safe_speed_kkl}}

In the model, the safe speed $v_{{\rm s},n}$ in (\ref{final}) is chosen in the form 
\begin{equation}
v_{{\rm s},n}=
\min{(v^{\rm (safe)}_{ n},  g_{ n}/ \tau+ v^{\rm (a)}_{ \ell})},
\label{safe_kkl}  
\end{equation}
$v^{\rm (a)}_{ \ell}$  is an $\lq\lq$anticipation" speed of the preceding vehicle
that will be considered below,
the function    
\begin{equation}
v^{\rm (safe)}_{ n}=\lfloor v^{\rm (safe)} (g_{n}, \ v_{ \ell,n})  \rfloor
  \label{Safe_Speed3}  
\end{equation}
 in (\ref{safe_kkl}) is related to the safe speed $v^{\rm (safe)} (g_{n}, \ v_{ \ell,n})$ 
in the model by Krau{\ss} {\it et al.}~\cite{Kra}, 
 which is a solution of   the  
 Gipps's equation~\cite{Gipps}  
   \begin{equation}
v^{\rm (safe)} \tau + X_{\rm d}(v^{\rm (safe)}) = g_{n}+X_{\rm d}(v_{\ell, n}),
  \label{Gipps_Safe_Speed}
\end{equation}
  where 
$X_{\rm d} (u)$ is the braking distance that should be passed by the vehicle
 moving first with the speed $u$ before the vehicle can come to a stop.

The condition (\ref{Gipps_Safe_Speed})   enables us to find the safe speed $v^{\rm (safe)}$ as a function
of the space gap $g_{ n}$ and    speed $v_{ \ell,n}$ provided $X_{\rm d} (u)$ is a known function.
In the case when the vehicle brakes with a constant deceleration $b$,
the change in the vehicle speed for each time step  is $-b\tau$ except the last time step
before the vehicle comes to a stop. At the last time step, the vehicle decreases its speed at the value
$b\tau\beta$,
where $\beta$ is a fractional part of $u/b\tau$. 
According to formula (\ref{next_x}) for
the  displacement of the vehicle
for one time step,
 the braking distance $X_{\rm d} (u)$ is~\cite{Kra}
\begin{equation}
X_{\rm d} (u)=\tau \big(u-b \tau+u-2b \tau+...+\beta b \tau \big).
\label{Braking}
\end{equation}
From (\ref{Braking}), it follows~\cite{Kra}
    \begin{equation}
X_{\rm d} (u)=b \tau^{2} \bigg(\alpha \beta+\frac{\alpha(\alpha-1)}{2}\bigg),
  \label{Gipps_Safe_Speed2}
\end{equation}
$\alpha=\lfloor u/b\tau \rfloor$ is an integer part of $u/b\tau$.

The safe speed $v^{\rm (safe)}$ as a solution of   equation (\ref{Gipps_Safe_Speed})
 at the distance $X_{\rm d} (u)$
 given by (\ref{Gipps_Safe_Speed2}) 
has
been found by Krau{\ss} {\it et al.}~\cite{Kra}  
\begin{equation}
v^{\rm (safe)}(g_{n}, v_{\ell, n})=b \tau (\alpha_{\rm safe}+\beta_{\rm safe}),
\label{SafeKr}
\end{equation}
where
\begin{equation}
\alpha_{\rm safe}=\left \lfloor{\sqrt {2\frac{X_{\rm d}(v_{\ell, n}) + g_{ n}}{b \tau^{2}}+\frac{1}{4}} -\frac{1}{2}}\right \rfloor,
\label{alpha}
\end{equation}
\begin{equation}
\beta_{\rm safe}=\frac{X_{\rm d}(v_{\ell, n}) +g_{ n}}{(\alpha_{\rm safe}+1)b \tau^{2}}-\frac{\alpha_{\rm safe}}{2}.
\label{beta}
\end{equation}

The safe speed in the model by Krau{\ss} {\it et al.}~\cite{Kra,Kra_PhD} provides   collision-less motion of  vehicles
if the time gap $g_{ n}/v_{ n}$ between two vehicles is greater than or equal to the time step $\tau$,
 i.e., if $g_{n} \geq v_{n}\tau$~\cite{Kra_PhD}. In the  model, it is assumed that
in some cases, mainly due to
lane changing or merging of  vehicles onto the main road within the  merging region of bottlenecks, 
the space gap $g_{n}$ can become less  than  $v_{n}\tau$. In these critical situations,
the collision-less motion of  vehicles in the model is a result of the 
 second term in (\ref{safe_kkl}) in which some prediction ($v^{\rm (a)}_{\ell}$)
 of the speed of the preceding vehicle 
 at the next time step is used. The related $\lq\lq$anticipation" speed $v^{\rm (a)}_{\ell}$  at the 
next time step   is given by formula
    \begin{eqnarray}
v^{\rm (a)}_{\ell}=   
\max(0, \min(v^{\rm (safe)}_{ \ell, n}, v_{ \ell,n}, g_{ \ell, n}/\tau)-a\tau),
  \label{Safe_Speed2}
\end{eqnarray}
where $v^{\rm (safe)}_{ \ell, n}$ is the safe speed (\ref{Safe_Speed3}), (\ref{SafeKr})--(\ref{beta}) for the preceding vehicle,
$g_{ \ell, n}$ is the space gap in front of the preceding vehicle.
Simulations have shown that formulas (\ref{safe_kkl}), (\ref{Safe_Speed3}),  (\ref{SafeKr})--(\ref{Safe_Speed2}) lead to   
collision-less vehicle motion over a wide range of parameters of  the merging region 
of on-ramp bottlenecks (Appendix~\ref{Models_Bott_Sec}). 
  
   In hypothetical steady states of traffic flow (Fig.~\ref{KKl_Steady} (a)), 
	the safe space gap   $g_{\rm safe}$  is determined
from condition $v=v_{\rm s}$. In accordance with Eqs. (\ref{safe_kkl})--(\ref{Gipps_Safe_Speed}),
 at a given $v$ in steady traffic states $v=v_{\ell}$
 for the safe speed $v_{\rm s}$ we get
     \begin{equation}
v_{\rm s}=g_{\rm safe}/ \tau,
  \label{Safe_Speed_g}
\end{equation}
 and, therefore, 
      \begin{equation}
 g_{\rm safe}=v \tau.
  \label{Safe_Speed_g2}
\end{equation}
   Thus, for hypothetical steady states of traffic flow $\tau_{\rm safe}=\tau=1$ s.

\begin{table}
\caption{Model parameters of vehicle motion in road lane  used in simulations of the main text}
\label{table_parameters}
\begin{center}
\begin{tabular}{|l|}
\hline
$\tau_{\rm safe}   = \tau=$ 1 s, $d = 7.5 \  \rm m/\delta x$, \\
$\delta x=$ 0.01 m, $\delta v= 0.01 \  {\rm ms^{-1}}$, $\delta a= 0.01 \  {\rm ms^{-2}}$, \\
$v_{\rm free}= 30 \ {\rm ms^{-1}}/\delta v$, $b = 1 \ {\rm ms^{-2}}/\delta a$, 
$a=$ 0.5 ${\rm ms^{-2}}/\delta a$, \\
$k=$ 3, $p_{1}=$ 0.3,  $p_{b}=   0.1$,  $p_{a}=   0.17$, $p^{(0)}= 0.005$, \\
$p_{\rm 0}(v_{n})=0.575+ 0.125\min{(1, v_{n}/v_{01})}$, \\
$p_{\rm 2}(v_{n})=0.48+ 0.32\Theta{( v_{n}-v_{21})}$, \\
$v_{01} = 10 \ {\rm ms^{-1}}/\delta v$, $v_{21} = 15 \ {\rm ms^{-1}}/\delta v$, \\
  $a^{(0)}= 0.2a$,   $a^{(\rm a)}=a^{(\rm b)}= a$. \\  
   \hline
\end{tabular}
\end{center}
\end{table}
\vspace{1cm}

In~\cite{Kerner2017A} it has been shown that the  Kerner-Klenov   model
(\ref{final})--(\ref{gap_formula}), (\ref{Syn_Gap}), (\ref{Syn_Gap2}), 
(\ref{noise_CA})--(\ref{final_P_1}), (\ref{safe_kkl}), (\ref{Safe_Speed3}), 
(\ref{SafeKr})--(\ref{Safe_Speed2})
is  
 a Markov chain:
  At time step $n+1$, values of model variables  $v_{ n+1}$, $x_{ n+1}$,   and  $S_{ n+1}$
 are calculated based   only on their values $v_{ n}$, $x_{ n}$,   and $S_{ n}$
 at step $n$.

It should be noted that after   the Kerner-Klenov car-following model with indifference zones in car-following
based on the three-phase theory (dashed region in Fig.~\ref{Eco_ACC}) 
has been introduced~\cite{KKl} (as mentioned, in this review article a discrete version
of this model of Ref.~\cite{KKl2009A} has been used),
a number of different traffic flow   models incorporating some of the hypotheses of the three-phase  
 theory  have been developed and many   results with the use of these models
have been found (e.g.,~\cite{Davis2004B9,Davis2014C,KKl2003A,Kerner_Review2_Int1,KKH,KKH1,KKHR}, \cite{Davis2003B,Davis2006A,Davis2006B,Davis2007A,Davis2008A,Gao2007,Wu2009,Kimathi2012B}, \cite{Hausken2015A,He2009,Hoogendoorn20088,Jiang2004A,Jiang2005A}, \cite{Jiang2005B,Jiang2007A,JiangHu2014B,Rui2015C}, \cite{Jiang2017A_1,Jiang2007B,Rui2015D,Jin2011,Jin2010,Kerner1998D,KKl2006C,KKHS2013_Int1,KKS2011_Int1},
\cite{KKS2014A_Int1,KKW,Klenov,Klenov2,Kokubo,KnorrSch2013A,Lee_Sch2004A}, 
\cite{LeeKim2011,Neto2011,Qian2017A,Pott2007A,ReKl,ReKl_new},  \cite{Rehborn2011A,Rehborn2011B,Rehborn2014B,Rehborn2008A,Rempe2016A,Rempe2017A}, \cite{Siebel2006,Tian2009,Tian2012,Tian2016AA_Int1,Rui2015B_Int1,TianLi2016B_Int1},  \cite{Jiang2007C,Wu2008,XiangZhengTao2013A,YangLu2013A,Yang2018A,Han-Tao_Zhao2020A,Xiaojian_Hu2020A,Jun-Wei_Zeng2019A,Xiaojian_Hu2019A}).
A review of some of these models that are related to the class of
 cellular automation models  
has been recently done by Tian et al.~\cite{Tian2018A}.

\section{Discrete Version of Classical ACC Model \label{Cla_ACC_S}}

In simulations of the classical ACC-model (\ref{ACC_General}), 
as in the model of human driving vehicles (Appendix~\ref{Identical_KKl_Up}) we use 
  the discrete time $t=n\tau$, where $n=0,1,2,..$; $\tau=$1 s is   time step.
Therefore, the space gap to the preceding vehicle is equal to $g_{n}=x_{\ell, n}- x_{n}-d$
and the relative speed is given by $\Delta v_{n}= v_{\ell, n}-v_{n}$,
  where $x_{n}$ and $v_{n}$ are coordinate and speed of the ACC-vehicle, $x_{\ell, n}$ and $v_{\ell, n}$ are coordinate
and speed of the preceding vehicle, $d$ is   the vehicle length that is assumed the same one
for  autonomous driving and human driving vehicles.
Correspondingly, the classical model of the dynamics of ACC-vehicle (\ref{ACC_General}) can be rewritten as follows~\cite{Kerner_Review3_Int1,Kerner2017A}:  
 \begin{equation}
 a^{\rm (ACC)}_{n} = K_{1}(g_{n}-v_{n}\tau^{\rm (ACC)}_{\rm d})+K_{2} (v_{\ell,n}-v_{n}).
\label{ACC_dynamics_Eq}
\end{equation}

The ACC vehicles move in accordance with Eq.  (\ref{ACC_dynamics_Eq}) where, in
addition, 
 the following  formulas are used:
\begin{equation}
\label{next1_ACC}
 v^{\rm (ACC)}_{{\rm c}, n} = v_{n}+\tau \max(-b_{\rm max},
\min(\lfloor a^{\rm (ACC)}_{n} \rfloor, a_{\rm max})),  
\end{equation}
\begin{equation}
 v_{n+1} = \max(0, \min({v_{\rm free}, v^{\rm (ACC)}_{{\rm c},n}, v_{{\rm s},n} })),
\label{next2_ACC}
\end{equation} 
 $\lfloor z \rfloor$ denotes the 
integer part  of $z$.
Through the use of formula (\ref{next1_ACC}),  acceleration and   deceleration
of the ACC vehicles are limited by 
some maximum   acceleration $a_{\rm max}$ and  maximum deceleration
$b_{\rm max}$, respectively.
 Owing to the formula  (\ref{next2_ACC}),
the speed of the ACC vehicle $v_{n+1}$
 at   time step $n+1$ is limited by the 
maximum speed in free flow $v_{\rm free}$  and by the safe speed
 $v_{{\rm s},n}$  to avoid
 collisions between vehicles.
The maximum speed in free flow $v_{\rm free}$ and the safe speed
 $v_{{\rm s},n}$ are chosen, respectively, the same as those in the
 microscopic model of human driving vehicles (Appendix~\ref{KKl_Model_Ap}).
It should be noted that the model of ACC-vehicle merging from the on-ramp
onto the main road is similar to that for  human driving vehicles 
(see Appendix~\ref{On_ACC_Bott_Sec}).

 Simulations show that the use of the safe speed in formula
  (\ref{next2_ACC}) does not influence on the dynamics
of the ACC vehicles (\ref{ACC_General}) in {\it free flow} outside  the bottleneck.
However, due to vehicle merging from the on-ramp onto the main road,
  time headway of the vehicle to the preceding vehicle
can be considerably smaller than $\tau^{\rm (ACC)}_{\rm d}$. Therefore,
formula
  (\ref{next2_ACC}) allows us to avoid collisions
of the ACC vehicle with the preceding vehicle in such dangerous situations. Moreover,
very small values of time headway can occur in  congested traffic; formula
  (\ref{next2_ACC}) prevents vehicle collisions in these cases also.

 \section{Discrete Version of   TPACC Model \label{App_TPACC_Model}} 

The model for human driving   
vehicles~\cite{KKl,KKl2003A,KKl2009A} used in all simulations is discrete in time
(Appendix~\ref{KKl_Model_Ap}). Therefore, we simulate TPACC-model (\ref{TPACC_main5})
with    discrete time $t_{\rm n}=\tau n, \ n=0,1,...$.
Respectively, TPACC-model (\ref{TPACC_main5}) should be rewritten as follows:
\begin{equation}
a^{\rm (TPACC)}_{n}=  
  \left\{
\begin{array}{ll}
K_{\rm \Delta v}(v_{\ell, n}-v_{n}) &  \textrm{at $g_{n} \leq G_{ n}$} \\ 
K_{1}(g_{n}-v_{n}\tau_{\rm p})+K_{2} (v_{\ell,n}-v_{n}) &  \textrm{at $g_{n}> G_{ n}$}, \\ 
\end{array} \right.
\label{TPACC_main}  
 \end{equation}
where  $G_{ n}=v_{n}\tau_{\rm G}$ and  $\tau_{\rm p}<\tau_{\rm G}$.

 \subsection{Safety Conditions \label{Safe_TPACC_Sec}}

When $g_{n} < g_{\rm safe, n}$, 
the TPACC-vehicle should move in accordance with some safety conditions
 to avoid collisions between vehicles (Fig.~\ref{Eco_ACC}). A collision-free  
TPACC-vehicle motion is  described as made in~\cite{Kerner_Review3_Int1} 
for the classical model of ACC:  (\ref{next1_TPACC}) (\ref{next2_TPACC})
\begin{equation}
\label{next1_TPACC}
 v^{\rm (TPACC)}_{{\rm c}, n} = v_{n}+\tau \max(-b_{\rm max},
\min(\lfloor a^{\rm (TPACC)}_{n} \rfloor, a_{\rm max})),  
\end{equation}
\begin{equation}
 v_{n+1} = \max(0, \min({v_{\rm free}, v^{\rm (TPACC)}_{{\rm c},n}, v_{{\rm s},n}})),
\label{next2_TPACC} 
\end{equation}
 where   the TPACC  acceleration and deceleration 
  are limited by   $a_{\rm max}$ and   
$b_{\rm max}$, respectively;
the speed   $v_{n+1}$  (\ref{next2_TPACC})
 at   time step $n+1$ is limited by the 
maximum speed   $v_{\rm free}$  and by the safe speed
 $v_{{\rm s},n}$ that have been chosen, respectively, the same as those in the
  model of human driving vehicles; $\lfloor z \rfloor$ denotes the 
integer part  of $z$. 

\subsection{$\lq\lq$Indifference Zone'' in Car-Following}

In accordance with Eq.~(\ref{next2_TPACC}),  condition
		$v^{\rm (TPACC)}_{{\rm c},n} \leq v_{{\rm s},n}$ is equivalent to condition
	$g_{n}\geq g_{\rm safe, n}$.    
		Under this condition,  
  from the TPACC model (\ref{TPACC_main})--(\ref{next2_TPACC}) it follows
 that when      time headway $\tau^{\rm (net)}_{n}=g_{n}/v_{n}$
 of the TPACC-vehicle   to the preceding vehicle is within the range
\begin{eqnarray}
\tau_{{\rm safe},n}\leq \tau^{\rm (net)}_{n}\leq \tau_{\rm G}, 
\label{TPACC_main_range_app} 
 \end{eqnarray}
the acceleration (deceleration) of the TPACC-vehicle does not depend on
  time headway. In (\ref{TPACC_main_range_app}),
	\begin{eqnarray}
\tau_{{\rm safe},n}=g_{{\rm safe}, n}/v_{n} 
\label{TPACC_main_tau_s_app} 
 \end{eqnarray}
	 is a safe time headway
	and it is assumed that $v{_n}>0$.    
	The safe gap $g_{{\rm safe}, n}$ in formula (\ref{TPACC_main_tau_s_app}) is found as follows.   
The value $g_{{\rm safe}, n}$ is found as a  solution of equation	 
			\begin{eqnarray}
v^{\rm (TPACC)}_{{\rm c},n} = v_{{\rm safe}}(g_{{\rm safe}, n}, v_{\ell,n}, v^{(\rm a)}_{\ell}),
\label{TPACC_main_v_s_app} 
 \end{eqnarray}
in which   the value
	$v^{\rm (TPACC)}_{{\rm c},n}$ is given by	(\ref{next1_TPACC}) and 
	the function $v_{{\rm safe}}(g_{{\rm safe}, n}, v_{\ell,n}, v^{(\rm a)}_{\ell})$ is determined
	by formula (\ref{safe_kkl})  
	where $g_{n}$ is replaced by $g_{{\rm safe}, n}$.

	In accordance with
	(\ref{TPACC_main_range_app}), for the TPACC model (\ref{TPACC_main})--(\ref{next2_TPACC})
there is {\it no} 
fixed 	desired time headway to the preceding vehicle     
  (Fig.~\ref{Eco_ACC}). This means that in the TPACC model (\ref{TPACC_main})--(\ref{next2_TPACC})
		there is 	$\lq\lq$indifference zone'' in the choice of time headway in car-following.
This is in contrast with 
  the classical ACC model (\ref{ACC_General})  
	for which there is a fixed desired time headway  in car-following.
 It should be noted that formula
	(\ref{TPACC_main_range_app}) for the indifference zone in time headway of TPACC
	is a discrete version of
	formula (\ref{TPACC_main_range}) for the indifference zone in time headway of TPACC discussed in the main text.
	
	\subsection{Operating Points}
	
\begin{figure}
\begin{center}
\includegraphics*[scale=.55]{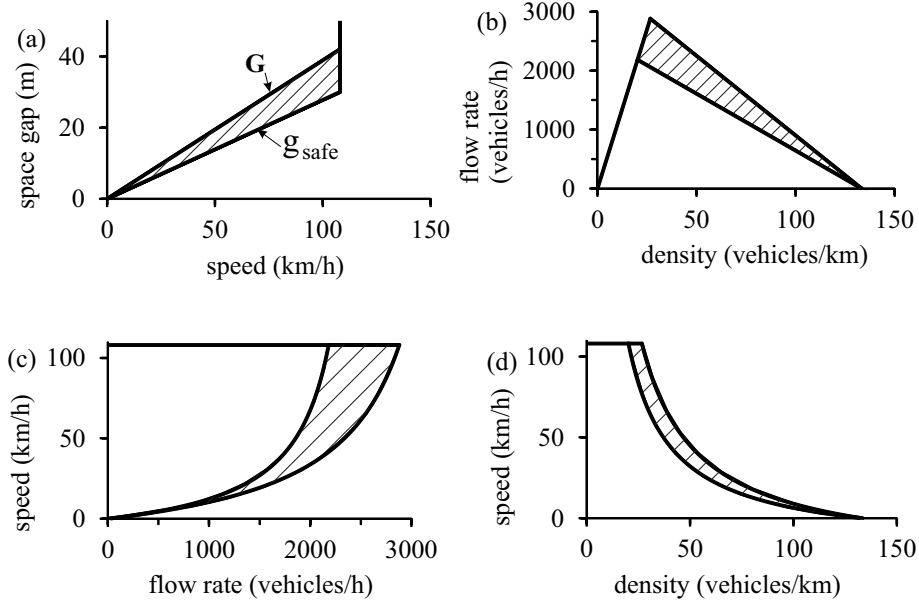}
\end{center}
\caption[]{Operating points of  TPACC model (\ref{TPACC_main})--(\ref{next2_TPACC})    
	presented in the space-gap--speed   (a), flow--density (b),
speed--flow (c), and speed--density (c) planes.  
 Model parameters
$\tau_{\rm G}=$1.4 s, $\tau_{\rm p}=$1.3 s, $\tau_{\rm safe}=$1 s,
$v_{\rm free}=30$m/s (108 km/h),
vehicle length (including the mean space gap between vehicles stopped
 within a wide moving jam) $d=$7.5 m.
}
\label{Operating_TPACC}
\end{figure}

	From formula for the safe speed $v_{{\rm s},n}$ in
(\ref{next2_TPACC}) that is  given in  Appendix~\ref{Safe_speed_kkl}, we find that
 the safe time headway $\tau_{{\rm safe},n}$
in (\ref{TPACC_main_range}) for the operating points  of TPACC model (\ref{TPACC_main})--(\ref{next2_TPACC})
is a constant value that
is equal to $\tau_{\rm safe}=\tau=$1 s. In operating points of TPACC
model (\ref{TPACC_main})--(\ref{next2_TPACC}),  
	$a^{\rm (TPACC)}=0$; respectively, $v=v_{\ell}$,    $g_{\rm safe}(v) \leq g \leq G(v)$,
	and $v=v_{\rm free}$ at $g > G(v)$, where  $g_{\rm safe}(v)=v \tau$,
	$G(v)=v \tau_{\rm G}$.
The operating points of the   TPACC model (\ref{TPACC_main})--(\ref{next2_TPACC}) 
   cover a 2D-region in the space-gap--speed  plane (dashed 2D-region in Fig.~\ref{Operating_TPACC} (a)).
  The inequalities
 $v \leq  v_{\rm free}$, $g\leq G(v)$, and $g \geq   g_{\rm safe}(v)$
 define a 2D-region in  
the space-gap--speed plane (Fig.~\ref{Operating_TPACC} (a))
 in which the operating points exist for the
 discrete version of TPACC model (\ref{TPACC_main})--(\ref{next2_TPACC}).

It should be noted that the speed $v$ and space gap $g$
   are integer in the discrete version of  TPACC model (\ref{TPACC_main})--(\ref{next2_TPACC}).
	Therefore,
   the operating points do not form a continuum in the space-gap--speed  plane
   as they do in the continuum version of TPACC model (\ref{TPACC_main5}).

From Fig.~\ref{Operating_TPACC}, we can see that under conditions $g_{\rm safe}(v) \leq g \leq G(v)$
 for each given speed $v>0$ of TPACC 
 there is no fixed time headway 
 to the preceding vehicle in   operating points of the TPACC model 
	 (dashed 2D-regions in Fig.~\ref{Operating_TPACC}),
	as explained in Sec.~\ref{TPACC_St_S} (Fig.~\ref{Eco_ACC}).
	
	The discretization interval of 
	TPACC acceleration (deceleration) made in TPACC model (\ref{TPACC_main})--(\ref{next2_TPACC})
	is chosen to be an extremely small value that is 
	equal to $\delta a=$0.01 ${\rm m/s^{2}}$ (see   Appendix~\ref{KKl_Model_Ap}).
	 Therefore, the  maximum value of a small round down of $a^{\rm (TPACC)}_{n}$ in 
	Eqs.~(\ref{next1_TPACC}), (\ref{next2_TPACC})
	through the application of the floor operator  
	$\lfloor a^{\rm (TPACC)}_{n} \rfloor$   is less than 0.01 ${\rm m/s^{2}}$ and it 
	is, therefore, negligible.  We have tested that {\it
	no} conclusions  about physical features of TPACC dynamic behavior have been changed, when the continuum in space model of human driving vehicles of Ref.~\cite{KKl2003A}  and, respectively,
	the continuum in space  TPACC model version (without the the floor operator) 
	is used (the reason for the use of the discrete in space
	model for human driving vehicles of Ref.~\cite{KKl2009A}
	that leads to Eqs.~(\ref{next1_TPACC}), (\ref{next2_TPACC}) 
	has been explained in~\cite{KKl2009A} as well as in Appendix~A of the book~\cite{Kerner2017A}).
	
	Simulations show that
 the use of the safe speed in formula  (\ref{next2_TPACC})
	does not influence on the dynamics
of the TPACC vehicles (\ref{TPACC_main}) in   free flow  outside  the bottleneck.
However, formula  (\ref{next2_TPACC}) allows us to avoid collisions
of the TPACC vehicle with the preceding vehicle in   dangerous situations
that can occur at the bottleneck as well as in   congested traffic.

\section{Model of ACC with Combination of   Dynamic Features of Classical ACC and TPACC \label{Models_ACC_TP_Sec}} 

The ACC-acceleration   $a^{\rm(C)}_{n}$ in a discrete in time version of the model of ACC
(\ref{Combain01})--(\ref{Combain05}) of Sec.~\ref{Dyn_Rules_S}  
reads as follows: 
\begin{eqnarray}
a^{\rm(C)}_{n}=
\left\{\begin{array}{ll}
\tilde a^{\rm(C)}_{n} &  \textrm{at $g_{n} \leq G^{\rm(C)}_{ n}$,} \\
a^{\rm(ACC)}_{n} &  \textrm{at $g_{n}> G^{\rm(C)}_{ n}$}, \\
\end{array} \right.
\label{Combain01_App}
\end{eqnarray}
where 
\begin{equation}
\tilde a^{\rm(C)}_{n}=a^{\rm(2D)}_{n}(1-p^{\rm(C)})+a^{\rm(ACC)}_{n}p^{\rm(C)},
\label{Combain02_App}
\end{equation}
\begin{equation}
a^{\rm(2D)}_{n}=K_{\rm \Delta v}\Delta v_{ n},
\label{Combain03_App}
\end{equation}
\begin{equation}
a^{\rm(ACC)}_{n}=K_{1}(g_{n}-v_{ n}\tau_{\rm p})+ K_{2}\Delta v_{ n},
\label{Combain04_APP}
\end{equation}
\begin{equation}
G^{\rm(C)}_{ n}=G_{ n}(1-p^{\rm(C)})+v_{ n}\tau_{\rm p}p^{\rm(C)}.
\label{Combain05_App}
\end{equation}

When $g_{n} < g_{\rm safe, n}$, 
the ACC-vehicle (\ref{Combain01_App})--(\ref{Combain05_App}) should move in accordance with some safety conditions
 to avoid collisions between vehicles. A collision-free  
ACC-vehicle motion is  described as made  
for the classical model of ACC:  
\begin{equation}
v_{ n+1}=\max(0, \min({v_{\rm free}, v^{\rm(C)}_{{\rm c}, n}, 
v_{{\rm s},n} })).
\label{Combain07_App}
\end{equation}
where
\begin{equation}
v^{\rm(C)}_{{\rm c}, n}=v_{ n}+\tau \max(-b_{\rm max}, \min( 
\lfloor a^{\rm(C)}_{n} \rfloor , a_{\rm max})).
\label{Combain06_App}
\end{equation}

\section{Model of On-Ramp Bottleneck \label{Models_Bott_Sec}}

   \begin{figure}
\begin{center}
\includegraphics*[scale=.7]{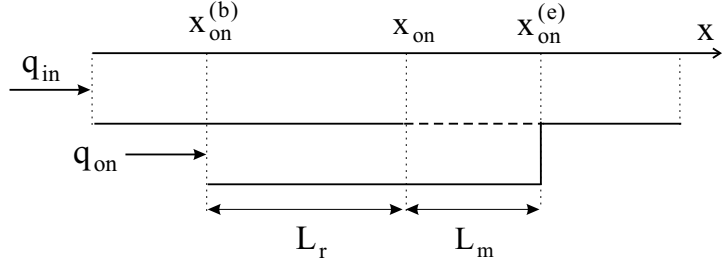}
\end{center}
\caption[]{Model of on-ramp   bottleneck on   single-lane   road.  
}
\label{KKl_Bottlenecks2}
\end{figure}

An on-ramp bottleneck consists of two parts (Fig.~\ref{KKl_Bottlenecks2}):
\begin{description}
 \item [(i)] The merging region of  length  $L_{\rm m}$ 
 where vehicle can  
merge onto the main road from the on-ramp lane.
 \item [(ii)] A part of the on-ramp lane of length $L_{\rm r}$
 upstream of the merging region 
where vehicles
move in accordance with  the model of Appendix~\ref{KKl_Model_Ap}. The maximal speed of vehicles is 
$v_{\rm free}=v_{\rm free \ on}$. 
\end{description}
At the beginning   of the on-ramp lane ($x=x^{\rm (b)}_{\rm on}$) the flow rate to the on-ramp $q_{\rm on}$ 
is given through boundary conditions
that are the same   as those that determine the flow rate $q_{\rm in}$ at the beginning of the main road
(see Appendix~\ref{Boundary_Ini_Con_S} below).

 \subsection{Model of Vehicle Merging at Bottleneck \label{MB_merging_Sec}}

  \subsubsection{Vehicle Speed Adaptation withing Merging Region of Bottleneck \label{MB_merging_SA_Sec}} 
  
 For the on-ramp bottleneck, when a vehicle is within the merging region of the bottleneck, the vehicle takes into account 
 the space gaps to the preceding vehicles and their speeds
both in the current and target lanes. Respectively, instead of formula (\ref{delta}), 
  in (\ref{final2}) for the speed $v_{{\rm c},n}$ the following formula is used:
      \begin{equation}
v_{{\rm c},n}=\left\{
\begin{array}{ll}
v_{ n}+\Delta^{+}_{ n} &  \textrm{at $g^{+}_{n} \leq G(v_{n}, \hat
v^{+}_{n})$} \\
v_{ n}+a_{ n}\tau &  \textrm{at $g^{+}_{n}>G( v_{n}, \hat
v^{+}_{n})$}, \\
\end{array}\right.
  \label{Lane_Change2_before}
  \end{equation}
       \begin{equation}
\Delta^{+}_{ n}=\max(-b_{ n}\tau, \min(a_{ n}\tau, \ \hat v^{+}_{n}-v_{
n})),
  \label{Lane_Change2_before2}
  \end{equation}
         \begin{equation}
\hat v^{+}_{n}=\max(0, \min(v_{\rm free}, \  v^{+}_{n}+\Delta
v^{(2)}_{r})),
  \label{Lane_Change2_before3}
  \end{equation}
$\Delta v^{(2)}_{r}$ is  constant (see Table~\ref{table_parameters_bottlenecks}). 

Superscripts    $+$   and  $-$  in variables, parameters, and functions 
denote the preceding vehicle and the trailing vehicle 
in the $\lq\lq$target" (neighboring) lane, respectively.
The target lane is the 
lane into which the vehicle wants to change.

The safe speed $v_{{\rm s},n}$ in (\ref{final}), (\ref{final2}) for the vehicle that is the closest one
to the end of the merging region  
is chosen in the form 
\begin{equation}
v_{{\rm s},n}=\lfloor v^{\rm (safe)}(x^{\rm (e)}_{ \rm on}- x_{n}, \ 0) \rfloor
\label{safe_on}
\end{equation}
  (see Table~\ref{table_parameters_bottlenecks}).

  \subsubsection{Safety Conditions for Vehicle   Merging   \label{MB_merging_Safe_Sec}}
 
  Vehicle merging at the bottleneck
 occurs, when     safety conditions ($\ast$) {\it or}   safety conditions  ($\ast \ast$) are satisfied.

   Safety conditions ($\ast$) are as follows:
  \begin{equation}
  \begin{array}{ll}
g^{+}_{n} >\min(\hat  v_{n}\tau , \ G(\hat  v_{n}, v^{+}_{n})), \\
g^{-}_{n} >\min(v^{-}_{n}\tau, \ G(v^{-}_{n},\hat  v_{n})),
\label{merging_a}
\end{array} 
\end{equation}
\begin{equation}
 \hat v_{n}=\min(v^{+}_{n},  \ v_{n}+\Delta v^{(1)}_{r}), 
 \label{A2}
\end{equation}
 $\Delta v^{(1)}_{r}>0$ is constant (see Fig.~\ref{KKl_Bottlenecks2}
and  Table~\ref{table_parameters_bottlenecks}).
 
 Safety conditions  ($\ast \ast$) are as follows:
\begin{equation}
x^{+}_{n}-x^{-}_{n}-d > g^{\rm (min)}_{\rm target},
\label{merging_b}
\end{equation}
where
\begin{equation}
g^{\rm (min)}_{\rm target}=\lfloor \lambda_{\rm b}  v^{+}_{n} +d \rfloor,
\label{merging_b2}
\end{equation}
 $\lambda_{\rm b}$ is constant. In addition to conditions (\ref{merging_b}), the safety condition  ($\ast \ast$) includes the condition that
the vehicle should pass the midpoint 
\begin{equation}
x^{\rm (m)}_{n}=\lfloor (x^{+}_{n}+x^{-}_{n})/2 \rfloor
\label{midpoint_f}
\end{equation}
between two neighboring vehicles in the target lane, i.e.,   conditions 
  \begin{equation}
 \begin{array}{ll}
x_{n-1}< x^{\rm (m)}_{n-1} \  \textrm{and} \
 x_{n} \geq x^{\rm (m)}_{n} \\
\ \textrm{or} \\
x_{n-1} \geq x^{\rm (m)}_{n-1} \  \textrm{and} \
 x_{n} < x^{\rm (m)}_{n}.
\end{array} 
\label{mid2}
\end{equation}
should also be satisfied.

  \subsubsection{Speed and Coordinate of Vehicle  after Vehicle Merging   \label{MB_merging_After_Sec}}
 
 The vehicle speed after vehicle merging is equal to
     \begin{equation}
v_{n}=\hat v_{n}.
  \label{Lane_Change2_after}
  \end{equation}
  
  Under conditions ($\ast $),
the vehicle coordinates $x_{n}$  remains the
same.
 Under conditions ($\ast \ast$), the vehicle coordinates $x_{n}$ is equal to 
     \begin{equation}
x_{n} = x^{\rm (m)}_{n}.
  \label{Lane_Change2_after2}
  \end{equation}

\subsection{Merging of ACC-Vehicle or TPACC-Vehicle  at On-Ramp Bottleneck \label{On_ACC_Bott_Sec}}

 Here we consider rules of the merging of an ACC-vehicle  at the on-ramp bottleneck
presented in~\cite{Kerner2017A} and used in simulations.   The same rules have also been used in simulations
of the merging of an TPACC-vehicle from the on-ramp lane onto the main road
at the bottleneck.

In the on-ramp lane, an ACC-vehicle or an TPACC-vehicle 
moves in accordance with  the ACC model
(\ref{ACC_dynamics_Eq})--(\ref{next2_ACC}) or in accordance with
  the TPACC-model (\ref{TPACC_main})--(\ref{next2_TPACC}), respectively. The maximal speed of the
ACC vehicle or the TPACC-vehicle in the on-ramp lane is
$v_{\rm free}=v_{\rm free \ on}$. 
The safe speed $v_{{\rm s},n}$ in (\ref{next2_ACC}) for the ACC-vehicle and in (\ref{next2_TPACC})
for the TPACC-vehicle
that is the closest one
to the end of the merging region  
is the same as that for human driving vehicles that is
given by formula (\ref{safe_on}).

  An ACC-vehicle or an TPACC-vehicle merges from the on-ramp lane onto the main road,
   when   some   safety conditions ($\ast$) {\it or}   safety conditions  ($\ast \ast$)
	are satisfied for the ACC-vehicle or the TPACC-vehicle.
Safety conditions ($\ast$) for ACC-vehicles and TPACC-vehicles are as follows:
  \begin{equation}
  \begin{array}{ll}
g^{+}_{n} >\hat  v_{n}\tau, \quad
g^{-}_{n} >v^{-}_{n}\tau,
\label{merging_a_ACC}
\end{array} 
\end{equation}
where $\hat v_{n}$ is given by formula (\ref{A2}).
Safety conditions  ($\ast \ast$) 
are given by formulas
(\ref{merging_b})--(\ref{mid2}), i.e., they are the same as those for human driving vehicles. Respectively,
as for human driving vehicles,
the ACC-vehicle speed and its coordinate  or
the TPACC-vehicle speed and its coordinate after the ACC-vehicle or the
TPACC-vehicle has merged from the on-ramp onto the main road are determined by formulas
  (\ref{Lane_Change2_after})
	and (\ref{Lane_Change2_after2}).

 \begin{table}
\caption{Parameters of model of on-ramp bottleneck   used in simulations of the main text}
\label{table_parameters_bottlenecks}
\begin{center}
\begin{tabular}{|l|}
\hline
$\lambda_{\rm b}=$ 0.75,  
   $v_{\rm free \ on}=22.2 \ {\rm ms^{-1}}/\delta v$,   \\
   $\Delta v^{\rm (2)}_{\rm r}=$ 5 \  ${\rm ms^{-1}}/\delta v$, \\
   $L_{\rm r}=1 \ {\rm km}/\delta x$,   $\Delta v^{\rm (1)}_{\rm r}=10 \ {\rm ms^{-1}}/\delta v$,  \\
   $L_{\rm m}=$ 0.3  \ ${\rm km}/\delta x$. \\
   \hline
\end{tabular}
\end{center}
\end{table}
\vspace{1cm}

 \section{Boundary   Conditions for Mixed Traffic Flow \label{Boundary_Ini_Con_S}}

Open boundary conditions are applied.
 At the beginning of the road new vehicles are generated one after another in each of the lanes
of the road at time instants
\begin{equation}
t^{(m)}=\tau \lceil m \tau_{\rm in}/\tau \rceil, \ m=1,2,....
\label{t_new_KKl}
\end{equation}
In (\ref{t_new_KKl}), $\tau_{\rm in}=  1/q_{\rm in}$,
 $q_{\rm in}$ is the flow rate
in the incoming boundary flow per lane, $\lceil z \rceil$ denotes
 the nearest integer greater than
or equal to $z$.
Human driving and autonomous driving vehicles are  randomly generated  at the beginning of the road
with the rates  related to  chosen values of the flow rate $q_{\rm in}$ and 
the percentage $\gamma$ of the autonomous driving vehicles in mixed traffic flow:
(i) At time instant $t^{(m)}$ (\ref{t_new_KKl}), a new vehicle is human driving vehicle, when 
condition  
\begin{equation}
r_{2}\geq \gamma/100 
\end{equation}
is satisfied where $r_{2}={\rm rand}(0,1)$.
(ii) At time instant $t^{(m)}$ (\ref{t_new_KKl}), a new vehicle is autonomous driving vehicle, when 
the opposite condition  
\begin{equation}
r_{2}< \gamma/100 
\end{equation}
is satisfied. 
The same procedure of random generation of human driving and autonomous driving vehicles is applied at
the beginning of the on-ramp lane. In this case, however, in (\ref{t_new_KKl})
the flow rate $q_{\rm in}$ should be replaced by the on-ramp inflow rate $q_{\rm on}$.

A new vehicle appears on the road 
only if the distance from the beginning of the road ($x=x_{\rm b}$)
to the position $x=x_{\ell, n}$ of the farthest upstream vehicle on the road
is not smaller than the safe distance $v_{\ell, n} \tau+d$:
\begin{equation}
x_{\ell, n}-x_{\rm b} \geq v_{\ell, n} \tau+d,
\label{Coordinate_preceding_KKl}
\end{equation}
where $n=t^{(m)}/\tau$. Otherwise,  condition (\ref{Coordinate_preceding_KKl}) is checked
at  time   $(n+1)\tau$ that is the next
 one
 to time  $t^{(m)}$ (\ref{t_new_KKl}), and so on, until
the condition (\ref{Coordinate_preceding_KKl}) is satisfied.
Then the next vehicle appears on the road. After this occurs, the number  $m$ in (\ref{t_new_KKl})
is increased by 1.

The speed  $v_{ n}$ and coordinate $x_{n}$
of the new vehicle
 are  
\begin{eqnarray}
\begin{array}{ll}
v_{ n}= v_{\ell, n}, \\
x_{ n}={\rm max}(x_{\rm b}, x_{\ell, n}-\lfloor{v_{ n}\tau_{\rm in}}\rfloor).
\end{array}
\label{Coordinate_in_KKl}
\end{eqnarray}
The flow rate $q_{\rm in}$ is chosen to have the value $v_{\rm free}\tau_{\rm in}$ integer.
In the initial state ($n=0$), all  vehicles have the  free flow speed $v_{ n}=v_{\rm free}$
 and they are positioned at  space intervals  $x_{\ell, n}-x_{n}=v_{\rm free}\tau_{\rm in}$.

After a vehicle has reached  the end of the road 
it is removed.
Before this occurs, the farthest downstream vehicle maintains its speed and lane.
For the vehicle following the farthest downstream one, the $\lq\lq$anticipation" speed 
$v^{\rm (a)}_{ \ell}$ in (\ref{safe_kkl})  is  equal to the speed of the farthest downstream vehicle.

{\bf Acknowledgments:}

I would like to thank Sergey Klenov for help and useful suggestions.
We thank our partners for their support in the project
   $\lq\lq$MEC-View -- Object detection for autonomous driving based on Mobile Edge Computing",
    funded by the German Federal Ministry of Economic Affairs and Energy.

\end{document}